\documentclass{article}
\usepackage[utf8]{inputenc}
\usepackage[margin=1in]{geometry}
\usepackage[sort,square,numbers]{natbib}
\usepackage{hyperref}
\usepackage{graphicx}
\usepackage[dvipsnames]{xcolor}
\usepackage{soul}
\usepackage{tabularx}
\usepackage{booktabs}
\usepackage{subcaption}
\usepackage{lscape}
\usepackage{amsmath}
\usepackage{bm}
\usepackage{xfrac}
\usepackage{algorithm}
\usepackage{algpseudocode}
\providecommand{\keywords}[1]
{
  \small	
  \textbf{\textit{Keywords---}} #1
}

\title{Carthago Delenda Est: Co-opetitive Indirect Information Diffusion Model for Influence Operations on Online Social Media}
\author{Jwen Fai Low, Benjamin C. M. Fung, Farkhund Iqbal, and Claude Fachkha}
\date{}

\begin{document}

\maketitle

\begin{abstract}
For a state or non-state actor whose credibility is bankrupt, relying on bots to conduct non-attributable, non-accountable, and seemingly-grassroots-but-decentralized-in-actuality influence/information operations (info ops) on social media can help circumvent the issue of trust deficit while advancing its interests. Planning and/or defending against decentralized info ops can be aided by computational simulations in lieu of ethically-fraught live experiments on social media. In this study, we introduce \textit{Diluvsion}, an agent-based model for contested information propagation efforts on Twitter-like social media. The model emphasizes a user's belief in an opinion (stance) being impacted by the perception of potentially illusory popular support from constant incoming floods of indirect information, floods that can be cooperatively engineered in an uncoordinated manner by bots as they compete to spread their stances. Our model, which has been validated against real-world data, is an advancement over previous models because we account for engagement metrics in influencing stance adoption, non-social tie spreading of information, neutrality as a stance that can be spread, and themes that are analogous to media's framing effect and are symbiotic with respect to stance propagation. The strengths of the \textit{Diluvsion} model are demonstrated in simulations of orthodox info ops, e.g., maximizing adoption of one stance; creating echo chambers; inducing polarization; and unorthodox info ops, e.g., simultaneous support of multiple stances as a Trojan horse tactic for the dissemination of a theme.
\end{abstract}

\keywords{agent-based modeling, information diffusion, information propagation, information operations, influence operations, psychological operations, rumor spreading, rumor management, social cybersecurity, persuasive computing}

\section{Introduction}
\label{sec:introduction}
\par Artificial intelligence (AI), alternatively known as machine learning, has advanced on all fronts --- behavioral, textual, audio, visual, and video --- to the point where comprehensive and convincing digital personas can be crafted from mere lines of code (Section \ref{sec:digital-personas}). As a result, the detection of information operations (defined in Section \ref{sec:info-ops-definition}) as a method for combating the spread of undesirable narratives by adversarial agents on online social networks, while still important, becomes increasingly non-viable as the predominant means of achieving information security. A \citeyear{courchesne2021reviewsocialscience} review of research papers on countering influence operations that went as far back as 1972 noted that all 223 studies they found were on user-focused countermeasures (users being consumers of disinformation) \citep{courchesne2021reviewsocialscience}. The authors concluded that ``[b]eyond fact-checking, the research base is very thin'', and called for increased focus on key areas such as ``studying countermeasures that target the creators of disinformation content in addition to studying consumer-facing policies''. In response to this call for action, we propose complementing existing detect-and-correct methods that conduct struggles against undesirable information on a tactical, low-level per-message or per-user basis with a method that facilitates waging war against disinformation from a high-level strategical perspective, where users and messages are seen as a part of a synergistic whole meant to achieve a combined arms outcome. 
\par To that end, we designed \textit{Diluvsion}, an agent-based model for simulating how information operations can perturb opinion evolution in an online social media platform. The model relies on parameters derived from publicly available data on a social media platform to forecast the outcomes of info ops on similar platforms. As such, it can serve as a tool to be used by planners in crafting campaigns to attack disinformation or defend against it without needing to first test the interventions in the real world --- tests that can have unsavory ethical and even geopolitical implications \citep{goel2014facebooktinkersusers, grassegger2018facebooksaysits}. Like many other information diffusion models, \textit{Diluvsion} is inspired by epidemiology but updated to consider new insights on transmission dynamics from the COVID-19 pandemic (Section \ref{sec:classical-models}). But, unlike many models, \textit{Diluvsion} is developed with info ops and bots in mind, meaning that unconventional routes for expanding influence are accounted for and that optimistic assumptions for information diffusion in other arguably marginally more benign contexts are discarded, for example, the ability to select seed nodes to maximize propagation in product advertising \citep{li2015realtimetargetedinfluence, nguyen2016costawaretargetedviral}.
\par While models are often designed to adhere closely to the real world, no model is meant to be a perfect recreation of the real world; design choices reflect aspects that the modelers wish to emphasize. In our case, \textit{Diluvsion} brings into prominence the role played by indirect information, namely social cues \citep{epstein2022howmanyothers} --- engagement metrics and the perception of mass adoption --- in opinion propagation. Repetition is the key to the operation of our model. Much like Cato's brilliant tactic of incessantly playing up Carthage's threat to Rome \citep{kiernan2004firstgenocidecarthage} and calling for Carthage's destruction in the Roman Senate led to the genocide \citep{kiernan2004firstgenocidecarthage} of Carthage, much like repeating a lie can let it take on the mantle of truth \citep{hassan2021effectsrepetitionfrequency, fazio2015knowledgedoesnot}, so does the persistent saturation of an opinion in the media landscape can drive its adoption among the populace \citep{zollmann2019bringingpropagandaback, macleod2018manufacturingconsentvenezuela, fuchs2018propagandahermanchomsky, mullen2010hermanchomskypropaganda, herman2010manufacturingconsentpolitical, herman2000propagandamodelretrospective, parenti1993inventingrealitypolitics}.
\par The contributions of this paper are:
\begin{itemize}
    \item An agent-based co-opetitive information diffusion model for information operations on social media that is novel in its incorporation of concepts not considered by existing models: an infectious neutral stance; themes that must piggyback on stances to infect users and have an enhancing or diminishing effect on the stance's infectivity; non-social tie-based transmission of information; parity in the treatment of bots backing different stances; perception of mass support inferred through engagement metrics such as the number of retweets influencing stance adoption; and agents with heterogeneous roles and activity levels populating the simulation.
    \item In-depth rationale behind the design/modeling choices, demonstrating the empirical evidence that our model is predicated upon and consequently the real-world applicability of our model.
    \item Initial ``validation'' of the model by drawing simulation and graph structure generation parameter values from distributions found in real-world data and further validation that succeeded in reproducing via simulations the distributions from a different empirical dataset. 
    \item Simulations that are in part ablation studies and in part experimentation with info ops tactics. The maneuvers examined include bot-backed minority stances facing off against a non-bot-backed majority stance, polarization/depolarization, and maximizing theme diffusion. The emergent outcomes from implicit cooperation and competition found in these simulations are examined and discussed in the context of their implications for decentralized info ops.
\end{itemize}

\section{Background}
\label{sec:background}
\par As our work occurs at the intersection of different domains, we split our discussion of relevant works into two major sections: information diffusion and influence maximization (Section \ref{sec:historical-models}), and information operations (Section \ref{sec:info-ops}).

\subsection{Information diffusion models}
\label{sec:historical-models}
\par The study of information diffusion --- the process by which information spreads from one entity to another --- has a long history. In the particular case of diffusion on online social networks, the history is sufficiently extensive to allow for a survey \citep{guille2013informationdiffusiononline} of existing work to be published in 2013. The closely related domains of influence maximization and consensus formation have similarly rich histories, which can be seen in a 2021 survey \citep{azaouzi2021newtrendsinfluence}. Influence maximization (IM) studies the problem of locating influential nodes that maximize the spread of information to the entire network. Consensus formation is concerned with determining the manner in which a group of agents can all arrive at the same opinion. For an in-depth discussion of these topics, we refer the readers to the aforementioned surveys. Here, we restrict ourselves to discussing the most prominent and relevant models and the elements from those models that influenced the development of our model, \textit{Diluvsion}. The information presented is also meant to help situate \textit{Diluvsion} among the constellation of predecessor models.

\subsubsection{Information}
\par In the space of diffusion/propagation models, information goes by many different names, including information \citep{liu2016shircompetitiveinformation}, influence \citep{azaouzi2021newtrendsinfluence}, rumor \citep{kostka2008wordmouthrumor}, opinion \citep{quattrociocchi2014opiniondynamicsinteracting}, memes \citep{weng2012competitionmemesworld}, and news \citep{piqueira2020daleykendalmodels}, depending on the context. While we will use these terms interchangeably throughout this paper, we favor the term ``opinion'' due to the fact that the major component of the information being spread in networks within our model, the stance, is polarized and homophilic. Together with non-polarizable themes, they form a unit of information in our model (Section \ref{sec:stance-themes}). That is not to say that other forms of information are not present in the model. Engagement metrics, for instance, are exposed to agents and can influence their decision to adopt a piece of information. 
\par We are aware of works that have studied how real news and fake news spread differently on social media \citep{vosoughi2018spreadtruefalse, zhao2020fakenewspropagates}. The information of interest in this paper, i.e., the information being disseminated, is not generally regarded as ``fact'' (e.g., quantum mechanics and general relativity applied within their respective restricted contexts and not outside) but ``opinion'' (e.g., the dispute over the rarity of scale-free networks in the real-world \citep{holme2019rareeverywhereperspectivesa}), which does not have a clear delineation between true and false. In our model, ``disinformation'' is simply a term of convenience for undesirable information; messages are indistinguishable in terms of veracity, so veracity cannot influence a message's spread. Only stances and themes have influence.

\subsubsection{Uncontested information diffusion}
\label{sec:classical-models}
\par Based on the typology introduced in \citep{guille2013informationdiffusiononline}, information diffusion models can be separated into two categories: explanatory and predictive. Explanatory models aim to ``infer the underlying spreading cascade, given a complete activation sequence.'' \textit{Diluvsion} belongs to the other category, predictive models, that aim to predict ``how a specific diffusion process would unfold in a given network, from temporal and/or spatial points of view by learning from past diffusion traces.'' Predictive models are further divided into graph- and non-graph-based approaches. Graph-based approaches assume that information spreads along the edges in the graph. Non-graph-based approaches ignore network topology and model information diffusion as a process represented by a system of differential equations, in emulation of similar equations in epidemiology \citep{kermack1927contributionmathematicaltheory, kermack1932contributionsmathematicaltheory}. 
\par The two most well-known graph-based models are the independent cascade (IC) \citep{goldenberg2001talknetworkcomplex} and linear threshold (LT) \citep{granovetter1978thresholdmodelscollectivea} models, and they serve as the base models for most influence maximization research. IC models diffusion as a chain reaction in which a node that has adopted a piece of information, designated as being activated, attempts to activate inactive neighboring nodes in the graph with a certain probability. There are no retries and an activated node remains activated forever. In LT, an inactive node becomes activated if the sum of influence from activated neighboring nodes exceeds the node's threshold. Influence takes the form of weighted directed edges between nodes and the threshold is generally drawn from a uniform random distribution. An activated node remains forever activated in LT as well. IC can be seen as sender-centric in contrast to the receiver-centric LT.
\par In non-graph-based models, the model names usually take after the sequence of state transitions permissible within the model, where the states are susceptible (S), infected (I), and recovered (R), which are based on the earliest epidemiological model, SIR \citep{kermack1927contributionmathematicaltheory}. The first explicit link between information and disease was established by \citeauthor{daley1964epidemicsrumours}, who were inspired by the SIR model to create the Daley--Kendall (DK) model for rumor propagation \citep{daley1964epidemicsrumours}. Notable variants of DK include the Maki--Thompson model \citep{maki1973mathematicalmodelsapplications}. Rumor diffusion models have states such as ignorant (I, susceptible), spreader (S, infected), and stifler (S/R, recovered) that mirror their epidemiological counterparts, although these specific terms were not used in the initial work by \citeauthor{daley1964epidemicsrumours}. Later reviews \citep{hethcote2000mathematicsinfectiousdiseases, newman2003structurefunctioncomplex} showed the cross-pollination of ideas between rumor-specific and biological epidemiological models. Subsequent studies on epidemiological diffusion models also innovated upon the transition sequence, e.g., SIRS \citep{shaji2018innovatedsirsmodel} or added additional states, e.g., SIHR \citep{zhao2012sihrrumorspreading} rumor spreading model where H stands for hibernator and SHIR \citep{liu2016shircompetitiveinformation} model for competitive information diffusion on online social media where H stands for hesitated. DK descendants innovated in a similar fashion, e.g., ISCR \citep{muhlmeyer2020modelingsocialcontagion} which added a counter-spreader (C) state.
\par In \textit{Diluvsion}, network topology matters, as in IC and LT, and information generally travels based on the agents' connections. The influence of epidemiological models, SIR, SIS, etc., are in inclusion of non-graph edge routes (circumventing topology) for information diffusion to our model, as well as in the agents' non-durable immunity (SIS). However, agents are never in an uninfected state within our model, as they only ever transition between the three mutually exclusive infected states (stances), negative, neutral, and positive, with infection by non-exclusive themes providing an additional differentiating mechanism akin to viral variants (Section \ref{sec:stance-themes}).
\par The SARS-CoV-2 pandemic has revealed the shortcomings of some graph-based models. In information diffusion, the common assumption is that a link/edge exists between two people \textit{only} if they have social ties (e.g., friends, family, followers, and colleagues). However, in epidemiology, such an assumption is not always made because ties can be formed with complete strangers in many situations. A passerby breathing, coughing, or sneezing in a crowded cafe spreads aerosolizable infectious disease almost as effectively as staying under the same roof with a diseased family member \citep{kang2020internetcafekaraoke}. On social media, many posts are publicly viewable and lively discussions can occur in the replies section of those posts, analagous to a forum or a cafe, and a user may be exposed to the opinions of strangers, i.e., people who are neither followers nor followees, who replied to the post of the user's followee (this idea will be explored in Section \ref{sec:indirect-influence}). Another issue is the strong assumption inherent in designating the recovered/stifler state as an agent's end state in an information epidemic. SARS-CoV-2 immunity is transient. It is not permanent. Reinfection by the same or different variants is a real possibility \cite{pulliam2022increasedrisksarscov2}. Non-permanent resistance to contagions that increases or decreases depending on the infection type is a mechanic implemented in \textit{Diluvsion} to offer an alternative perspective to the strong assumptions of permanent immunity in the existing literature (Section \ref{sec:humans}).

\subsubsection{Competitive information diffusion}
\label{sec:contested-models}
\par Classical information diffusion does not account for information possessing polarity and existing in a contested space. ``Vaccines are safe'' and ``vaccines are dangerous'' are contradictory opinions which a single entity is unlikely to hold simultaneously. There are actors, such as healthcare authorities, with a vested interest in not just rapidly disseminating the opinion that vaccines are safe and having people adopt that opinion but also stemming the adoption of the opposing opinion and having the opposing opinion be dislodged from those who have adopted it. Competitive information diffusion models, also known as opinion formation models, were developed to simulate such scenarios. In biology, competition can be thought of as cross-immunity contagions \citep{wang2019coevolutionspreadingcomplex}, where developing immunity for one decreases the likelihood of being infected by the other.
\par Two broad categories of competition exist based on the type of information being diffused: continuous or discrete. Certain consensus formation mechanisms (e.g., taking the mean or median) are not possible for an information type (e.g., selecting a material for constructing an object, although the problem can be reformulated such that selection is based on continuous-valued attributes such as tensile strength). Elements of continuous and discrete models can be made to work together, as demonstrated by the Biased Voter model \citep{das2014modelingopiniondynamics} that combines DeGroot and voter models.
\par For continuous-valued information, the DeGroot/French--DeGroot \citep{frenchjr.1956formaltheorysocial, degroot1974reachingconsensus} model is widely used for competitive diffusion. In the model, every incoming edge for a node/agent has a weight denoting another agent's influence (a weight of zero is equivalent to not having an edge). There is also a weighted self-loop for the extent to which an agent values its own opinion. At every time-step, an agent updates its opinion by taking the weighted average of all neighboring opinions, including its own. This process is repeated until the values (opinions) are stable. In a strongly connected graph, perfect consensus (all agents share the same opinion) is guaranteed. The Bounded Confidence model, alternatively known as the Hegselmann--Krause model \citep{hegselmann2002opiniondynamicsbounded}, also updates an agent's opinion by averaging others' opinions, except that instead of the opinions of all neighbors, the opinions are those who share similar opinions, with similarity being bounded by some arbitrary value, hence the model's name. Clustering (a form of polarization), instead of perfect consensus, is the likely outcome of the Bounded Confidence model. A review of continuous-valued models, which includes DeGroot, Weighted-Median, Bounded Confidence, and Quantum Game, can be found in \citep{devia2022frameworkanalyzeopinion}.
\par For discrete opinions, one popular model is the voter model, independently developed by \citeauthor{clifford1973modelspatialconflictb} \citep{clifford1973modelspatialconflictb} and by \citeauthor{holley1975ergodictheoremsweaklya} \citep{holley1975ergodictheoremsweaklya}. At every step, the model first chooses an agent and its neighbor, both at random. The agent then adopts the opinion of its neighbor. Another popular model is the social influence model by political scientist \citeauthor{axelrod1997disseminationculturemodel}, designed to explain how global polarization arises due to local homophily. In this model, each agent is defined by its cultural traits, which is a fixed-length string of discrete numbers. Similar to the voter model, the Axelrod model randomly chooses an agent and its neighbor at every step. Then, with a probability equal to their cultural similarity, the agent will update one of its traits that differs from its neighbor's. The Axelrod model can be considered an extension of the voter model.
\par One natural extension to voter models is to have an agent's opinion updates be guided by the majority instead of just any random neighbor. The majority rule model by \citeauthor{chen2005majorityruledynamics} selects at random a flippable spin (a changeable binary opinion) and have the spin adopt the majority state of its interaction neighborhood \citep{chen2005majorityruledynamics}. \citeauthor{fu2008coevolutionarydynamicsopinions} incorporated minority avoidance into the majority rule model by having agents occasionally sever ties with those holding a minority opinion \citep{fu2008coevolutionarydynamicsopinions}. Other information diffusion models studied a different form of minority avoidance \citep{takeuchi2015publicopinionformation, ross2019aresocialbots}, specifically the ``spiral of silence'' \citep{noelle-neumann1974spiralsilencetheory}, which theorizes that people are reluctant to disagree with the majority, hence restricting themselves from expressing a non-conforming opinion, ultimately leading to the continued domination of the majority voice. The majority voice can be illusory, created by a vocal minority.
\par There are many ways in which competition can be modeled using epidemiological models. For two competing information contagions, a winner-takes-all situation emerges, where the stronger contagion dominates completely instead of just gaining a majority across all graph types, given the conditions of no permanent immunity (SIS-like, where S is susceptible and I is infected), mutually exclusive infections, and nodes that are homogeneous in their resistance towards both contagions \citep{prakash2012winnertakesall}. Non-graph-based epidemiological models can also account for competition. Two of the proposed epidemiological models in \citep{muhlmeyer2020modelingsocialcontagion} accounted for the presence of competing discrete information differently: the ISCR model features an additional class of user, counter-spreaders (C), in addition to the classical ignorant (I), spreader (S), and recovered (R), while the ISSRR model divides the spreader and recovered classes based on the which side of the bipolar information they were infected by or recovered from.
\par Extensions to IC and LC (Section \ref{sec:classical-models}), which are discrete models, allow them to model competitive influence maximization. The simplest modification is to start the diffusion of two different pieces of information from two different origin/seed nodes. The different IC and LT derivatives distinguish themselves based on the restrictions or conditions that the competing factions are subject to: selecting the most influential starting nodes for one faction given that the competing faction's starting nodes are known under IC \citep{carnes2007maximizinginfluencecompetitive}; best response and first mover strategies under IC \citep{bharathi2007competitiveinfluencemaximization}; proving that being the first mover is not always an advantage under a special case of IC \citep{kostka2008wordmouthrumor}; different forms of competitive diffusion under LT, e.g., different tie-breaking rules \citep{borodin2010thresholdmodelscompetitive}; and two competing cascades on a signed network under the IC model \citep{srivastava2015socialinfluencecomputation}.
\par The limitations and unrealistic assumptions of IC and LT have been recognized, with \citep{aral2018socialinfluencemaximization} highlighting three misspecifications of classical IC and LT: not modeling assortativity of influence, ignoring non-uniformity of the joint distribution of influence and susceptibility (due to peer effects, homophily, and social influence), and heterogeneity in influence and susceptibility. However, \citep{aral2018socialinfluencemaximization} has misspecifications of its own. In the problem of IM, \citep{aral2018socialinfluencemaximization} defaults to seed selection: a faction (representing a stance) has the opportunity to pick and choose seed nodes from which to begin its diffusion. However, this seed selection assumption that non-competitive and competitive IM operate under is alien to info ops (Section \ref{sec:info-ops}). Influencers on social media can be paid to promote a stance, but not all can be bought off. Info ops are also often conducted in a decentralized manner to make them resilient against the decapitation strike tactic of silencing/compromising influential nodes (Section \ref{sec:decentralization}), so info ops planners are unlikely to even consider the problem of seed selection.

\par Traces of LT's threshold mechanism combined with DeGroot's weighted influence (despite our model being designed for discrete instead of continuous opinion), and the voter/majority rule model are present in our model, grounding it in the extant competitive diffusion literature. A threshold (resistance) must be crossed for opinion adoption/infection events. Unlike classical majority rule models, in which an agent's adoption of the majority opinion is deterministic (the neighborhood's majority opinion will definitely be adopted), ours is probabilistic, with each opinion weighted by perceived popularity, as measured through the proxies of engagement metrics. While agents do not have weighted self-loops as in DeGroot, they are influenced by how others judge the opinions that they express (Section \ref{sec:indirect-influence}). The mechanics make it so that human agents in \textit{Diluvsion} are more inclined to convert to a stance that is perceived to be held by the majority in the quasi-global neighborhood (Section \ref{sec:indirect-influence}). Adopting an unpopular opinion is possible, though unlikely. An agent's discrimination against minority opinions is therefore encoded in the probability distribution. This form of minority avoidance in our model, \textit{Diluvsion}, despite sharing conceptual roots with those encountered in the literature, is considerably different its technical implementation.
\par The design of our mechanics for themes can be considered a special restricted case of Axelrod's cultural traits. The example of traits given in Axelrod's original paper allows for $10^5$ possible states since there are five positions filled with numbers ranging from zero to nine. The number of positions in \textit{Diluvsion} is variable, from one to an arbitrary maximum, and a theme can fill a position only once. If the maximum was five and there were 12 themes, $\sum^{5}_{k=1}\frac{12!}{k!(12-k)!}=1585$. Peer effects, homophily, and social influence all affect an agent's decisions in \textit{Diluvsion}, heightening realism (see Section \ref{sec:indirect-influence}). One manifestation of homophily is in opinions sharing similar stance and themes as the agent itself possessing greater weights during an infection event, and this homophily is comparable to Axelrod's and Hegselmann-Krause's homophily, though less extreme.

\subsubsection{Cooperative and competitive information diffusion}
\label{sec:coopetitive-models}
\par Cooperation can also be an outcome of the interactions between multiple information contagions present in a system \citep{wang2019coevolutionspreadingcomplex}. Immunological dysfunctions caused by the infections of viruses such as HIV \citep{claudeameisen1991celldysfunctiondepletion}, Ebola virus \citep{wiedemann2020longlastingsevereimmune}, and SARS-CoV-2 \citep{phetsouphanh2022immunologicaldysfunctionpersists} that make the infected more susceptible to other viruses are examples of cooperative biological contagions. Relative to competitive diffusion, the many forms of cooperative information diffusion have been less frequently studied.
\par Cooperation can occur at the information/contagion level. The diffusion model in \citep{myers2012clashcontagionscooperation} learns from empirical data the affinity between different contagions (URLs), finding that news stories of the same event boost the adoption of each other's URLs, implying cooperation. The spread of a highly infectious news story's URL suppresses the spread of URLs from other unrelated, less infectious news stories, implying competition. Different models \citep{chang2018codiffusionsocialcontagions, zhang2022interactionmultipleinformation} have recreated the synergistic (e.g., complimentary products such as iPhone and Apple Watch) and antagonistic relationships between contagions propagating in heterogeneous and homogeneous multilayer networks where simultaneous infection by multiple contagions (coinfection) is possible. 
\par Cooperation can occur at the agent level. The implicit form of such cooperation is common in the study of multi-agent systems in the context of robotics (vehicular information diffusion and consensus formation \citep{adler2006selforganizedcontextadaptiveinformation, saldana2017resilientconsensustimevarying}, drone swarms \citep{arnold2019whatrobotswarm, wang2022optiswarmopticalswarm}, and search-and-rescue robots \citep{anderson2008implicitcooperationstrategies}). Ideas found in (not necessarily originating in) robotics can be adapted to information diffusion, e.g., swarm intelligence algorithms for influence maximization under IC and LT models \citep{simsek2018usingswarmintelligence, olivares2021multiobjectivelinearthreshold}.
\par Finally, the cooperative effect can be found in the links/edges connecting nodes in a network; a node $u$ being infected by a contagion through another node $v$ makes $u$ more susceptible to future infection by a different contagion via $v$ \citep{wang2023linkcooperationeffect}, reminiscent of the effects of trust and affection in social relations.
\par The implicit cooperation for \textit{Diluvsion} happens at the agent-level, where all bots targeting the most vulnerable for stance propagation \ref{sec:bots}) and when bots' independent actions unknowingly contribute towards a common objective (Section \ref{sec:result-bot-impact}). At the information level, stances and themes induce homophilic conversions, and engagement metrics induce bandwagoning (Sections \ref{sec:stance-themes} and \ref{sec:indirect-influence}).

\subsubsection{Agent-based models}
\par Agent-based models (ABMs) aim to infer macro-level outcomes from micro-level interactions and can be viewed as a further reduction of abstraction compared to models based on interconnected homogeneous nodes and models constructed only from differential equations. Nothing stipulates that agent-based models have to incorporate a network topology in modeling diffusion (e.g., agents can be placed on a grid and the diffusion can be governed by spatial proximity), but in practice, agent-based models of information diffusion on social media invariably use network topology, as networks are an effective approximation of social organization in human communities, online and offline. Being freed from the constraints of stringently adhering to the precepts of classical IC and LT, and their many extensions, enables more customized and accurate models to be built based on empirical data in sociological and psychological studies, albeit costlier to run computationally. An example is the ABM for the highly specific context of online anti-vaccination movements \citep{sobkowicz2021agentbasedmodel} which contained three different agent types: doctors (similar but not equivalent to external authority/mass media in other models), patients (humans), and initiators (stubborns/bots).
\par A comprehensive review of the various design aspects of modeling information diffusion in social networks as multi-agent systems (MAS) can be found in \citep{jiang2015diffusionsocialnetworks}. This review focuses on the trade-off between the simplistic assumptions found in classical MAS and the complexity of real-world social networks. Thinking in terms of complexity versus abstraction can help establish boundaries between classical and agent-based models of diffusion. This demarcation is used by some ABMs, e.g., \citep{ross2019aresocialbots}. Thus, while agency is arguably present in some models discussed earlier due to individualization/heterogeneity of nodes, e.g., DeGroot \citep{frenchjr.1956formaltheorysocial, degroot1974reachingconsensus}, those models never fully embraced the possibilities of agency and the resulting emergent behavior (e.g., John Conway's Game of Life) in the interest of keeping the models analytically tractable, which is why we do not discuss them in this section.
\par \textit{Diluvsion} is an agent-based model. It borrows many concepts from existing models that are relevant to accurately simulating the impact of info ops on information diffusion on social media.
\par One borrowed concept is that agents have a set amount of energy, with every action depleting their energy reserves. In \citep{kaligotla2020diffusioncompetingrumours}, agents have a fixed energy level and interactions do not cost energy; however, after some arbitrary point in time, every interaction decreases the agent's energy level. This setup enables the study of the correlation between agent energy and rumor propagation intensity. In \citep{onuchowska2019usingagentbasedmodelling}, the simulations were populated by different classes of agents, each class with its own range for how frequently its members tweet and retweet. Our implementation of the energy level is based on empirical data from Twitter that separated users into three tiers (Sections \ref{sec:humans} and \ref{sec:real-params}). Our agents' actions are additionally restricted by a diurnal cycle (Section \ref{sec:resolution}).
\par Agents are also commonly constrained by having a memory and attention (also uncommonly, and perhaps confusingly, called vigor \citep{li2018modellingmultipleinfluences}) limit placed upon them, e.g., \citep{weng2012competitionmemesworld, wang2017rumorspreadingmodel, li2018modellingmultipleinfluences, li2019multiagentsystemmodelling, beskow2019agentbasedsimulation}, reflecting the non-omniscient nature of humans. We discuss this mechanic in greater detail in Section \ref{sec:memory}.
\par Linked to the concept of memory is the concept of repetition driving information adoption. In the agent-based SIR rumor diffusion model introduced in \citep{ikeda2014multiagentinformationdiffusion}, agents can receive the same information multiple times, with agents likelier to adopt information that has been seen more frequently, contributing to the spread of the information as agents eventually share the information that they have adopted. In the information entropy-based rumor spreading model in \citep{wang2017rumorspreadingmodel}, a rumor's information salience is measured by the frequency of the rumor's occurrence in the agent's memory, with greater frequency linked to higher salience. The most salient rumor is selected for transmission during the spreading phase of the model, and this selection mechanism constitutes the conformity effect put into practice \citep{wang2017rumorspreadingmodel}. In an ABM for studying astroturfing by bots under the spiral of silence framework, conformity takes the form of an opinion climate, which is composed of the opinions expressed by neighboring agents with silent agents excluded \citep{ross2019aresocialbots}. Another model \citep{li2019multiagentsystemmodelling} of conformity by has an agent slowly revising its opinion to become more similar to its neighbors'. Yet another model, AMID, features users who are likely to adopt opinions supported by the majority of adjacent neighbors in \citep{li2018modellingmultipleinfluences}. Additionally, under AMID, trust between agents was modeled as the frequency with which a truster shares the trustee's messages, with users likelier to adopt messages from trusted sources \citep{li2018modellingmultipleinfluences}. 
\par Repetition's influence similarly suffuses every part of \textit{Diluvsion}. A stance that occurs more frequently in an agent's memory has a greater chance of being adopted. But, we introduced additional complications into the mechanism. Conformity effects are extended to account for the engagement metrics prominently displayed alongside every message on social media, which serve as cues on how others perceive the credibility of the message (Section \ref{sec:indirect-influence}). Conformity effects are also balanced against homophily for stances and themes, which is based on the real-world tendency for people to prefer opinions that are similar to their own.
\par A divide exists among ABMs on the matter of opinions being exclusionary or capable of co-existing with others. Exclusionary opinions, as conceptualized in \citep{hu2017competingopiniondiffusion}, permit a user to only adopt one discrete opinion at any one time. Models in which opinions are modeled as a point along a continuous spectrum, as is common in counter-disinformation ABMs, e.g., \citep{ikeda2014multiagentinformationdiffusion, beskow2019agentbasedsimulation, averza2022evaluatinginfluencetwitter}, naturally feature exclusionary opinions as opinions cannot simultaneously occupy two positions along the spectrum. Studies using the spiral of silence framework, e.g., \citep{takeuchi2015publicopinionformation, ross2019aresocialbots}, require polarized/exclusionary opinions by design; users would not be hesitant to voice opinions that ran counter to the perceived majority opinion climate if an oppositional relationship did not exist. For non-exclusionary opinions, there is also a myriad of conceptualizations. In SMSim, developed for studying non-polarized information diffusion \citep{decgatti2013simulationbasedapproachanalyze}, bit vectors denote a message's association with the set of possible topics in the simulation. For the rumor spreading model introduced in \citep{wang2017rumorspreadingmodel}, infection by an opinion means that the opinion goes into the agent's memory bank. There is no incompatibility mechanism that forces a stored opinion to be ejected from memory; removal is only due to insufficient space in the bank. And with some models, the distinction between exclusionary and non-exclusionary is less clear, e.g., \citep{myers2012clashcontagionscooperation} where agents can only adopt one contagion but the agent's infection history continues to influence future contagion adoption. In the AMID model \citep{li2018modellingmultipleinfluences}, each message is associated with all possible topics in the simulation with varying strengths (fuzzy sets) and there is an additional binary number for each topic denoting the message's positive or negative opinion towards the topic, a formulation reminiscent of \textit{Diluvsion}'s stance and themes (Section \ref{sec:stance-themes}). Users in AMID are capable of caring about multiple topics simultaneously, but the opinion (polarity) for each topic is an exclusionary binary value.
\par On the matter of co-existing opinions, \textit{Diluvsion} straddles the middle as there are exclusionary and non-exclusionary components of an opinion within \textit{Diluvsion}, corresponding to polarized stances and non-polarizable themes respectively (Section \ref{sec:stance-themes}).
\par A common feature of ABMs is the segregation of the agents into distinct classes. This practice is intuitive in studies investigating the impact of bots on social media, e.g., \citep{beskow2019agentbasedsimulation, onuchowska2019usingagentbasedmodelling, ross2019aresocialbots, averza2022evaluatinginfluencetwitter}, as bots have to behave differently than the rest of the users to distinguish themselves and for their influence to be measurable. In \citep{onuchowska2019usingagentbasedmodelling}, both humans and bots were further subdivided into additional categories, e.g., amplifier and commentator, based on extant research on social media user behavior. Some works \citep{beskow2019agentbasedsimulation, averza2022evaluatinginfluencetwitter} divided agents based on stance, with ``truth'' propagandists, neutral/uninformed people, and ``disinformation'' peddlers exhibiting different behavioral patterns; bots are associated solely with the undesirable stance, disinformation. While most models keep an agent's role static throughout the simulation, some allow agents to radically alter their behavior upon meeting certain conditions, e.g., radicalization turning passive users into active disseminators  \citep{sobkowicz2021agentbasedmodel, averza2022evaluatinginfluencetwitter}.
\par Under \textit{Diluvsion}, three main attributes define an agent: its activity level, its action-role, and whether the agent is a human or a bot. There are three activity levels, five action-roles (15 unique action distributions as the distributions for the same role differ based on activity level), and a binary bot/human membership. As such, a total of 30 possible classes of users exist in \textit{Diluvsion}. Agents do not change roles. Our conception of bots (Section \ref{sec:bots}) hews closer to \citep{onuchowska2019usingagentbasedmodelling, ross2019aresocialbots} in that their \textit{behavior} is stance-agnostic for the reasons specified in Section \ref{sec:digital-personas}, even though \citep{onuchowska2019usingagentbasedmodelling, ross2019aresocialbots} also only associate bots with malicious behavior.
\par Models found in the literature that most closely resemble our \textit{Diluvsion} are those developed for a purpose similar to ours: the ones by \citeauthor{beskow2019agentbasedsimulation} and \citeauthor{averza2022evaluatinginfluencetwitter} which dealt with information operations involving bots shaping discourse on online social media platforms \citep{beskow2019agentbasedsimulation, averza2022evaluatinginfluencetwitter}. Unlike \citep{ross2019aresocialbots} which also studied the impact of bots on online discourse, \citep{beskow2019agentbasedsimulation, averza2022evaluatinginfluencetwitter} did not integrate the spiral of silence into their models. And unlike \citep{onuchowska2019usingagentbasedmodelling} which was interested in the descriptive statistics (e.g., tweets and retweets per day) that a mixed population of bots and humans would generate, \citep{beskow2019agentbasedsimulation, averza2022evaluatinginfluencetwitter} are concerned with competitive diffusion of three stances, just like our study. Although to be accurate, our model allows all three stances to be propagated, but in \citep{beskow2019agentbasedsimulation, averza2022evaluatinginfluencetwitter}, only the two stances that are antithetical to each other can be propagated while the neutral stance is a static default state. The crucial differences between our work and \citep{beskow2019agentbasedsimulation, averza2022evaluatinginfluencetwitter} are the stance-agnosticism of our bots (Section \ref{sec:bots}), neutrality being an infectious stance in our model, as previously mentioned (Section \ref{sec:stance-themes}), the inclusion of indirect influence mechanics (engagement metrics and non-edge information transmission, Section \ref{sec:indirect-influence}), and heterogeneous activity levels and roles for agents (Section \ref{sec:real-params}).

\subsection{Information operations}
\label{sec:info-ops}
\subsubsection{Terminology}
\label{sec:info-ops-definition}
\par Within the existing literature, information operations (info ops), influence operations, and psychological operations have been used interchangeably to describe ``activities that target information bases of an adversary, protect one's own information assets and make efficient use of one's own information''\citep{luiijf1999informationassuranceinformation} even though each term has a specific meaning, e.g., psychological operations are originally defined as falling under the umbrella of info ops \citep{emery2014terroristuseinformation}. In this paper, we adhere to the loose usage of these terms as other researchers have, e.g., authors of \citep{hollis2018influencewarwar, courchesne2021reviewsocialscience} switching between the terms within the same document; Cambridge Analytica varyingly described as engaging in psychological operations \citep{bakir2020psychologicaloperationsdigital}, information operations, \citep{dowling2022cyberinformationoperations}, and influence operations \citep{wilson2019cambridgeanalyticafacebook}.
\par Although info ops is ``a military defined concept and doctrine'' originating from the post-Cold War US and NATO military establishments, it ``interacts largely with society as the majority of the targets either attacked or to be defended by Information Operations means are not military but civil targets'' \citep{luiijf1999informationassuranceinformation}. As pervasive computing has become a reality in a world where interconnected mobile computing devices are increasingly commonplace among civilians, there has never been a richer, more fertile ground for info ops \citep{campbell2003securityprivacypervasive}, especially when the rise of social media normalizes rapid information sharing.

\subsubsection{Social media}
\label{sec:socmed}
\par Social media has always been a tool for information warfare. As early as 2003 \citep{herrera2014revolutionagesocial} / 2006 \citep{bayat2020revolutionrevolutionariesmaking}, the cyberdissident diplomacy (CDD) initiative from the US State Department reached out to ``tech-savvy youth in the sixteen-to-thirty-five age range with the aim of training them in particular forms of cyberdissidence and online campaigning'', leading ``young Muslims'' to ``serve as the proxies to the US'' in a ``war of ideas'', an ``ideological war''--- influence operations, in other words --- ``to maintain its global hegemony'' \citep{herrera2012youthcitizenshipdigital, herrera2014revolutionagesocial}. As the provocative title of one journal paper neatly summarized, \textit{``Facebook to Mobilize, Twitter to Coordinate Protests, and YouTube to Tell the World'': New Media, Cyberactivism, and the Arab Spring} \citep{arafa2016facebookmobilizetwitter}. In the 2009 protests in Iran, Twitter was deemed so crucial to the protesters that the US State Department intervened in Twitter's scheduled server upgrade to ensure that Iranian users did not suffer from service interruption \citep{burns2009twitterfreeiran}. In Russia where the user base of native social media platforms VKontakte and Odnoklassniki dwarfs that of the US-developed Facebook, it is Facebook that mobilized anti-regime forces in 2011 \citep{white2014didrussianearly}. The clearest example of social media as an instrument of war is the USAID-backed ZunZuneo, which is a Twitter-like service designed to run solely in Cuba through mobile text messaging for the express purpose of covertly inciting regime change, according to the reporting done by the US news agency Associated Press \citep{butler2014ussecretlybuilt, ap2014ussecretlycreated}.
\par But as the flames of social unrest fanned in faraway ``jungle'' \citep{majumdar2022eutopdiplomat} countries escaped their confines and started engulfing First World/developed ``garden'' countries, the dangerous potential of social media was finally recognized by the research community and alarms were raised. In a review of academic publications on countermeasures against influence operations, the authors found a ``dramatic increase in the number of studies since 2016'' which is ``[i]n line with other research trends in the broader field of influence operations and disinformation'' \citep{courchesne2021reviewsocialscience}; 2016 was the year when Russia was accused of meddling in the US presidential election in favor of Trump via social media campaigns \citep{aj2018russiausedall}. Threats can be homegrown too: Amazon, Apple, and Google suspended the US social media platform Parler for its role in the 2021 Capitol Hill riot \citep{aj2021amazonsuspendparler}. The panic in the research community has culminated in the birth of a new field, social cybersecurity, which seeks to ``characterize, understand, and forecast cyber-mediated changes in human behavior and in social, cultural, and political outcomes'' as one of its objectives \citep{carley2020socialcybersecurityemerging}.
\par \textit{Diluvsion} is a tool that seeks to aid in the fulfillment of the aforementioned objective by enabling the characterization of Twitter-like services and the forecast of info ops on such services.

\subsubsection{Twitter-like platforms}
\label{sec:twitter-clones}
\par Likely because of the ease of collecting data from Twitter, many researchers have focused on modeling information diffusion on Twitter such as \citep{weng2012competitionmemesworld, das2014modelingopiniondynamics, gatti2014largescalemultiagentbasedmodeling, beskow2019agentbasedsimulation, averza2022evaluatinginfluencetwitter}. This is one reason the simulation environment within \textit{Diluvsion} models itself after Twitter. A greater motivating factor is the profusion of Twitter-like social media platforms. Truth Social, Parler, and Gab cater to users chafing at the politically restrictive policies of Twitter \citep{stocking2022rolealternativesocial}. Mastodon lets users escape the whims of a single central authority by dispersing regulatory power among multiple central authorities \citep{zulli2020rethinkingsocialsocial}. Twitter-like platforms have also emerged to serve different regions and languages, such as China's Sina Weibo. Even social media platforms that have never sought to emulate Twitter have analogs to certain features of Twitter (which may not have originated from Twitter), e.g., Facebook's share and share with a comment being similar to Twitter's retweet and quote-tweet, respectively.
\par The mushrooming of Twitter alternatives also speaks to a credibility crisis, where users no longer trust that they can express their views free of suppression from a central authority. Decentralization is a natural antidote to this credibility crisis. Our conceptualization of decentralization within \textit{Diluvsion} does not take after Mastodon, since their federated instances still allow moderation power to rest in the hands of the few \citep{zulli2020rethinkingsocialsocial}. We speak of a decentralization where power is much more diffuse --- or at least appears so (Section \ref{sec:decentralization}).

\subsubsection{Decentralization and radicalization}
\label{sec:decentralization}
\par The most effective way to shut down conversation is the top-down approach of restricting free speech, as the Ukrainian experience of banning Russian social media VKontakte from 2017 onwards has shown \citep{golovchenko2022fightingpropagandacensorship}. Yet at the same time, the vast majority of both pro-Ukraine and pro-Russia Ukrainians circumventing the ban to continue accessing VKontakte \citep{golovchenko2022fightingpropagandacensorship} shows that the desire to escape overt, centralized censorship can transcend political affiliations. In liberal democracies, harsh speech policing measures are not well-tolerated by users, even if private corporations instead of governments are the ones doing the policing \citep{patty2019socialmediacensorship}, as evinced by the many alternatives to Twitter (Section \ref{sec:twitter-clones}). And if governments instead of corporations enforce censorship, the governments risk being viewed as draconian, tyrannical, totalitarian, authoritarian, etc., by liberty-minded people. A 2019 attitudinal survey of US internet users found that ``among political attitudes, opposition to censorship ($\beta = -.231$, $p < .001$) had a significant negative effect on support for government regulation of platforms'' \citep{riedl2022antecedentssupportsocial}. Hence the rise of a decentralized approach to information warfare: shaping discourse with an invisible hand, unbeknownst to the public.
\par Decentralization has proven effective on offline social networks. Decentralization allows for radicalization without attribution and accountability. Absent clear chains of command, membership simply involves claiming affiliation. Members can appear to be acting autonomously, when collectively those actions advance the interests of the movement \citep{medina2009geospatialanalysisdynamic}. Should the members' actions harm the movement, ``they can simply disavow the act without disavowing the ideology'' \citep{oneil2017rightwingterrorism} in the same vein as black ops and black sites \citep{linnemann2019blacksitesdark}. Without central authority, a movement can resist collapse from decapitation strikes \cite{sussman2008templaterevolutionsmarketing, arafa2016facebookmobilizetwitter, rueckert2023howjournalistmurder}. Ivan Marovic, a leader in the successful Serbian regime change organization, Otpor, which is funded by NED; IRI; USAID; Freedom House; and British Westminster Foundation, explained to National Public Radio's Bob Garfield that ``[l]eaders could have been blackmailed or bribed or even maybe killed. You can't do that with brands or ideas'' \citep{garfield2005revolutionincmedia}. This is similar to how a perpetual struggle can be waged against an idea, e.g., the Global War on Terrorism. Terrorist themselves from white supremacists \citep{caspi2012worstbadviolent} to al-Qaeda and Hamas \citep{medina2009geospatialanalysisdynamic} also favored decentralized structures for their resilience and ``tactical security''.
\par Success in the offline world has led to attempts to replicate decentralized movements online. The US cyberdissidence program is one such attempt at growing an online decentralized movement (Section \ref{sec:socmed}). Another is CENTCOM's Operation Earnest Voice (OEV), which empowered ``one US serviceman or woman to control up to 10 separate identities based all over the world'' so they can create ``false consensus in online conversations'' \citep{cobain2011revealedusspy, bennett2013exploringimpactevolving}. Graphika and Stanford Internet Observatory (SIO) reported in 2022 that their joint investigation with Twitter and Meta uncovered hundreds of accounts engaging in covert pro-Western info ops over the course of five years \citep{graphika2022unheardvoiceevaluating}. An Israeli company, Team Jorge, specializing in info ops that ran multiple campaigns influencing elections around the world \citep{megiddo2022israelisdestabilizingdemocracy, kirchgaessner2023revealedhackingdisinformation}, created Advanced Impact Media Solutions (AIMS), a one-click system for creating ``fake digital persona[s] built to mimic human behavior and avoid detection'' which operate ``on social media sites, disseminating rumors, harassment, defamation or praise --- whatever the client asked for'' \citep{benjakob2023hackingextortionelection, megiddo2023cambridgeanalyticaisraeli}. Other Israeli info ops companies Archimedes and Percepto also employed fictitious digital personas \citep{debre2019israelicompanytargeted, megiddo2023antisemiticinfluencersfake}. White supremacists \citep{oneil2017rightwingterrorism} and international Islamists \citep{aistrope2016socialmediacounterterrorism, awan2017cyberextremismisispower} took advantage of social media for recruitment \citep{awan2017cyberextremismisispower, aistrope2016socialmediacounterterrorism, oneil2017rightwingterrorism}, to transition to decentralized structures \citep{aistrope2016socialmediacounterterrorism, oneil2017rightwingterrorism}, and to conduct info ops of their own \citep{emery2014terroristuseinformation}. The ``IT cells'' manipulating Indian elections adopted a functionally leaderless structure, where the bottom layer ``creating and trending narratives'' maintains ``enough distance from the top to give leaders plausible deniability if the digital foot soldiers become too extreme'' \citep{rueckert2023howjournalistmurder}. Another example is NAFO (North Atlantic Fellas Organization), which Politico described as consisting of ``internet culture warriors taking on Russian disinformation and raising funds for Ukraine'' by adopting the same tactics as ``the Islamic State'' and ``the American far-right boogaloo movement'' \citep{scott2022digitalbridgenafo}. According to one freelance disinformation researcher and self-proclaimed member \citep{choe2023joohnchoedisinformation, choe2023jchoejoohnchoetwitter}, NAFO is a leaderless movement \citep{choe2022letmeexplain}. The Wall Street Journal describes NAFO as having ``no command structure'' \citep{michaels2022ukraineinternetarmy}.
\par Owing to the achievements and prevalence of decentralized info ops, the simulations we chose to show running in \textit{Diluvsion} (Section \ref{sec:simulations}) were developed with decentralized info ops in mind, with affecting change through uncoordinated coordination. The change agents being unaware of each other (Section \ref{sec:bots}) in \textit{Diluvsion} is also a deliberate design choice grounded in decentralization --- leaderless terrorist/freedom fighter cells \citep{kaplan1997leaderlessresistance}.
\par To affect change and appear organic at the same time, a decentralized movement needs a large number of warm bodies. But growing a movement is time consuming. Furthermore, each additional truly independent (human) agents within a movement is an additional security risk. Humans can turn into whistleblowers. Financial trails from hiring propagandists risk being traced back to the mastermind.
\par CENTCOM's OEV, Archimedes, Percepto, and Team Jorge's AIMS already hinted at a solution: autonomous artificial personas (Section \ref{sec:digital-personas}).

\subsubsection{Digital personas and bots}
\label{sec:digital-personas}
\par Based on all the markers accessible via cyberspace --- i.e., excluding anything that requires physical interaction in the real world --- bots can be made virtually indistinguishable from live human operatives. This is all thanks to progress in the field of machine learning making it easier than ever to deceive human senses and cognition \citep{kietzmann2020deepfakestricktreat}. On the behavioral and textual front, services such as Character.AI allow users to create chatbots with distinct personalities, such as Elon Musk or Marvel movie characters, and these chatbots are capable of carrying conversations for hours with humans \citep{rosenberg2023customaichatbots}. Chatbots can exert influence over humans; interacting with chatbots has a positive impact on people's attitude towards vaccines and also their intention to get vaccinated \citep{altay2023informationdeliveredchatbot}. Bot-run Twitter accounts are perceived as credible sources of information \citep{edwards2014thatbotrunning}. On the audio front, text-to-speech audio generation \citep{khanjani2023audiodeepfakessurvey} has led to convincing audio clips of Biden and Trump chatting over gaming sessions \citep{rosenberg2023aigeneratedaudiojoe}. On the video front, there are talking head video generation neural networks producing realistic depictions of people saying things that they have never said, e.g., \citep{wang2022oneshottalkingface}. A variety of techniques exist for transposing one person's face onto the body of another person in a video, e.g., \citep{xu2022mobilefaceswaplightweightframework}. High resolution videos can be created with only text prompts, e.g., \citep{blattmann2023alignyourlatents}. A navigable 3D scene can be created through neural radiance fields using just a few static images, e.g., \citep{yu2021pixelnerfneuralradiance, barron2022mipnerf360unbounded}. On the image front, text-to-image synthesis networks can hallucinate scenes that never existed in the real world purely by relying on text descriptions, e.g., \citep{zhu2019dmgandynamicmemory}. For realistic faces, one does not need to steal the identity of a human, as faces of non-existent people can be generated \citep{kammoun2022generativeadversarialnetworks}.
\par To tie it all together, AIs given an initial goal from humans can plan to accomplish the goal by giving itself a list of tasks and setting about accomplishing the tasks by itself using various tools (including itself and other AIs) at its disposal \citep{qin2023toollearningfoundation}, currently best exemplified by Auto-GPT. Pretending to be a human employed in a job will not be too difficult, as AIs can produce codes that run \citep{kashefi2023chatgptprogrammingnumerical} and also play at the role of a knowledgeable analyst since vast swaths of programming and data science work can be automated via ChatGPT \citep{hassani2023rolechatgptdata}.
\par While we are not aware of any entity that has made digital personas using a combination of all these bleeding edge AI capabilities\footnote{OEV predates the AI/neural network revolution. Graphika and SIO noted that assets in the pro-Western covert info ops they uncovered ``created fake personas with GAN-generated faces''\citep{graphika2022unheardvoiceevaluating}, but no details were provided on whether other content are AI-generated. AIMS bots do not rely fully on AI: profile pictures were ``stolen from a genuine profile'', although the ``creation of [textual] content'' is ``driven by AI'' \citep{megiddo2022israelisdestabilizingdemocracy, benjakob2023hackingextortionelection}.}, there is a high likelihood of such hypothetical human-like bots being created and deployed in service of info ops at present or in the near future. Relying on detection as the sole strategy to safeguard against them might be a dead-end. A collaborative report by the Stanford Internet Observatory, OpenAI, and Georgetown University’s Center for Security and Emerging Technology concluded that it was reasonable to assume that ``attributing short pieces of [textual] content as AI-generated will remain impossible'' \citep{goldstein2023generativelanguagemodels}. The same report warned that, based on their deliberately limited scope of ``generative language models'' producing ``persuasive text'', AI is ``likely to significantly impact the future of influence operation'' \citep{goldstein2023generativelanguagemodels}. The days in which bots are simplistic and easily detectable automatons, such as those found by BotHunter to be influencing Singaporean elections to little apparent success \citep{uyheng2021activeaggressivelittle} or found by Botometer (which already often mistakenly classifies humans as bots \citep{rauchfleisch2020falsepositiveproblema}), are drawing to a close with advancements in AI.
\par The promise of AI-enabled bots lies not just in rapidly raising an info ops army as Team Jorge did with its 39,000 avatars \citep{megiddo2022israelisdestabilizingdemocracy, kirchgaessner2023revealedhackingdisinformation}, but also in reducing the cost of info ops. ``The potential of language models to rival human-written content at low cost suggests that these models---like any powerful technology---may provide distinct advantages to propagandists who choose to use them'' \citep{goldstein2023generativelanguagemodels}. Accessibility to content generation capabilities, which is correlated with cost, has also been widened with the advent of AI. Image generation via diffusion models such as Dall-E-2, Midjourney, and Stable Diffusion are available as services ``for free or for a low price'' \citep{hachman2023bestaiart} to practically everyone. Consequently, social media platforms have been flooded by ``hyper-realistic fake images deceiving many online'' such as those of Trump resisting arrest in March 2023 \citep{saliba2023realworryhow}.
\par Bots can solve the problem of credibility crisis by appearing independent. This is not the same credibility crisis we spoke of earlier where social media platforms were not trusted to be non-interventional (Section \ref{sec:twitter-clones}). This is a credibility crisis among entities pushing the message --- usually users of the platform and not the platform itself. Messages from US-linked organizations are distrusted even for the seemingly benign (or even beneficial) cause of counter-terrorism \citep{aistrope2016socialmediacounterterrorism}. Attempts at laundering their reputation by using proxies such as ``community groups, non-governmental organisations and private enterprise'' may be unsuccessful because simply being affiliated with the US in any way taints the proxies' reputations \citep{aistrope2016socialmediacounterterrorism}. However, by conducting info ops through a collective of artificial online personas who cannot be linked back to any state or non-state actor, the info ops would not be subjected to any long-held animosity that the targeted population may have towards an actor, thus encountering less resistance. A scholar examining the US's credibility problem on social media in the issue of counter-terrorism similarly noted that ``producing the appearance of independence by hiding behind fictitious online identities is one option'' to address the credibility deficit \citep{aistrope2016socialmediacounterterrorism}.
\par Our conception of bots being nigh undetectable, as expounded in this section, plays a critical role in the design of information propagation within \textit{Diluvsion}. In \textit{Diluvsion}, we assume that bots are bound by many of the same limitations that a human agent would be, e.g., their activity level, as they would seek to blend in with the human population. We also assume that people would not be able to ascertain another social media user's true nature and affiliations, hence why a person's judgment on the credibility of a message/tweet relies on external social cues. And this human heuristic of assessing credibility by following the herd makes social media users uniquely vulnerable to decentralized info ops that can be trivially signal-boosted by bots, as we will discuss in Section \ref{sec:indirect-influence}.
\par However, at the same time, there exists in \textit{Diluvsion} a special class of agents separate from humans that are called ``bots'' (Section \ref{sec:bots}); within \textit{Diluvsion}, ``bots'' is a term of convenience for all forms of propagandists. If a human takes on the role of a propagandist (calculatingly penning tweets to maximize engagement, using Photoshop or a physical airbrush to alter images, etc.) on social media, they are rightly classified as ``bots'' since they are indistinguishable from purely algorithmic bots engaging in info ops. 
\par The reason we reject the terms ``propagandists'' or ``operatives'' in favor of ``bots'' is because we reject the current erroneous framing within the agent-based simulation community of segmenting agent types based on stance (defined in Section \ref{sec:stance-themes}) on an issue. In \texttt{twitter\_sim} \citep{beskow2019agentbasedsimulation}, there are ``three types of users: normal users (ignorants), bots/trolls (spreaders), and truth defenders (stiflers).'' In the model introduced in \citep{averza2022evaluatinginfluencetwitter}, there are ``three types of agents: Deceptive agents, representing bots or malicious human users; neutral agents, representing most Twitter users; and news agents, official sources who share accurate, verified information.'' A pattern emerges: bots are invariably associated only with the ``bad guys'', which is categorically false since NAFO, OEV, and the pro-Western covert info ops documented by Graphika and SIO are examples of bots working for ``good'' causes by disseminating ``truth''. Under \textit{Diluvsion}, all propagandists are bots regardless of their stance.
\par Reclaiming the term ``bots'' is necessary to ensure accuracy in analysis and simulation, and thus avoid errors in strategic judgment. Propaganda work and counter-propaganda work are often broadly equivalent in tactics. Persisting in the delusion that the ``good guys'' behave in a radically different manner in matters of war is detrimental to the formulation of good policy. Researchers will chase after the wrong solutions due to that initial error in judgment. Recall the example of the failure of US counter-terrorism info ops on social media \citep{aistrope2016socialmediacounterterrorism}. Foreign policy analysis and reporting, e.g., \citep{garfield2007counterpropagandafailureiraq, mclaughlin2014whyusgovernment, debusmann2015unitedstateslags, gerstel2016isisinnovativepropaganda}, will fixate on attributing the success of Islamist or ISIS propaganda over US counter-propaganda to their speed, sophistication, creativity, having a better narrative, etc., and rarely question whether the failure of counter-propaganda lies in the counter-propagandists' lack of credibility, in the message listeners' fear and loathing of the US. The State Department's Digital Outreach Team (DOT) accusing ISIS of terror tactics via social media accounts clearly affiliated with the US \citep{aistrope2016socialmediacounterterrorism} appears oblivious to the fact that the use of terror tactics by the US itself\footnote{In recent years, it has come to light, through the reporting by outlets such as Current Affairs, The Guardian, and The Wire \citep{abad-santos2013this13yearoldscareda, mcveigh2013dronestrikestears, byrne2022droneskies}, that indiscriminate bombing via drones has instilled terror in the hearts of the populace in many regions of the Middle East. Children have come to fear clear blue skies because those are the perfect weather for flying drones.} undermines such a message. The initial error in judgment in assuming that the US's hands are clean ensured US counter-propaganda efforts' continued failure \citep{aistrope2016socialmediacounterterrorism}.

\subsubsection{State and non-state actors}
\label{sec:actors}
\par Other notions that we need to disabuse ourselves of for the sake of strategic clarity is that info ops originating from the ``good guys'' --- democratic states --- are only sanctioned against so-called non-democratic states and that meaningful distinction exists between state and non-state actors. Democratic states attack each other and themselves. State and non-state actors lead a chimeric existence. 
\par For social media platforms, the distinction between state and non-state actors is increasingly blurred due to the vast reach that the platforms have in people's lives, leading governments to constantly intervene in the platforms' daily operations \citep{denardis2015internetgovernancesocial} or even prop up the platforms, e.g., the case of Twitter, Iran, and the US State Department \citep{burns2009twitterfreeiran} and the case of USAID-funded ZunZuneo \citep{butler2014ussecretlybuilt, ap2014ussecretlycreated}. So deeply and confoundingly intertwined is the public-private relationship that in \citep{patty2019socialmediacensorship}, the author explicitly set out to argue that it is simultaneously unlikely and likely that courts will conclude that a social media company is a state actor.
\par Distinguishing between private and public entities among those engaged in info ops would be similarly difficult. Former intelligence operatives naturally gravitate towards private info ops firms, e.g., \citep{toistaff2018trumpcampaignmined, haaretz2018israelihackersreportedly, bamford2023trumpcampaigncollusion, megiddo2023cambridgeanalyticaisraeli, megiddo2023antisemiticinfluencersfake}, and these operatives bring along their contacts in the government, skills, and knowledge into their new roles \citep{bamford2023trumpcampaigncollusion}. One remaining difference may be the scale of resources that state-level actors can muster compared to non-state actors, but even that disparity is vanishing as states outsource info ops to private entities, potentially as a form of legalized corruption via awarding lucrative contracts \citep{garfield2007counterpropagandafailureiraq, chayes2020strategiesareforeign} and as a way to shield themselves from culpability (Section \ref{sec:decentralization}). According to Oxford Internet Institute's 2020 Industrialized Disinformation report, despite the difficulty in ``tracking down contractual evidence of private contracting firms'', they found that ``almost US \$60 million was spent on hiring firms for computational propaganda since 2009''  \citep{bradshaw2021industrializeddisinformation2020}. Public-private joint efforts can be seen in CENTCOM awarding a \$2.76M contract to the private corporation Ntrepid to develop OEV's persona management software \citep{cobain2011revealedusspy}; in the US State Department awarding a \$500,000 contract to Cambridge Analytica's parent company SCL for ``target audience research'' \citep{tau2018israeliintelligencecompany}; in the many info ops companies manipulating democratic elections at the behest of clients from democratic states \citep{wilson2019cambridgeanalyticafacebook, bakir2020psychologicaloperationsdigital, dowling2022cyberinformationoperations, toistaff2018trumpcampaignmined, debre2019israelicompanytargeted, megiddo2022israelisdestabilizingdemocracy, benjakob2023hackingextortionelection, megiddo2023cambridgeanalyticaisraeli, megiddo2023antisemiticinfluencersfake, rueckert2023howjournalistmurder, kirchgaessner2023revealedhackingdisinformation, bamford2023trumpcampaigncollusion}; in Team Jorge running info ops through ``Demoman International, which is registered on a website run by the Israeli Ministry of Defense to promote defence exports'' \citep{megiddo2022israelisdestabilizingdemocracy, kirchgaessner2023revealedhackingdisinformation}; and in NAFO, which according to a Kyiv Post interview of its founder, is ``endorsed by U.S. Congressman Adam Kinzinger, as well as U.K. Minister of Defense Ben Wallace'' \citep{smart2022foundernaforeveals}.
\par Conducting info ops against other democracies is not beyond the pale for democracies. Many documented instances exist. The campaigns conducted by the UK-based Cambridge Analytica in support of Trump and in collaboration \citep{tau2018israeliintelligencecompany, haaretz2018israelihackersreportedly, megiddo2023cambridgeanalyticaisraeli} with the Israeli Team Jorge is a well-known example \citep{wilson2019cambridgeanalyticafacebook, bakir2020psychologicaloperationsdigital, dowling2022cyberinformationoperations}. The actions of Israeli info ops companies in meddling with non-US democracies --- e.g., Team Jorge's interference\footnote{Veracity of Haaretz's reporting: ``Journalists from The Guardian, Der Spiegel, Die Zeit, Le Monde, the international organization of investigative journalists OCCRP, Radio France, Haaretz, TheMarker and other media outlets worked in France, Kenya, Israel, the United States, Indonesia, Germany, Tanzania and Spain to examine the veracity of Jorge’s claims about his worldwide deeds. Shockingly, many of the allegations were corroborated.'' \citep{megiddo2022israelisdestabilizingdemocracy}} in ``33 presidential-level election campaigns, 27 of which were successful'' \citep{megiddo2022israelisdestabilizingdemocracy, kirchgaessner2023revealedhackingdisinformation}; Percepto's info ops against a Burkina Faso election \citep{megiddo2023antisemiticinfluencersfake}; Archimedes targeting a Nigerian election \citep{debre2019israelicompanytargeted} --- are no less impactful simply because they received comparatively little attention in mainstream media and scholarly circles. Israeli Prime Minister Netanyahu also attempted to influence the 2016 election in favor of Trump \citep{bamford2023trumpcampaigncollusion}. Info ops conducted by an outside actor may even be orchestrated by paymasters situated within the targeted state. Trump campaign official Rick Gates sought proposals from an Israeli company Psy Group to ``to create fake online identities, to use social media manipulation and to gather intelligence to help defeat Republican primary race opponents and Hillary Clinton''; the proposal, named Project Rome, was ultimately never acted upon \citep{psygroup2016projectromecampaign, mazzetti2018rickgatessought}. The US under the Obama regime in turn has employed covert info ops against Israel \citep{ahren2016usfundsaided}: a US Senate inquiry found that ``[i]n service of V15 [an anti-Netanyahu group], OneVoice deployed its social media platform, which more than doubled during the State Department grant period; used its database of voter contact information, including email addresses \ldots and enlisted its network of trained activists, many of whom were recruited or trained under the grant, to support and recruit for V15.'' Even a seemingly innocuous ``voter button'' displayed on Facebook to encourage turnout in UK elections can be construed as an attack because Facebook, as a US company, should not be influencing UK politics in any way \citep{grassegger2018facebooksaysits}.
\par All these examples are in essence the use of non-state intermediaries in information warfare involving three state-level actors widely considered to be democracies (according to the Economist Intelligence Unit's 2016 and 2022 Democracy Index \citep{eiu2017economistintelligenceunit, eiu2023economistintelligenceunit}, the US and Israel are flawed democracies and the UK is a full democracy), putting to rest the idea that info ops are only ever employed in the stalwart defense of peace, freedom, and democracy against states with autocratic, authoritarian, totalitarian, etc. tendencies\footnote{For purveyors of ``No True Scotsman'' and ``Democratic Peace'' \citep{kim2005democraticpeacecovert}: Some academics have argued that Israel is an apartheid state \citep{falk2017israelipracticespalestinian} and also a settler colonial state \citep{lloyd2012settlercolonialismstate}. The US has been characterized as an oligarchy \citep{gilens2014testingtheoriesamerican} and as a plutocracy \citep{gilens2005inequalitydemocraticresponsiveness, bartels2016unequaldemocracypolitical}, and some Americans prefer to define the US as a republic \citep{holt2016democracyrepublicanswering}. The UK may fit the definition of an ``electoral authoritarian'' regime \citep{schedler2015electoralauthoritarianism} due to a history of banning political parties \citep{bourne2017mappingmilitantdemocracy}, media censorship \citep{bitso2013trendstransitionclassical}, and electoral fraud \citep{farrall2022whoarevictims}, on top of inheriting a hereditary monarchical head of state and hereditary peers in the House of Lords who exercise indirect political influence \citep{kaiser2014houselordsmonarchy}. An argument can also be advanced that certain states within the triad, being client states, lack true sovereignty \citep{sylvan2003agentbasedmodelacquisition, freeman2023diplomatictaxonomynew} so conducting inter-state info ops does not constitute warfare as no one's sovereignty has been violated.}.
\par Info ops' use by democracies is not limited to state-to-state situations. Info ops can be used against a democracy's own populace. In India, the world's largest democracy, ``[t]he 2014 election of Narendra Modi catapulted the BJP to power thanks, at least in part, to an expansive network of `IT cells' aimed at spreading positive news about the BJP and attacking its detractors'' \citep{rueckert2023howjournalistmurder}. Cambridge Analytica is believed to have influenced the Brexit vote results \citep{reutersstaff2018whatarelinks, megiddo2023cambridgeanalyticaisraeli}, though its culpability is disputed by some \citep{bbc2020cambridgeanalyticanot}.
\par Having established that info ops originating from state and non-state actors would have no appreciable difference due to ever-present public-private entanglement, we can claim that incorporating the publicized details of info ops conducted by non-state actors into our modeling will not bias the model towards the form of info ops conducted by non-state actors and vice versa, and we can claim that findings from \textit{Diluvsion} simulations are equally applicable to info ops from or against either state or non-state actors. 
\par In \textit{Diluvsion}, we defaulted to bots starting out embedded in a social network prior to being activated, in contrast to some preceding models that started bots out with only a single connection (edge) to the network, e.g., \citep{beskow2019agentbasedsimulation}, or let bots have lower degrees than humans, e.g., \citep{ross2019aresocialbots}, as they assumed that communities have not been penetrated by bots prior to an info ops. Our choice is based on reports of uncovered info ops. Team Jorge estimated that about 1,000 bots are needed to ``postpone an election in Africa without good cause'' with half being newly created bots and the other half preexisting bots \citep{benjakob2023hackingextortionelection}; the ``inventory of fictitious accounts'' were categorized based on regions/identities (e.g., Arabs, Russians, Asians, and Africans) and reused for different info ops \citep{megiddo2022israelisdestabilizingdemocracy} --- segmentation by region and recycling bots for multiple campaigns strongly suggest that Team Jorge would equip bots with social relations (if the bots' positions within a social network did not matter, swapping in a campaign-appropriate regional identity on preexisting bots instead of maintaining an extensive inventory would have been more cost-effective). AIMS bots also exhibit meticulous attention to detail, with some of the bots even having bank accounts \citep{benjakob2023hackingextortionelection}. Taking that fact into consideration, Team Jorge are unlikely to neglect situating their bots inside a suitably believable network of social relations at creation time. Percepto's case adds further support to the likelihood of bots having relations. One of Percepto's bots was an entirely fictitious investigative reporter based in Paris that had a real news agency with actual humans working under the bot \citep{megiddo2023antisemiticinfluencersfake}. Even the simplistic bots that apparently failed in manipulating Singaporean elections ``appear relatively ubiquitous throughout the social network'' despite a ``failure to amass network influence'' \citep{uyheng2021activeaggressivelittle}. The strongest evidence comes from bots used in the covert pro-Western info ops studied by Graphika and SIO \citep{graphika2022unheardvoiceevaluating}, which exhibited ``a typical long-tail distribution in the follower footprints, with a few influential accounts followed by a descending list of accounts with progressively fewer followers'', showing that the distribution of node degrees among bots mimic that of humans on social media (in most \citep{barabasi1999emergencescalingrandom, martha2013studytwitteruserfollowera} but not all cases \citep{broido2019scalefreenetworksarea} of human networks). The probability of an info ops bot army starting out from the periphery of a network is low.

\section{The Model --- Diluvsion}
\label{sec:model}
\par In our information diffusion model for Twitter-like social media, \textit{Diluvsion}, information is composed of two elements: stance and themes. Stances are polarized and exclusionary. A message can have multiple but non-repeated themes. The dissemination of stances through the network is of greater concern to us than themes. Details on stances and themes are found in Section \ref{sec:stance-themes}.
\par There are two major classes of agents in our model, humans (Section \ref{sec:humans}) and bots (Section \ref{sec:bots}). While ``humans'' are exclusively humans, ``bots'' can encompass artificial algorithmic entities and natural organic entities so long as they play the role of propagandist. Our motives for using this terminology are explained in Section \ref{sec:digital-personas}.
\par Section \ref{sec:memory} explains the attention-memory mechanism that agents are subjected to. In brief, the mechanism aims to mimic the impact that real humans' limited information processing capability and retention capacity can have on information diffusion. Section \ref{sec:engagement-action} lists the range of actions that agents can perform within the simulation to propagate or boost the propagation of stance and themes. Section \ref{sec:engagement-metrics} describes statistical data --- engagement metrics --- that agents are exposed to and affected by. Section \ref{sec:indirect-influence} describes indirect influence mechanisms as well as the rationale behind our model's reliance on engagement metrics as proxy for indirect influence.
\par The chosen resolution for \textit{Diluvsion} simulations, i.e., the length of real-world time that a single step in the model is equivalent to, is justified in Section \ref{sec:resolution}. The starting parameters of the simulation are detailed in Section \ref{sec:params}.
\par An assumption under our model is that all agents are devoted to debating one topic. They may devote a differing amount of time per day on social media and their forms of engagement may differ (e.g., spending most of their time liking posts they agree with, primarily retweeting others, etc.). These additional traits of the agents are sufficiently divergent that each combination constitutes a sub-class of bots and humans (Section \ref{sec:real-params}). 
\par Private accounts are excluded from our model primarily because they have a limited ability to engage in public debates; a private account's tweets can only be seen by the account's followers. For instance, if a private account sees an opinion they disagree with, refutations that the private account $u$ posts in reply will not be seen by the author $v$ of the disagreeable tweet and the author's, $v$'s, followers if the author $v$ is not a follower of the private account $u$. This limited reach of private accounts makes them inappropriate for inclusion in simulations of public info ops. Additionally, private accounts constitute a small minority of Twitter users so their exclusion would not drastically affect our results. According to a Pew Research Center survey on adult US Twitter users conducted in 2019, 13\% of Twitter accounts are private \citep{remy2019howpublicprivate}.
\par We also exclude the ``silent majority''/``no activity'' users (Section \ref{sec:real-params}) from our model because their lack of activity means that there is little data to collect on how they exert influence and how they, in turn, are influenced. Their lack of activity also means that they would have no discernible impact on the evolution of a Twitter discourse.
\par \textit{Diluvsion} is built on the Python agent-based modeling framework Mesa, which in turn uses the Python network analysis package NetworkX as a backbone.

\subsection{Stances and themes}
\label{sec:stance-themes}
\par Stances are opinions on an issue. To ease interpretability, we allow only three discrete stances in our model: two located at the poles and one at the exact midpoint between the two. We let the distance from the midpoint to either pole be one. The three stances are known as negative, neutral, and positive. Alternate names used throughout this paper for the negative stance include ``con'', ``against'', and ``opposition'' and the positive stance is also known by ``pro'', ``favor'', and ``support''.  While not all opinions can be modeled under this system, e.g., opinions on the actual worth of a used car, many can be, e.g., opinions on voluntary abortion. \par Our model differs from others in the literature, e.g., \citep{li2019multiagentsystemmodelling}, in that we allow neutrality as a stance to be propagated. Bipolar ideological battles have been consistently framed as drawing users to either pole, but not taking a stand can be as principled a stance as taking a stand, at least in principle, although perhaps not always in practice (some may argue that true equidistant neutrality in the strictest sense does not exist as everything exhibits a bias\footnote{Academics, e.g., \citep{mullen2010hermanchomskypropaganda, zollmann2019bringingpropagandaback}, have argued that mass media always represent ruling class interests, making impartiality impossible. To maintain a facade of impartiality, mass media ``strictly limit the spectrum of acceptable opinion, but allow very lively debate within that spectrum'' \citep{bahamonde2018powerstructurechilean}. Others have argued that journalism, ``in its pretense of ideological neutrality'', created ``a particularly pernicious myth'' as the presence of filters in both the news reporter and news consumer prevents neutrality \citep{meyers2020partisannewsmyth}. Spiral of silence \citep{noelle-neumann1974spiralsilencetheory} predicts that some people are unwilling to voice out an opinion to avoid contradicting what is perceived to be a majority opinion. Such inaction can be interpreted as neutrality, yet it does not contribute to the emergence of neutrality but instead contributes to the (potentially false) majority's continued dominance. Central banks invoking market neutrality to depoliticize corporate security purchase decisions are biased because ``[a]s critical political economists, sociologists and anthropologists have so frequently argued, markets embody a specific, political vision of society, which is not shared by every member of the polity'' \citep{vantklooster2020mythmarketneutrality}.}). An example of a manifestation of neutrality is the Non-Aligned Movement \citep{vieira2016understandingresilienceinternational} in geopolitics.
\par Stances are different from sentiments. Sentiments are concerned with the emotive content of a message, whereas stances are concerned with the message's position with regards to an issue of interest. For example, both statements ``I hate wearing masks'' and ``I hate doctors not wearing masks around immunocompromised patients'' show a negative sentiment due to their negative tone and presence of words such as ``hate'' in them, but in terms of stances, the first statement is against mask-wearing while the second is in favor.
\par Themes are the non-polarizable aspects of an opinion, and they are non-polarizable only with respect to the issue of interest. We illustrate the concept with an example where the issue of interest is mask-wearing: a message decrying a store for banning masked patrons for fear of criminal intentions and another message praising a store for banning unmasked patrons out of health concerns, despite differences in stance (anti- vs. pro-mask) and sentiment (negative in condemning vs. positive in complimenting), are united under the common theme of ``non-medical authority'' being exercised. And although people may have opinions on authority (favor or against), those stances have no direct bearing on the issue of concern, mask-wearing. Themes are comparable to the concept of frames in the field of communications \citep{levin1998allframesare, vreese2005newsframingtheory, chong2007framingtheory} but is \textit{not} equivalent to frames \footnote{To underscore the importance of themes/frames: Incidentally, the concept of framing introduced by \citeauthor{tversky1981framingdecisionspsychology} was first motivated by the ``Asian disease problem'' \citep{tversky1981framingdecisionspsychology}, where people were surveyed on their preferences to two solutions to an ``unusual Asian disease'', with one solution framed as saving lives and the other as causing deaths. Framing the disease as Asian betrays a problematic inclination harbored by the authors in their supposition that a disease terrible enough to force people to face a moral dilemma could have only originated from Asia or Asians, much like SARS and SARS-CoV-2 are exoticized and racialized as Asian \citep{zhou2016acceleratedcontagionresponse, decook2021kungfluroof}. But perhaps the framing \textit{of} framing's motivating problem can be excused as it dates from an era when people have yet to learn to better hide their prejudices. In the own words of \citeauthor{kahneman2011thinkingfastslow}, ``[s]ome people have commented that the `Asian' label is unnecessary and pejorative \ldots but the example was written in the 1970s, when sensitivity to group labels was less developed than it is today'' \citep{kahneman2011thinkingfastslow}. The term ``Asian disease problem'' remains in common usage in scholarly circles today.}. A frame is envisioned as a wrapper, a method of carrying messages. Different perspectives (wrapper) on an issue (message) changes one's perception on an issue. For us, themes function as more than wrappers. They are capable of being transmitted to (infecting) a target, just like stances. There can be multiple themes per message just as there can be multiple frames/wrappers per message. Multiple themes are similar to opinions encoded as binary strings in the rumor spreading model introduced in \citep{wang2017rumorspreadingmodel} and to a lesser degree the discrete-numbered cultures in the Axelrod model \citep{axelrod1997disseminationculturemodel}, SMSim's single-topic bit vector \citep{decgatti2013simulationbasedapproachanalyze}, and AMID's fuzzy set multi-topic messages \citep{li2018modellingmultipleinfluences}. 
\par The usefulness of explicitly modeling themes and having multiple themes per tweet/user can be demonstrated with an example involving an unpopular theme (adopted by a low percentage of agents). To make things easier to understand, we name this theme ``conspiracy''. By wrapping every message a bot army tweets out with a layer of conspiracy and by having more popular themes accompany the conspiracy theme since multiple themes are allowed, we can ensure that the conspiracy theme is more willingly adopted by others because of its constant presence and its association with popular themes. This ultimately results in a conspiratorial mindset slowly taking over the targeted human populace on social media. We indeed simulated such a scenario under \textit{Diluvsion} (Sim 9, Section \ref{sec:result-theme}). A less nefarious theme can always take the place of ``conspiracy'', e.g, ``affordability''.
\par Each message, i.e., tweet, in our model is composed of a single stance and one or more themes. Each agent has a stance and a set of themes as well. Agents can switch stances and can adopt themes over time. To keep the agents' themes from homogenizing as the simulation progresses because of the adoption mechanic, the maximum number of themes that an agent can possess at any one time is always lower than the set of all available themes in a simulation. This limitation also reflects that not all aspects (themes) of an issue can be the core interests of an agent; people have a limited capacity to care. All tweets by an agent will have the same stance as the one held by the agent at the time of tweeting. Themes, however, can differ between tweets (Sections \ref{sec:humans} and \ref{sec:bots}), e.g., a human agent will use the same themes as the tweet it is replying to, even if the themes are not part of the agent's theme set. Under \textit{Diluvsion}, compatibility in both the stance and the themes is required to improve the persuasiveness of a message to its recipient (Section \ref{sec:approx-params}). Once again using a mask-wearing example, a message that attempts to proselytize mask-wearing through the themes of safety and medical authority will find no purchase on an agent whose themes reflect a concern for comfort, e.g., masks being difficult to breathe through. Stance plus themes can also be thought of as viral variants, except that the transmission of stance and themes is dependent on homophily, but viral variants do \textit{not} necessarily operate under this mechanic\footnote{There is some superficial similarity in how newer variants of SARS-CoV-2 such as Delta and Omicron evolve greater binding affinity to ACE2 receptors in human cells than the ancestral wild type \citep{shah2021omicronheavilymutated, wu2022sarscov2omicronrbd}. Antibodies and some therapeutics either aim to present a more attractive binding target for the virus or bind to ACE2 receptors first before the virus \citep{sokolowska2020outsmartingsarscov2empowering, fiedler2021antibodyaffinitygoverns, wang2023predictionantibodybinding}, essentially outcompeting the virus. Themes are analogous to binding affinity, where a tweet exhibiting a greater number of matching themes (homophily) with the targeted user increases the chance of the tweet's stance being accepted by the targeted user, just as a the strength of binding affinity can influence whether an antibody (a stance) or a virus (a different stance) binds to a host cell.}.

\subsection{Memory and attention}
\label{sec:memory}
\par Memory and attention mechanisms are common features in many models of information diffusion, e.g., \citep{weng2012competitionmemesworld, wang2017rumorspreadingmodel, li2018modellingmultipleinfluences, li2019multiagentsystemmodelling, beskow2019agentbasedsimulation} and the motivations behind incorporating such mechanisms are largely similar. Modeling agents as having limited attention and limited memory is meant to reflect the reality where humans are incapable of digesting all the information presented to them and retaining the digested information over the long term. \textit{Diluvsion} features an attention and memory mechanism. An attention-memory limit governs the number of tweets that an agent sees per activation (when an agent wakes during a time-step and performs actions), the number of its own most recently tweeted tweets that the agent remembers, and the number of users that the agents remember seeing. Memory observes the first-in first-out (FIFO) queue, with the newest item displacing the oldest should the limit of a memory bin is reached. Since it is a combination attention and memory limit, it is possible for items that have entered an agent's memory to be forgotten before the items have a chance to be seen by the agent. For instance, if there are 31 new tweets directed to an agent whose memory is empty and has a limit of 30, the earliest/oldest 31\textsuperscript{st} tweet will be ``forgotten'' before ever being read by the agent. The memory of seen users is duplicative and cumulative: a user can appear more than once in an agent's history and users that appear more frequently are given greater weight in calculations (e.g., when deciding which user to follow).
\par While we keep track of the history of \textit{all} users seen by an agent (this history is different and separate from the limited memory seen user history) and the history of \textit{all} influential users (users that infected/changed the stance of the agent), they do not affect agent behavior and are only used by us to diagnose how influence spreads in the social network (Section \ref{sec:simulations}). These histories are also duplicative. 
\par Bots are constrained by a memory limit though theirs is several times higher than that of the humans. Bots are assumed to be more dedicated to spreading stances and/or themes than regular humans whose social media usage is less singularly goal-driven. Therefore, bots would dedicate more resources, such as greater attention and memory, to accomplish their goals. 

\subsection{Engagement}
\label{sec:engagement-action}
\par The ways in which a user can engage with the social media platform Twitter are referred to as actions in our model. The set of actions available to an agent in our model --- following, unfollowing, like, tweet, retweet, quote-tweet, and reply--- is different from the set of engagement metrics, which we will discuss in Section \ref{sec:engagement-metrics}.
\par Following is a one-way relationship where the follower $u$ indicates their interest in receiving updates on the activities of the followee, $v$, e.g., when $v$ tweets, retweets, etc. Unfollowing is the sundering of the follow relationship. A follower following a followee indicates that the follower is receptive to information flowing out from the followee. A loss of interest in the followee and/or the followee's tweets losing their informativeness often leads to the demise of the follow relationship \citep{kwak2011fragileonlinerelationship, xu2013structuresbrokenties}. In our model, a potential followee having matching stance and themes with a user increases the candidate's chance of being followed by the user.
\par Liking (previously known as ``favoriting'') is the most voluminous \citep{antelmi2019characterizingbehavioralevolution} action performed by Twitter users. Likes on social media can be considered a ``numeric representation of social acceptance'' \citep{rosenthal-vonderputten2019likessocialrewards}. The bandwagon effect --- the heuristic of following the crowd used by humans when other signals of authenticity are weak or absent --- can guide a person's decision to like, as people were found to be more likely to like photos with many likes over those with few likes \citep{sherman2016poweradolescenceeffects}.
\par Tweeting is the act of composing and posting a message on the social media platform. Since our model excludes private accounts, all tweets are assumed to be publicly viewable and searchable. While there are a number of factors (Twitter's algorithm, a human user's own preferences, etc.) affecting whether these viewable tweets are actually seen by users, our model assumes that an agent's tweets are always seen by the agent's followers so long as the tweet remains in the memory (Section \ref{sec:memory}) of the followers. While not the most common action, the production of tweets is the basis for all other actions, except for following and unfollowing, because liking, retweeting, etc., require tweets to first exist.
\par Retweeting is the sharing of another's tweet with one's followers without any embellishment. According to \citep{metaxas2015whatretweetsindicate}, retweets indicate interest, trust, and agreement with the message. As credibility and trust are important to having a tweet retweeted, our model factored trust into an human agent's decision on which tweet to retweet. Rather than using a randomly generated number to quantify trust relationship between agents, we relied on indirect influence, namely engagement metrics (like, retweet itself, and reply), as a proxy for trust (detailed in Section \ref{sec:engagement-metrics}, justified in Section \ref{sec:indirect-influence}). 
\par Replying is responding to another's tweet with one's own tweet, not as an independent tweet but as a tweet that will be added to the original tweet's reply thread, easily accessible by others should they tap or click on the original tweet. While not all replies challenge the stance of the parent/source tweet, replies are strongly weighted towards being negatory in nature. The Stance in Replies and Quotes (SRQ) dataset \citep{villa-cox2020stancerepliesquotes} shows that 44.0\% of the replies are against the stance expressed in the tweet being replied to (parent/source tweet), and that approximately half, 22.9\%, are supportive. The remaining 33.2\% are comments and queries about the parent tweet's stance, which can be interpreted as being neutral in nature. In the problem of detecting contradiction for rumorous claims \citep{lendvai2016contradictiondetectionrumorous}, there is a special category for detecting disagreeing replies (as opposed to expressing disagreement in independent tweets), highlighting replies as vehicles for disagreement.  Our model adheres to the conception of replies being primarily, though not exclusively, negatory in nature.
\par Quote-tweeting, alternatively known as quoting, is sharing another's tweet with one's followers while adding extra content from the quote-tweeter. Quote-tweeting is relatively rare, even when compared against replies. In a dataset of COVID-19 tweets that consisted of just retweets, quote-tweets, and replies, quote-tweets' share is 3.82\% while replies have 4.91\% \cite{dinh2020covid19pandemicinformation}. A study on the roles assumed by Twitter users on the platform \citep{antelmi2019characterizingbehavioralevolution} showed that quote-tweeting is seldom practiced, with Repliers as a role and replying as an action outnumbering Quoters and quoting across all roles and user activity levels save for Quoters themselves. The motivation behind quoting appears equally divided between refuting and affirming the original tweet. In the SRQ dataset \citep{villa-cox2020stancerepliesquotes}, 46.8\% of quote-tweets are negatory with respect to the quoted tweet and 43.5\% are affirmatory, with the remaining 10.7\% being comments and queries regarding the quoted tweet's stance. Stance ambivalence in motivating quote-tweets, as evinced by the even split between negatory and affirmatory, is modeled in \textit{Diluvsion} as compatibility in stance having a lower weight than themes when an agent picks which tweet to quote-tweet.
\par Ratios can also offer clues on whether agreement or disagreement drives an action (Section \ref{sec:engagement-metrics}). Actions as modeled in \textit{Diluvsion} is generally in agreement with findings from ratiometrics.
\par In our model, an agent's decision to like, retweet, or reply to a tweet is partially based on the agent's agreement with the tweet's stance and themes, and partially on whether the tweet's popularity merits bandwagoning. Whether popularity or compatibility is the dominant factor depends on which action is being taken, as detailed in the explanation of this mechanism in Section \ref{sec:approx-params}. The existence of a follower-followee (or inverse) relationship between agents does not influence an agent's action; the ``text semantics and contents'' of a tweet, ``regardless of tweet author'', are the most important factors in determining if a user engages with a tweet through liking, retweeting, replying or quoting \citep{toraman2022understandingsocialengagements}.
\par Agents who received replies to their tweets or have had their tweets quoted usually receive notifications, so we assume that information from replies and quote-tweets diffuse to the targeted agents. Additionally, replies to a reply-tweet are read by a fixed percentage of the followers of the author of the tweet being replied to, which is a form of non-social tie information diffusion (Section \ref{sec:indirect-influence}).
\par Like Twitter and other real-world social media platforms, we only allow a user to like a tweet once and retweet a tweet once. The practice of including a link to a tweet, $a$ in a reply-tweet, which increases the retweet counter of $a$ and can be performed multiple times, is excluded from our model. And although it is possible to reply to the same tweet multiple times on Twitter, we restrict the number of times an agent can reply to a tweet to just one. Any information that a replier wishes to convey through multiple replies to a parent tweet can be thought of as having been consolidated into one reply.
\par Tweeting is the action that all agents, bot and human, default to if the agent's attention-memory contains no seen tweets or if the chosen action for a particular time-step of the simulation is invalid. Many actions, e.g., liking and retweeting, require tweets to be present before they can act on those tweets, so this default behavior helps populate the simulation with tweets. Invalid actions are common around simulation start due to a confluence of three factors (1) the lack of tweets, (2) limiting each tweet to one like, retweet, quote-tweet, and reply each per user, and (3) liking and retweeting being common actions. A consequence of tweet scarcity is that the action distributions generated from the simulation deviate slightly from empirical data, despite the model using parameters from empirical data, as we shall discuss in Section \ref{sec:validation}.

\subsubsection{Engagement metrics and ratioing}
\label{sec:engagement-metrics}
\par There are three prominent engagement metrics --- likes, retweets, and replies --- visible to Twitter users. While we allow for quote-tweeting as an action in our model, its metrics are not visible to agents in our model and are therefore incapable of influencing the agents (Section \ref{sec:indirect-influence}). In the context of stance conversion, likes are positive, retweets are moderately positive, and replies are moderately negative.
\par Excluding the number of quote-tweets from engagement metrics is because of this metric being hidden throughout much of Twitter's existence. In 2020, quote-tweets were still semi-hidden, with Twitter forcing users to tap/click on a tweet then on the tweet's retweets before they could see the number of retweets broken down into retweets and quote-tweets \citep{twitter2020donmisstweets, peters2020twitterwillnow}. Besides quote-tweets, we also exclude additional metrics that have been surfaced since Elon Musk's Twitter acquisition, such as the counts of views and bookmarks a tweet has received. This is because the foundations of our model --- engagement data we and others have collected (Section \ref{sec:real-params}) and the vast majority of scholarly literature on the impact of engagement --- date from the era when these statistics were hidden. On Twitter's current timeline interface where users scroll through tweets, only viewcounts have been added alongside the counts of likes, retweets, and replies. The newly surfaced engagement metrics are also rarely displayed on other platforms, e.g., Instagram and YouTube only show the number of likes and replies that a post has received while the Twitter-like Mastodon shows retweets, replies, and likes.
\par The number of followers is an engagement metric that does not influence human agents. Although a user's follower count is publicly visible to anyone who visits the user's profile page, it is not visible in the context of other actions --- liking, retweeting, replying, and quote-tweeting. Visiting a user's profile page is an extra step from the typical Twitter usage behavior of ``infinite scrolling'', breaking from a user interface (UI) designed to entrap users in perpetually consuming media \citep{noe2019identifyingindicatorssmartphone, lyons2022designdevelopmentmobile}. For following and unfollowing a user, these actions are available as part of a dropdown menu accessible by clicking an icon on a tweet; visiting a user's profile page and being made aware of the user's follower count are entirely avoidable. Bots, however, do use follower count as information for some of their decisions, as they are motivated agents of stance/theme propagation.
\par Ratioing on Twitter is a phenomenon that best expresses the logic behind engagement when Twitter conversations involve contested issues. When a user expresses an objectionable opinion, others may express their disagreement by replying to the user's tweet. Should the number of replies (regardless of whether they all express disagreement) or the number of likes of any one of the disagreeeing tweets outnumber the likes of the parent tweet, a ``ratio'' has been achieved, i.e., the ratioed user has been publicly and quantitatively mocked. A study on an info ops campaign waged by social media users opposed to the fossil fuel industry used ratioing as a means to identify tweets targeted by the operatives as the operatives employed the strategy of questioning fossil fuel organizations’ legitimacy with large numbers of negative comments \citep{troy2022getratioedquestioning}. A broader definition of ratioing includes cases in which the number of replies exceeds the number of retweets. This definition was employed in a study of the different ratios of engagement metrics that tweets from Presidents Trump and Obama received \citep{minot2021ratioingpresidentexploration}, which found that Trump consistently achieve higher reply-retweet --- more controversial --- ratios compared to Obama.

\subsubsection{Indirect influence and trust}
\label{sec:indirect-influence}
\par Our conceptualization of influence departs from some works in the literature that regard all influences as having a positive impact. These works include \citep{rath2018utilizingcomputationaltrust}, which posits that the high trustingness of an acting agent and the high trustworthiness of the acted-upon agent are the determinants behind the acting agent's decision to act (retweet, reply). Another work is \citep{zhang2019learninginfluenceprobabilities}, which deemed one user as having influenced another if the user engaged with the tweet in any way (retweet, like, reply, quote, and follow). From the perspective of stance diffusion, such conceptualizations of influence clash with empirical observations of why people choose to respond to messages. As discussed earlier in Section \ref{sec:engagement-action}, replies and, to a lesser extent, quote-tweets are often used to make rebuttals. Ratioing (Section \ref{sec:engagement-metrics}) demonstrates that not all interactions constitute affirmative influence. The disconnect between the models \citep{rath2018utilizingcomputationaltrust, zhang2019learninginfluenceprobabilities} and the data is a result of ignoring polarity/stance. \citeauthor{zhang2019learninginfluenceprobabilities} explicitly stated that their method of finding matching tokens between two tweets ``identify whether their topics are similar''; nothing is said about the user's stance on the topic.
\par Direct influence comes from direct attempts at persuasion, which in our model consists of the contents of a tweet --- its stance and themes. Direct influence travels along the prescribed lines of influence, social ties, from followee to follower. Indirect influence is information outside of a tweet's content that can influence an agent's adoption of the contents. Indirect influence can either increase or decrease the persuasiveness of tweets. Indirect influence can operate outside of observable follower-followee relationships.
\par One form of indirect influence modeled under \textit{Diluvsion} is the ability for replies to circumvent social ties when transmitting information, where a certain percentage of the followers of the parent tweets' authors will read the replies (Section \ref{sec:approx-params}). Tweets in most cases can be replied to by any user, not just followers or followees. More than just transmitting information to the user targeted by the reply, replies can also transmit information to the targeted user's followers. A browse through Twitter will reveal that many replies to tweets that have gained a lot of attention (high number of likes, retweets, etc.), colloquially known as ``going viral'', are advertisements or posts about topics tangential to those of the original tweets, e.g., fake BTC giveaway scams \citep{mazza2022readytousefake} and shilling for cryptocurrency. They are attempts to take advantage of the tweets' popularity. The hope is that some of the attention can be diverted to profitable or persuasive ends. 
\par Attempts have been made to model such indirect routes of information diffusion. The model proposed in \citep{averza2022evaluatinginfluencetwitter} allows for tweets that have crossed a certain threshold of engagement numbers to go viral, meaning that the tweet will be ``shown to all agents in the network that share the same topics with the sender agent, even if an edge does not connect them''. From our perspective, the indirect propagation of information captured in their implementation appears limited, as most information still travels along the network edges in their model. The sole exceptions were viral tweets. In \textit{Diluvsion}, virality would naturally emerge from the propensity for agents to favor sharing or otherwise engaging with popular tweets over unpopular ones (bandwagoning) since engagement metrics, themselves a form of indirect influence, factor into an agent's decisions on which tweet to engage.
\par The engagement metrics discussed in Section \ref{sec:engagement-metrics} constitutes a form of indirect influence. The metrics serve as both \citep{lin2016socialmediacredibility, shin2022twitterendorsedfake, epstein2022howmanyothers} social cues for the bandwagon effect \citep{nadeau1993newevidenceexistence, barnfield2020thinktwicejumping} and social cues for computing source trustworthiness \citep{lin2016socialmediacredibility, li2018modellingmultipleinfluences, rath2018utilizingcomputationaltrust} and message credibility \citep{shin2022twitterendorsedfake}.
\par Engagement metrics taking the place of credibility is supported by the existing research literature. One study found that ``bandwagon heuristics---such as the number of likes, comments, and retweets---increased the credibility of news'' \citep{shin2022twitterendorsedfake}. The automated activity of social probing by non-human-like bots with zero trust, i.e., strangers, which in the case of \citep{aiello2012peoplearestrange} consists of visiting the profile page of others, can boost popularity. That popularity translated successfully to influence, which was getting others to follow accounts recommended by the bots. The number of retweets, despite its noisiness and bias, has a positive correlation with expert-assessed credibility of messages, and it is effective at predicting ground-truth credibility when combined with credibility ratings from a different crowd, Mechanical Turk annotators \citep{sikdar2013understandinginformationcredibility}. Cues in the form of retweets (information sharing) boosted the credibility of a tweet's author regardless of whether the tweet originated from a peer or a stranger, and even if the retweets were performed by strangers \citep{lin2016socialmediacredibility}. Conversely, source credibility was found to play a major role in the information sharing behavior of users on Twitter \citep{ha2011whyareyou}. Credibility and influence are intertwined, feeding off each other and creating a rich-get-richer effect, where the wealth is social capital. 
\par The need to impute a separate trust value for a social tie shared between two agents is also obviated by engagement metrics that can serve as proxies. We improve upon the trust mechanism found in \citep{li2018modellingmultipleinfluences, rath2018utilizingcomputationaltrust}, where the proxy variables for trust are the number of times a user has retweeted \citep{li2018modellingmultipleinfluences, rath2018utilizingcomputationaltrust} and replied \citep{rath2018utilizingcomputationaltrust} to another user, by factoring in the relationship between engagement type, stance distance, and themes; weighing the different engagement metrics; and counting indirect engagement actions instead of direct ones. The authors of \citep{rath2018utilizingcomputationaltrust} themselves noted that replies as a proxy can be unreliable because replies may contain users expressing a stance different from the parent tweet, lending support to our decision to account for stance when using engagement metrics. Whether an engagement type is driven by heterophilia or homophilia in stances and themes (and consequently the metric generated for the engagement type) under our model is explained in Section \ref{sec:engagement-action}. 
\par As for using indirect engagement numbers over direct ones as proxies for trust, indirect engagement can be more reliable than direct engagement despite appearing counter-intuitive.
\par Traditional indicators of sources as being authoritative are less visible online \citep{metzger2013credibilitytrustinformation} because of a flattening of hierarchies. Surmounting the accessibility barrier (e.g., radiocommunication spectrum licensing, capital costs of print media) and the expertise barrier (e.g., sufficiently credentialed to be invited onto mass media to comment on an issue) denoted authoritativeness pre-social media, but under social media where such barriers towards broadcasting a message has disappeared, so too has their function as indicators of trustworthiness. Distrust in authoritative sources has also been on the rise, further weakening traditional signals of credibility. Distrust in authority being a widespread phenomenon is evident in Twitter discourse on vaccine and climate change where ``science'' (science as a source of authority, not science as practiced) is regarded with suspicion \citep{jacques2016hurricaneshegemonyqualitative, haltinner2021howbelievingthat, quintana2022polarizationtrustevolution}. While an authoritative source can still engender trust on social media, a stranger retweeting information from an authoritative source actually \textit{reduces} the trustworthiness of the source, whereas a stranger retweeting peers or other strangers increases trustworthiness \citep{lin2016socialmediacredibility}. The rising distrust in authorities was an impetus for the development of bot-augmented decentralized info ops (as discussed in Sections \ref{sec:decentralization} and \ref{sec:bots}). On social media, the audience has also been flattened through the phenomenon of context collapse \citep{marwick2011tweethonestlytweet}, where users cannot tailor messages to pander to specific subgroups because messages posted on social media are viewable by all, making it easy to offend others, with the offense causing diminished trust in the source \citep{fernandez2021wordsmatterwhat}.
\par With direct trust in a message's source significantly dampened by flattened hierarchy and context, message credibility and user trustworthiness online are strongly coupled. Social media users become only as trustworthy as the credibility of their messages. People judge the credibility of information by engagement metrics attached to that information \citep{lin2016socialmediacredibility, shin2022twitterendorsedfake}. Consequently, crowdsourcing trust is an established practice, e.g., computation of a Twitter user's credibility based on the degree of agreement between the user's tweets and their respective replies \citep{wijesekara2020sourcecredibilityanalysis}. Engagement-based source credibility indiscriminately boosts the credibility of peers \textit{and} strangers \citep{lin2016socialmediacredibility}. These engagement metrics are susceptible to the bandwagon effect, in which people base their opinion on what they perceive to be the opinion of the collective herd \citep{lin2016socialmediacredibility, shin2022twitterendorsedfake}, e.g., liking what others liked and retweeting what other have retweeted. Another factor at play is minority-avoidance and the spiral of silence, as discussed in Section \ref{sec:contested-models}. Trust quantified by indirect engagement numbers is the overwhelmingly dominant form of trust online.
\par That is not to say that a message source under an indirect trust system, e.g., as implemented in \textit{Diluvsion}, is wholly incapable of affecting the message's credibility, and by extension, the source's trustworthiness. A source that has amassed a large following of users sharing similar stances is likelier to receive more likes for its message than a user with zero followers.
\par The pervasive indirect influence of engagement metrics extends beyond how one perceives others and into how one perceives oneself. Every additional like or retweet that a user receives for their own tweets serves as a small validation, affirmation, and acceptance of the online self \citep{rosenthal-vonderputten2019likessocialrewards, brubaker2020digitalhyperconnectivityself}. Under \textit{Diluvsion}, an agent's memory of its own recently tweeted tweets also influences stance adoption, and the agent's perception of those recently tweeted tweets are shaped by their respective engagement metrics. Because an agent's tweet's stance is always the same as the agent's (unless they recently changed their stance) under our model, if an agent's recent tweets have received a lot of affirmation (e.g., a high number of likes), the agent will be more likely to retain their stance even when they encounter a lot of opposition to their stance in their memory of seen tweets (memory of others' tweets that they saw).
\par
\begin{figure}
    \centering
    \includegraphics[width=0.95\textwidth]{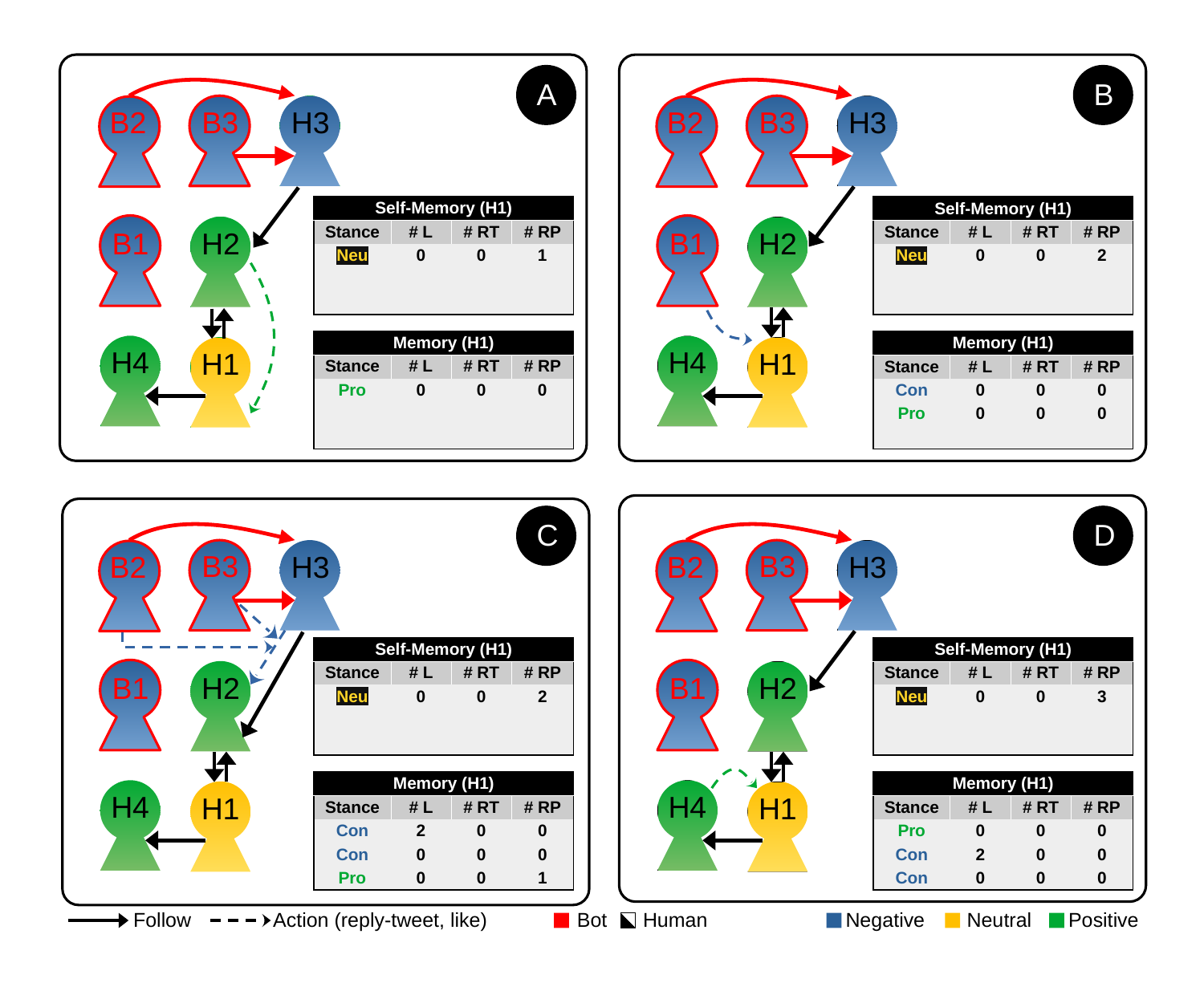}
    \caption{At time-step 1, a neutral human agent (H1) makes a tweet (action not depicted in this figure) expressing a neutral stance and themes of medical authority and individual liberty over collective good, e.g., ``My doctor tells me wearing masks is a personal preference.'' \textcolor{White}{\hl{\textbf{(A)}}}  While still at time-step 1, a second human agent (H2), H1's follower, concocts a reply-tweet with matching themes but a positive stance, e.g., ``The CDC said masks protect everyone, not just the wearer.'' \textcolor{White}{\hl{\textbf{(B)}}} A negative bot (B1) calls \texttt{poll\_tweets}, which is equivalent to doing a search of existing tweets, finds H1's neutral tweet, and replies with a negative tweet. \textcolor{White}{\hl{\textbf{(C)}}} A third human agent (H3) makes a reply-tweet with a negative stance to H2's earlier reply to H1. Two bots (B2 and B3) liked H3's reply-tweet. H3's reply happens to be seen by H1 because H1, as a follower of H2, occasionally reads the replies to H2's tweets. \textcolor{White}{\hl{\textbf{(D)}}} A fourth human agent (H4) makes a tweet with a positive stance that is seen by H1, which pushes the very first tweet seen out of H1's limited memory. \textcolor{White}{\hl{\textbf{(E, not shown)}}} At the start of time-step 2, when \texttt{read\_conversion} is called, H1 observes one positive-stanced tweet without any engagement, two negative-stanced tweets, one which received two likes, and finally its own neutral-stanced tweet that received three replies. To determine if H1 will experience a stance conversion after reading these tweets, we check if a generated uniform random number clears H1's resistance threshold. The number does. To see which stance from the seen tweets infect H1, a weight is calculated for each stance. The negative stance has a heavier weight than neutral or positive because the negative tweets are greater in number and have affirmative engagement instead of negatory (likes instead of replies). Weighted random choice picks the negative stance, H1 changes its stance to negative.}
    \label{fig:infection}
\end{figure}

 Figure \ref{fig:infection} demonstrates how indirect influence operates to infect an agent. This example omits the influence of stance and theme compatibility between the tweets and the agent in an infection event. For an example that does include it, see Section \ref{sec:result-embed-v-fringe}.

\sethlcolor{Black}

\subsection{Humans}
\label{sec:humans}
\par An agent designated as ``human'' in our model is an idealized representation of a typical social media user. All human agents follow the same heuristics, and they also have the same maximum memory size and maximum number of concurrent themes. Human agents may differ in their starting stance and themes, starting infection resistance, activity level, and roles. Over the course of the simulation, their stances and themes may change. Resistance evolves with every infection (stance change). Their activity level and roles remain fixed. The heuristics and maximums remain fixed as well.
\par The roles taken on by human agents, as well as their activity levels, are based on empirical Twitter data (Section \ref{sec:real-params}). Roles are the five groups that users were placed into based on their favored engagement action on Twitter, e.g., they were known as Likers if they exhibited a tendency to like others' tweets rather than compose their own tweets. Users were also separately placed into three activeness categories depending on the number of observable actions taken per week. Each role, with respect to its activity level, will have the highest frequency for the engagement action that the role is named for, e.g., moderately active Quoters will have the highest frequency of quote-tweeting compared to other moderately active roles. However, the same role will have different action frequency distributions between different activity levels, e.g., a highly active Quoter will be likelier to like tweets than they are to quote-tweet, but a moderately active Quoter will quote-tweet more often than like.
\par At every time-step, a check is performed to determine whether a human agent wakes. If it wakes, it will read tweets and potentially be infected by the tweets it has read. If it is awake, a second check is performed to determine whether it has used up its allotted quota of weekly actions (this quota is based on its assigned activity level). If it has not, the agent will definitely act during this time-step. If it has, the agent will have a very small fixed random chance of acting. The actions that an agent can perform are discussed in Section \ref{sec:engagement-action}.
\par For all actions, the part that involves broadcasting a stance --- i.e., \textit{writing} a reply, a tweet, or a quote-tweet --- the broadcasted stance is always the same as the agent's. Themes for replies and quote-tweets always match the themes of the parent tweet; the assumption is that when responding to another, normal users would not go off-topic. Themes for an independent tweet are drawn from the agent's own pool of themes, with themes that have been recently used by an agent having greater probability of being chosen. For the part of an action that does not explicitly broadcast a stance, i.e., \textit{selecting} a tweet to like, retweet, quote, or reply to, the agent will favor tweets with stance and themes that align with its own, but the agent is also influenced by engagement metrics to pick tweets with different stances and themes.
\par To be infected, a uniform random number must first exceed an agent's resistance threshold. A high resistance results in fewer infection events being triggered. If the threshold is cleared, one stance among the tweets an agent has read will be selected as the infecting stance. The selection is random, but each stance is weighted by the engagement metrics (Section \ref{sec:engagement-metrics}) and by the compatibility of the tweets representing the stance. The compatibility calculation factors in themes as well, e.g., for a neutral agent (equidistant from both poles), a group of tweets representing the positive stance that has many matching themes will influence the agent more strongly than a similar-sized group of negative tweets with non-matching themes, all else being equal. A successful infection results in the agent's adoption of not just the infecting stance but also a portion of the most popular themes found in the tweets representing the infecting stance. Examples of the infection mechanism at work can be found in Figure \ref{fig:infection} (the influence of stance and theme compatibility is omitted in this example) and in Section \ref{sec:result-embed-v-fringe}.
\par Although we have called infection as stance ``change'' or ``conversion'' throughout this paper, an infection event can end with an agent being infected by the \textit{same} stance. This situation can be thought of as witnessing strong arguments that affirms one's faith in one's stance.
\par Resistance is altered by infections. If the infecting stance is the same as the one held by an agent at the start of an infection event, the infection will improve resistance. Infections by non-matching stances degrade resistance. This mechanism is inspired by certain viruses such as the Hepatitis C virus (HCV) whose infection some people can clear, whereas in certain other population segments they induce an impaired immune response thereby allowing reinfections or chronic infections \citep{burke2010hepatitisvirusevasion, farci1992lackprotectiveimmunity}. More domain-appropriate inspirations are the observed instances of people slowly doubting their worldview if some strongly held beliefs have been shattered (e.g., South Korean feminists ``taking the red pill''\citep{jeong2018wetakered}, spiral of mistrust/cynicism \citep{garland2021surrender24news}). 


\subsection{Bots}
\label{sec:bots}
\par Bots are propagandists, biological (Section \ref{sec:digital-personas}) or electronic, dedicated to promoting a stance on a social network. The key differences between bots and humans are: a bot's stance is fixed, a bot never unfollows another user, a bot constantly surveys the conversation landscape, and a bot has a greater memory capacity. The set of heuristics that a bot follows also differ from that of humans as they are designed to spread their stance as widely as possible. None of the differences should be interpreted as bots having superhuman abilities or access to tools on social media platforms that regular users do not. The difference is purely a result of motivation and prioritization. Bots are just as limited as humans in terms of their activity levels and roles.
\par A bot surveying the conversation landscape, \texttt{poll\_tweet}, at the start of every time-step is equivalent to running a search or covertly reading tweets from accounts that it does not follow. A fixed number of the most recent tweets are returned with each survey attempt and these tweets are saved into the bot's memory of seen tweets. Surveying allows bots to take note of tweets/conversations that may have escaped the attention of human agents. If necessary, bots can jump into the conversations to influence other agents. This form of non-social tie information acquisition is exclusive to bots. For human agents, non-social tie information is acquired if they happen to be randomly selected as the followers who have read a reply to their followee's tweet (see Section \ref{sec:engagement-action}; the mechanism applies to bots too).
\par Bots do not unfollow so as to avoid triggering reciprocal unfollowing among their followers. Bots want as large an audience as possible to maximize the reach of their messages. This desire for a large audience is also manifested when choosing which tweets to quote, retweet, or reply to. Unlike humans who will only consider engagement and compatibility, bots will additionally consider a tweet author's follower count, favoring those who have large counts.
\par Bots not changing their stance is simply a result of them being bots, i.e., unyielding propagandists. Their greater memory capacity (Section \ref{sec:memory}) is also due to them being motivated agents of change.
\par However, bots do adopt new themes. Since bots are conceptualized as being interested in altering \textit{only} the stances with regard to the issue of interest, theme adoption does not compromise the bots' mission as themes are non-polarizable with regard to the issue of interest (Section \ref{sec:stance-themes}). It may even aid in the bots' mission as bots can better empathize with the rest of the agent population, e.g., composing tweets with themes that resonate with other agents and are able to infect others more because of greater theme compatibility.
\par When bots have to choose which tweet to reply to, they have two tactics --- targeting vulnerable users and targeting high engagement tweets --- that they alternate between. A probability distribution governs how often one tactic is used over the other. To target vulnerable users is to target tweets from users who have a low number of followees. The reasoning for this tactic is that such users have low number of information sources (followees) and are therefore more easily swayed by a flood of information from non-followees. To target high engagement tweets is to target tweets whose engagement metrics are the highest among those that a bot has witnessed in its memory and are authored by users with high follower counts. This tactic maximizes the visibility of a tweet as high engagement tweets are likelier (due to bandwagoning) to be targeted by agents for actions that trigger information sharing such as replying, retweeting, and quote-tweeting. This tactic has real-world precedence, e.g., BTC scams \citep{mazza2022readytousefake}.
\par Replies and quote-tweets authored by bots have the same stance as each bot's respective stance, just like humans, but the process of themes selection differs from humans. While humans would use the same themes as the parent tweets, bots would only use half of the themes from parent tweets, while the other half is drawn randomly from each bot's pool of themes. In the case where a bot's themes deviate significantly from the population, this tactic serves to subtly inject the bot's themes into the information environment to facilitate infection in the future, while infection chances at present are improved by adopting half of the same themes as the parent tweet. Theme selection for independent tweets also differs from humans, as a bot does not select based on a history of its own tweets. Instead, following the same selfish/motivated reasoning as when replying and quote-tweeting, half of the themes are the most common themes found in a bot's seen tweet memory while the remaining half are randomly chosen from the bot's pool of themes.
\par Like humans, bots start out embedded in a network instead of starting out in the network's periphery. Section \ref{sec:actors} explains the reasoning behind this positioning. The bots in our simulation are bound by the same diurnal activity schedule as the one humans follow. This is informed by the activity data of the covert pro-Western info ops' assets, which have peaks and troughs \citep{graphika2022unheardvoiceevaluating} that mimic a human's. Motivated by decentralization (Section \ref{sec:decentralization}), the bots act independently (just like human agents) and are not conscious of the presence of other bots. Naturally, this excludes explicit coordination between the bots.
\par Bots operate the same across stances within \textit{Diluvsion}, unlike other works that split propagandists based on stance. For instance, the ``bad guys'' bots/trolls class of agents in \texttt{twitter\_sim} actively spread disinformation while the ``good guys'' stiflers in \texttt{twitter\_sim} \citep{beskow2019agentbasedsimulation} are reactionary by design, only countering disinformation that they see and never proactively spreading truth.

\subsection{Temporal resolution}
\label{sec:resolution}
\par Humans follow a diurnal cycle, and their inclination towards being active on social media follows a similar cycle \citep{grinberg2013extractingdiurnalpatterns}. Therefore, the human and bot agents in our model also observe this cycle. The mean number of tweets from a dataset of mask-wearing conversations on Twitter, binned by the day of the week and the time of the day, gave us an action-probability table. This table approximates the likelihood of an average Twitter user using the service at a particular time-step. Within each time-step, the order in which the agents activate is random. A consequence of simulating peaks and troughs in platform activity levels is that during periods of high activity, information dissemination and actions to boost dissemination will reach the highest number of awake users. However, at the same time, the great volume of information flowing during high activity periods means that disseminated information risks being pushed out of the memory (Section \ref{sec:memory}) of agents with high numbers of incoming links.
\par The resolution of the simulation is 30 minutes. This resolution is sufficient and has empirical support: the Graphika and SIO investigation found that the info ops ``assets in the Afghanistan and Central Asia groups typically posted at roughly 15-minute or 30-minute intervals in any given hour'' \citep{graphika2022unheardvoiceevaluating}. Our decision was also influenced by the mean weekly frequency (the number of actions taken by a user in a week) for the high-activity group in the study by \citeauthor{antelmi2019characterizingbehavioralevolution}, which was found to be 133.84 \citep{antelmi2019characterizingbehavioralevolution}. If we took 133.84 actions and divided them over seven days, this leaves approximately 19 actions per day. At a resolution of 30 minutes, 19 actions will take 9.5 hours, including the interval time between actions. This wakefulness period does not violate typical sleep-wake cycles within human circadian rhythms \citep{ferrazcosta2015rscminingmodeling}.
\par As validation, when we generated 1,000,000 sets of 336 random numbers ($2 \text{ periods per hour} \times 24 \text{ hours} \times 7 \text{ days}$) with a 0.9 damping factor\footnote{When multiplied by probabilities in the action-probability table (Section \ref{sec:real-params}), which can range from 0 to 1, the factor shifts the graph down/reduces the amplitude, i.e., lowers the action probability.} and tested them against the activity-probability table (derived from our mask-wearing dataset), the maximum number of periods/time-step in which an agent is awake was $153.45 \pm 8.46$ per week, a range which can accommodate the maximum number of actions the most active class of agents can take per week within our model (we let this maximum be the rounded mean number of actions of high activity users, 134, from the dataset in \citep{antelmi2019characterizingbehavioralevolution}, which is different from the mask-wearing dataset used to generate the action-probability table). To allow for variability, once an agent exhausts its allotted action quota, if a random number check against the action-probability table says that the agent is awake and can therefore take an action, the agent has a fixed 1\% chance of exceeding the quota.

\subsection{Graph structure}
\label{sec:graph-generation}
\par Directed scale-free graphs were used as the simulation environments. Specifically, we used the improved version \citep{schweimer2022fastgenerationsimple} of the algorithm first introduced in \citep{schweimer2022generatingsimpledirected} by \citeauthor{schweimer2022generatingsimpledirected} for generating random directed social network (SN) graphs for use in information diffusion studies. The hyperparameters used for graph generation comes from the hyperparameters file in the open-source code repository shared by the author \footnote{\url{https://github.com/Buters147/Social_Network_Graph_Generator}} which themselves are derived from crawling real-world social networks on Twitter. The set of hyperparameters used were named ``Covid'', the largest of the 14 networks crawled with 50,133 nodes and 4,832,226 edges.
\par Desirable network properties such as a high amount of clustering and reciprocity are reproduced by the SN graph generation algorithm. Clustering describes users aggregating into homophilic clusters when interacting online on Facebook and Twitter \citep{cinelli2021echochambereffect}. Reciprocity describes the small but still significant amount of reciprocal edges on social media (22.1\% found on Twitter \citep{kwak2010whattwittersocial}). If we used non-SN scale-free graph generation algorithms, e.g., the one proposed in \citep{bollobas2003directedscalefreegraphs} by \citeauthor{bollobas2003directedscalefreegraphs} which improved upon the Barabasi-Albert method, certain properties of the network deviate further from expectations than if an SN graph generation algorithm was used (Section \ref{sec:validation}).
\par A total of 20 graphs with 1,000 nodes and 20 graphs with 900 nodes were generated. Because of an inherent flaw \citep{schweimer2022generatingsimpledirected} of the graph generation algorithm itself, approximately 2.5\% of the nodes for each graph are isolates, i.e., the nodes have no incoming or outgoing edges. The graphs were saved and reused for different scenarios to allow for reproducible results and also because graph generation is time consuming. The graphs with 900 nodes were used for the ``peripheral''/``fringe'' scenarios where bots (10\% of the network population, equivalent to 100 agents) do not start out embedded in the network but is added to the graph later and given only one outgoing edge to a random node, i.e., a bot's only social tie is being a follower of an existing human user.

\subsection{Parameterization}
\label{sec:params}
The same set of parameters are used for validation (Section \ref{sec:validation}) and simulations (Section \ref{sec:simulations}).

\subsubsection{Empirical data}
\label{sec:real-params}
\par To ensure that agents populating the network mimic the behavior of those found in real-world social media, we initialize the agents based on statistical data collected from Twitter users.
\par For the diurnal activity schedule, we took the mean number of tweets split by day of the week and into half-hour intervals (the resolution of the simulation) from the two million tweets we collected on COVID-19 masking debate between January 1--June 21, 2020 as a probability distribution for when an agent will be active.
\par The starting distributions of stances and themes, including the starting number of themes per agent, are taken from the February 27--March 2, 2020 period of the COVID-19 masking debate, which we have identified as the starting point for when the majority of the discussions occur within the appropriate context of masking as protection against COVID-19 instead of masking in other contexts, e.g., masks to protect against volcanic ashfall. A total of 12 themes are identified from the dataset. Examples of the themes are ``encourage'' (encouraging people to mask or to not mask) and ``appraisal-critize'' (praising others for wearing or for not wearing masks).
\par Although the proportions of stances and themes at the population level are maintained, the stance and the themes chosen for an individual human agent are entirely random, unaffected by the agent's neighbor's stance and themes. Some may take issue with our choice since homophily would predict that users sharing similar outlooks, political or otherwise, are more likely to be connected, so there should be homogeneous stance and theme clusters in the network. For emerging issues where strong narratives have not been established, the situation is not so clear-cut. Republicans are more anti-mask during the COVID-19-pandemic than Democrats \citep{kramer2020moreamericanssay, frankovic2020howtrumpbiden}, but the Twitter data on mask-wearing that we collected showed that right before the pandemic, many fans of Trump supporter Scott Presler were urging him to wear a mask to avoid catching disease from homeless people.
\par The counts of users found in the high, medium, and low activity categories are taken from \citep{antelmi2019characterizingbehavioralevolution}. For the range of actions available to the agents (Section \ref{sec:engagement-action}) except for following and unfollowing, the frequency distribution is based on the tables and figures in \citep{antelmi2019characterizingbehavioralevolution}. In \citep{antelmi2019characterizingbehavioralevolution}, the authors categorized users into five roles --- Tweeters, Quoters, Retweeters, Repliers, and Likers --- with a corresponding action frequency distribution. Each role is further sub-divided into high, medium, low, and no activity. Although almost 50\% of users belong to the no activity cluster (no visible activity but have done at least one hidden activity such as changing their screen name), we exclude this category because they would have no appreciable impact on the simulation dynamics owing to their inactivity. This leaves us with $5\times3=15$ human agent categories.
\par Statistics on following and unfollowing are not found in \citep{antelmi2019characterizingbehavioralevolution}. We relied on the descriptions provided in \citep{kwak2011fragileonlinerelationship, myers2014burstydynamicstwitter}. According to \citeauthor{kwak2011fragileonlinerelationship}, the ``average of unfollows is 15.4'' and ``average number of followees increases from 59.7 to 75.7'' in 51 days with ``about 30\% of users [having] unfollowed at least once'' \citep{kwak2011fragileonlinerelationship}. According to \citeauthor{myers2014burstydynamicstwitter}, ``[i]n a given month a user of degree 100 tends to gain 10 and loose [sic] 3 followers'' \citep{myers2014burstydynamicstwitter}). Their descriptions showed that while following and unfollowing are pervasive over a scale of multiple months, they are rare when one factors in all the possible actions that a user can take (tweeting, retweeting, etc.) at the scale of half-hour steps which our simulation runs at. Modeling the actions of following and unfollowing as suitably rare events translates to having follow and unfollow constituting $0.\overline{6}\%$ and $0.\overline{3}\%$ of all of an agent's actions, respectively. This applies to all 15 agent categories. Given that highly active and moderately active agents form 10.4\% and 22.2\% of the population (totaling $\sim30\%$) and their actions per week are 134 and 66, respectively, we expect highly active agents to follow $134 \times 0.66\% \times 4 = 3.5$ users and unfollow $1.8$ users over 28 days while moderately active users would attempt $1.7$ follows and $0.9$ unfollows, which is slightly lower than the frequency observed in \citep{kwak2011fragileonlinerelationship}.
\par The parameterization of activity levels and roles used statistics from \citep{antelmi2019characterizingbehavioralevolution} because it cannot be done using the mask-wearing dataset. Only tweet-level data were collected for the mask-wearing dataset; user-level data were absent. However, this does provide us an opportunity to validate a simulation using parameters from the dataset in \citep{antelmi2019characterizingbehavioralevolution} against an ``unobserved'' dataset (the mask-wearing dataset), as we shall see in Section \ref{sec:validation}, which is a more informative validation process than using the same dataset from which the parameters are drawn.

\subsubsection{Best approximations}
\label{sec:approx-params}
\par Certain parameters in our model require information that the existing literature does not provide and that we were unable to collect. The resistance to stance conversion is one such parameter. We opt to model the resistance as a truncated normal distribution with a low of 0.5, a mean of 0.75, a high of 1.0, and a standard deviation of 0.25. The size of a human agent's memory is fixed at $m = 30$, and this upper limit is used for three separate memory bins, which are the agent's own recent tweets, tweets by others that an agent has seen, and the author of those seen tweets. Bots have a memory of 120 and \texttt{poll\_tweet} returns 30 tweets each time. The number of themes that every human and bot agent can hold at once is five while the maximum percentage of new themes that an agent will adopt is 40\% for humans and bots. Humans adopt themes upon a successful infection, whereas bots adopt themes at every time-step. All of a user's tweets are assumed to propagate to all of the user's followers, although not all tweets are guaranteed to be seen as older tweets risk being pushed out of a follower's limited memory by newer tweets. The proportion of a user $u$'s followers who see a reply from another user $v$ to a tweet from $u$, which is the primary method of non-social tie information propagation in our model (there are other methods such as $u$ retweeting a seen tweet from a non-followee user), is set at 10\%, with $u$'s followers being chosen at random.
\par Every infection event by a stance matching a human agent's own increases the agent's resistance by 1\% while an infection by a non-matching stance decreases the agent's resistance, with the decrease being proportional to the distance between the stances, $-0.5\%$ for the closest neighboring stance and $-1\%$ for the stance on the opposite pole.
\par Stance compatibility is simply the distance between two stances divided by 2, $c_s = \sfrac{|s_{\text{user}} - s_{\text{tweet}}|}{2}$. Division by the maximum distance between stances, 2, ensures that the maximum value of $c_s$ is 1. Themes compatibility is the number of matching themes divided by the number of unique themes, $c_t = \sfrac{|t_{\text{user}} \cap t_{\text{tweet}}|}{|t_{\text{user}} \cup t_{\text{tweet}}|}$. Compatibility, $c = c_{s}w_{s} + c_{t}w_{t} $ where $w_{s}$ is the weight for stance compatibility and $w_{t}$ is the weight for themes compatibility. The default values for $\{w_{s}, w_{t}\}$ are $\{1, 0.8\}$ so having matching stances is more important than matching themes. Note that since incompatibility, not compatibility, in stances increases the chance of replying, the compatibility score when replying is calculated slightly differently, $c = ((1-c_s) \times w_{s}) + c_{t}w_{t} $. Weights are also modified when replying, $\{w_{s}, w_{t}\}$ are $\{0.8, 1\}$ to mildly de-emphasize stance compatibility, thus preventing replies from being exclusively an act of disagreement with a tweet's stance.
\par When deciding on which tweet to reply to, a human agent will prefer to reply to tweets that have the same themes but a different stance, giving less consideration to the amount of engagement that a tweet has received. When deciding which tweet to retweet, human agents will prefer tweets with matching stance and themes, with a tweet's engagement having slightly less weight on the decision. Quote-tweets emphasize theme compatibility over stance compatibility when choosing a target tweet to quote, and a slight importance is placed on compatibility over engagement just like when retweeting. The probability score of a tweet being liked, retweeted, quoted, or replied to, $p = gw_g + cw_c$ where $w$ stands for weight, $g$ for engagement, and $c$ for stance and theme compatibility. The weights $\{w_g, w_c\}$ are $\{0.5, 1\}$ when replying, $\{0.8, 1\}$ when retweeting and quote-tweeting, and $\{1, 1\}$ when liking. Engagement is a weighted sum of metric counts, $g = N_{\text{like}}w_{\text{like}} + N_{\text{RT}}w_{\text{RT}} + N_{\text{reply}}w_{\text{reply}}$. The weights used differ based on the action taken (e.g., in deciding whether one should like a tweet, the bandwagon heuristic would more pay more attention to the number of likes the tweet has instead of the number of replies). The weights for the engagement \textit{metrics}, $\{w_{\text{like}}, w_{\text{RT}}, w_{\text{reply}}\}$, for the engagement \textit{actions} of liking, retweeting, quoting, and replying are $\{1, 0.2, 0.2\}$, $\{0.2, 1, 0.2\}$, $\{0.2, 1, 0.2\}$, and $\{0.2, 1, 0.2\}$ respectively. Quote-tweeting and retweeting have the same weights as both actions expose one's followers to another's tweet. Both $g$ and $c$ are normalized by dividing over the values summed over all candidate tweets, e.g., for a tweet $x$, $g_{x} = g_{x}/\sum_{n=1}^{N_{\text{candidate}}}{g_{n}}$.
\par Bots differ from humans in that their probability for choosing a target, $p = g + c + f$, meaning that the probabilities are unweighted and an additional term $f$ has been added for the follower counts of the tweet authors. Like $g$ and $c$, $f$ is normalized. And instead of the probabilities being used in weighted random choice, they are used to sort the candidate tweets and the tweet with the highest probability is always selected. Bots also weigh each metric in engagement $g$ and the weights, $\{w_{\text{like}}, w_{\text{RT}}, w_{\text{reply}}\}$, for liking, retweeting, quoting, and replying are the same as those of humans.
\par Within the sub-component of the \texttt{read\_conversion} algorithm for computing a human agent's stance changes, the compatibility score, $r_{c}$, for each stance (e.g., $r_{c\text{-pro}}$) is the sum of the individual compatibility scores for the subset of an agent's seen tweets that exhibit the stance. The same applies to the engagement metrics' scores, $r_{g}$. The compatibility scores are normalized by dividing over the total score for messages of all stances. The engagement scores are normalized on a per-engagement type basis by dividing over the total of the frequencies for an engagement type. Prior to summation, the engagement metrics are weighted, $\{w_{\text{like}}, w_{\text{RT}}, w_{\text{reply}}\} = \{1, 0.5, -0.5\}$. The weighting reflects whether an engagement action is affirmatory or negatory in the context of stances. The summed compatibility scores and engagement metrics' scores are also weighted to avoid stance conversion from being entirely dominated by engagement metrics. The weight for compatibility is an agent's memory, $m$, while the weight for engagement is the reciprocal, $\sfrac{1}{m}$. As such, the final conversion score for a stance, e.g., pro, $w_{r-\text{pro}} =  m^{-1}r_{g\text{-pro}} + mr_{c\text{-pro}}$ where $r_{g\text{-pro}} = \sum_{x=1}^{N_{\text{pro}}} g_{x}$ and $r_{c\text{-pro}} = (\sum_{x=1}^{N_{\text{pro}}} c_{x}) / \sum{c}$. The engagement score for a tweet $x$ is $g_{x} = w_{\text{like}}N_{\text{like-}x}/\sum{N_{\text{like}}} + w_{\text{RT}}N_{\text{RT-}x}/\sum{N_{\text{RT}}} + w_{\text{reply}}N_{\text{reply-}x}/\sum{N_{\text{reply}}}$. The conversion scores are used as probabilities for selecting the stance which a human agent would convert to if its resistance threshold has been exceeded.

\section{Validation}
\label{sec:validation}
\par Validating information diffusion models can take the form of showing that the model is able to generate statistical distributions similar to those in the real world, e.g., recreating the percentages of tweets that receive zero engagement and tweets that go viral by adjusting parameters \citep{averza2022evaluatinginfluencetwitter}, and matching curve shapes for message volume \citep{decgatti2013simulationbasedapproachanalyze}. Another method is confirming that the parameters used in the model match those found in the real world, e.g., Twitter usage inter-arrival time in the \texttt{twitter\_sim} model \citep{beskow2019agentbasedsimulation} by \citeauthor{beskow2019agentbasedsimulation}.
\par As explained in Section \ref{sec:real-params}, a large number of key parameters in our model such as the distribution of a user's actions, a user's degree of activity (cognate to inter-arrival time), and the distribution of user roles are taken directly from real-world data. The underlying graph structure on which the simulation is run on (Section \ref{sec:graph-generation}) and the user's dirunal schedule (Section \ref{sec:resolution}) are also based on real-world data. Our model would naturally reproduce inter-arrival time, action distribution, etc., that are close (though not perfect matches) to empirical observations. Therefore, we have ``validated'' our model in the same sense that \citeauthor{beskow2019agentbasedsimulation} have. 
\par
\begin{figure}
    \centering
    \includegraphics[width=1.0\textwidth]{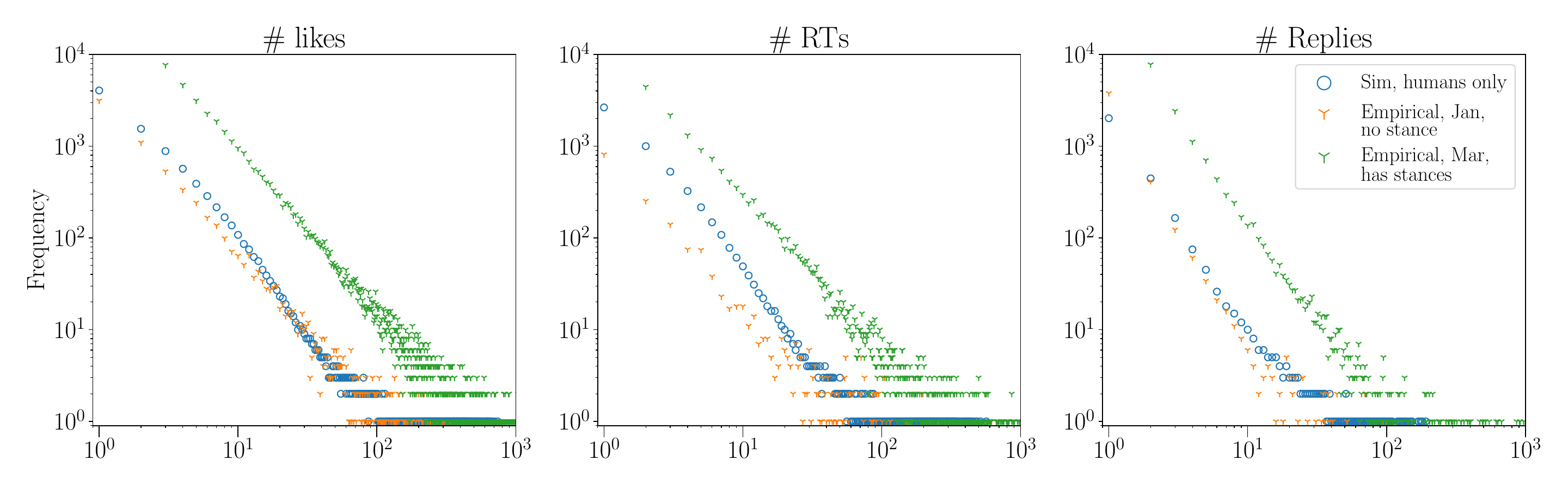}
    \caption{Engagement metrics: real-world data vs simulations with 1,000 agents.}
    \label{fig:validation-engagement}
\end{figure}

 For additional validation, we compared the engagement statistics obtained from simulations with those obtained from real-world data, namely the Twitter on mask-wearing conversations pertaining to the COVID-19 pandemic (Section \ref{sec:real-params}). Differences and similarities between statistics collected empirically and from the simulation are shown in Figure \ref{fig:validation-engagement}. The simulations are run for a period equivalent to 28 days and data was aggregated from 20 runs. Parameters for the agents are the same as those described in Section \ref{sec:params}. The configuration for the human-only simulation scenario is equivalent to the Sim 11 scenario described in Section \ref{sec:simulations}. 
\par There are two sets of real-world distributions from the mask-wearing dataset in Figure \ref{fig:validation-engagement} as we intend to illustrate how distributions can differ depending on how hotly a topic is contested. Each set of distributions is from the complete month-long period of the named month. The January 2020 data are from a less popular and less contentious period as it is composed of tweets that have been classified as out-of-topic/containing no stances, i.e., they are not offering any opinion on mask-wearing as it relates to COVID-19, but they are discussing masks in other contexts (e.g., protection against volcanic ash, fictional characters). In contrast, the March 2020 distribution is a subset of only tweets that have stances. March is also the period we have identified as when the mask-wearing conversation occured almost exclusively in the context of COVID-19.
\par The January distributions, from a subset where masks are a less contentious and less discussed topic, all contain lower values than the March distributions, i.e., shifted left. The distributions from the simulation in Figure \ref{fig:validation-engagement} nearly overlap the empirical January distribution for likes and replies. The simulation distribution is visibly shifted right compared to the January distribution when comparing retweets, although it is still bound to the right by the March distribution. For simulations, the within-dataset ratios $N_{\text{like}}:N_{\text{RT}}$ and $N_{\text{like}}:N_{\text{reply}}$ are on average 2.0 and 8.4 respectively, close to the numbers calculated directly from the data from which the simulation parameters are drawn (sans simulation), which are 2.1 and 8.5. If we had instead used a non-social network-specific graph generation algorithm (Section \ref{sec:graph-generation}), the discrepancy between the simulation and empirical ratios would have been starker, as the simulation ratios obtained would have been 1.7 and 5.2, validating our choice in the graph generation algorithm. January mask-wearing data shows ratios of 3.8 and 22.5, while March's ratios are 3.3 and 13.2. Retweets and especially replies are intensified when a topic is contentious (March), hence the lower ratios.
\par One factor involved in the discrepancy in distributions and ratios is the insufficient number of tweets at simulation start for agents to like, retweet, etc. We alleviated tweet scarcity by having an agent default to tweeting as an action if there are no incoming tweets in its memory and whenever a chosen action for a time-step is invalid. A consequence of this design choice is that the observed action distribution of agents would differ from the parameters they were given, as they tweeted more often than they should. However, the alternative of doing nothing if a chosen action for a time-step is invalid would have caused a similar issue of mismatch between parameters and observed distribution of actions taken, in addition to causing an agent to appear less active (due to skipped actions) and causing the starting phase of the simulation to last longer (due to tweet scarcity). We had considered seeding the simulation with a set of initial tweets but decided against it as the assumptions and ramifications --- heuristic for distributing the initial tweets to the nodes, the tweets causing stance changes in agents immediately at simulation start, etc. --- would complicate the analysis of the simulation results.
\par Another partial explanation for this difference is the size and structure of the networks. Simulations relied on a synthetic scale-free network as the graph structure in terms of the follower-followee network of the mask-wearing dataset is unknown to us. When compared with the January dataset, the across-dataaset ratio for the counts of likes, $N_{\text{like-Sim}}:N_{\text{like-Jan}}$, is 2.7, that of retweets, $N_{\text{RT-Sim}}:N_{\text{RT-Jan}}$, is 1.4, and that of replies, $N_{\text{reply-Sim}}:N_{\text{reply-Jan}}$, is 1.0. When compared with March, the respective ratios are 30.3, 18.0, and 19.2. If we infer the hidden size of the network through the engagement counts (hidden size because users in the mask-wearing dataset cannot represent the totality of network; metrics for a tweet can originate from a user outside the collected dataset), the number of users in January could be at least $2.7\times$ larger than in the simulation and for March, $20.3\times$ larger.
\par The across-dataset ratio of replies as well as the within-dataset ratios indicate that the simulations show a greater affinity to March than January data. For all engagement types, a constant multiplier of $\sim1.5\times$ reproduces March's ratios for within-dataset ratios, whereas while the across-dataset ratios for March's ratios are a constant $\sim20\times$ of the simulations except for $N_{\text{like-Sim}}:N_{\text{like-Mar}}$. January's multipliers differs depending on engagement type for across-dataset ($\sim2\times$ and $\sim3\times$) and within-dataset ($\sim3\times$, $\sim1.5\times$, and $\sim1\times$) cases. This makes the simulation a closer approximation of conditions where a topic is heavily debated, i.e., March's conditions. 
\par Overall, the distributions for the engagement actions strongly resemble real-world data, owing to the real-world statistics used to initialize simulation parameters and the robust SN-graph generation algorithm that also relied on empirical data. As the indirect influence from engagement actions and the metrics resulting from those actions are central to \textit{Diluvsion}, the findings in this section supports the model's validity.


\section{Simulations}
\label{sec:simulations}
\par We present a number of simulations, starting from the baseline scenario of bots representing two disadvantaged stances engaged in a co-opetitive influence operation and continuing with simulations where certain key functionalities of the bots are progressively altered/degraded, which can be thought of as ablation studies. Other scenarios tested the effect of varying the mixture of bot-backed stances. The results from the simulations are shown in Table \ref{tab:simulations} and we grouped our discussion of the results in Section \ref{sec:results} based on their implications for conducting real-world info ops. All scenarios featured 1,000 agents in scale-free networks and the results were averaged over 20 runs, with each run lasting 1344 steps (includes step 0, equivalent to 28 days). A bot's role determines its action distribution, i.e., the probabilities for liking, retweeting, etc. at each time-step. A role is named after the most frequent action the role takes, e.g., Likers like liking. Bot frequency is the number of times a bot is permitted to act per week. The definitions for both roles and frequencies (activity levels) are based on the Twitter data reported in \citep{antelmi2019characterizingbehavioralevolution}, and the definitions apply to bots and humans.
\par In classical influence maximization (Section \ref{sec:contested-models}), information that we wish to see widely propagated is injected via nodes that have been algorithmically identified as influential, i.e., positioned to maximize spread, but the bots in our simulations have no such advantage as their positions in the network are entirely random. Starting from the periphery of a network is the most disadvantaged position our bots are subjected to (Sims 6 and 7).
\par Inter- and intra-stance awareness of the presence of other bots is absent as the info ops is conducted in a decentralized fashion (Section \ref{sec:decentralization}). Competition is implied as each agent can only belong to one stance so an agent's conversion from one stance to another directly results in losing an agent to the other stance group; bots from one stance are not forbidden from converting agents belonging to a different bot-backed stance. Cooperation amongst agents arise naturally because agents gravitate towards performing affirmative actions, e.g., liking, for other agents that share similar a stance. Cooperation among two bot groups, each backing a different stance, can be implicit in that their uncoordinated and even competitive actions can accomplish a common goal better than they can individually, e.g., reducing support for a third non-bot-backed stance.
\par A high bot percentage, 10\%, is used in all scenarios so that the impact from bots is more easily apparent, not because this percentage is reflective of the actual bot population in the real world. Estimates of bot percentages on Twitter vary widely depending on the source, with Twitter's own analysis claiming 5.3\% in 2022, a 2016 Securities and Exchange Commission filing by Twitter claiming 8.5\%, Cyabra claiming 11\% in 2022, and a study on the 2020 Singaporean election claiming 26.99\% \citep{subrahmanian2016darpatwitterbota, uyheng2021activeaggressivelittle, duffy2022elonmuskcommissioned}. Bots defaulting to a high activity level also makes their influence more easily observable. Additionally, evidence exists that bots are prolific content creators on Twitter, with the Israeli web analytics company Similarweb finding that ``20\%--29\% of content in the US on Twitter is generated by bots'' \citep{varanasi2022twitterbotsappear}.
\par  To ensure that the overall stance ratio at simulation start matches the one at the start of March 2020 in the mask-wearing dataset, the stance ratio for humans is rebalanced to account for the bots' stance distribution. In other words, the number of human agents assigned to a particular stance is reduced if bots are assigned to that stance.
\par As Sim 1 is the baseline, Sim 1 is referenced by all other scenarios for comparison in the scenario descriptions below:
\begin{enumerate}
    \item Sim 1: Bots belonging to two minority stances, neutral and opposition, attempt to maximize the spread of their respective stances. The bot population is evenly split between neutral and opposition, and there are no bots in the camp of supporters, making the bot population's stance ratio $1:1:0$ ($\text{pro}:\text{neutral}:\text{con}$). This stance distribution enables us to observe synergistic and antagonistic effects between the two bot sub-populations. For all agent types, the number of themes an agent has as well as the distribution of the 12 themes are the same as in the March mask-wearing dataset. Bots start out embedded in the network, i.e., they have organic follower-followee relationships just as human agents do for reasons we have argued in Section \ref{sec:actors}. All the bots take on the role of a Replier to better demonstrate non-social tie spreading of information (Section \ref{sec:indirect-influence}). Bots are as active as highly active humans are. In choosing whose tweets to reply to, bots will favor those designated as vulnerable, which are people with a low number of information sources, i.e., low followee count. Bots will adopt commonly encountered themes to help the tweets they compose appeal to a wider audience.
    \item Sim 2: Bots will prioritize tweets that have high engagement metrics and that happen to also be authored by people with high follower counts when deciding whose tweet to reply to instead of targeting tweets from vulnerable agents as in Sim 1. 
    \item Sim 3: Bots are as active as moderately active humans rather than matching highly active humans' frequency.
    \item Sim 4: Instead of being Repliers, bots are Likers.
    \item Sim 5: Instead of being Repliers, bots are Retweeters.    
    \item Sim 6: Instead of being embedded in the network, each bot starts out isolated in the fringes of the network with only one followee relationship with a random human agent in the network (meaning that any tweet by the bot that is not a reply or quote-tweet will not be read by anyone in the network). This replicates the difficult starting condition that bots face at the start of an info-ops, as presumed by other researchers \citep{beskow2019agentbasedsimulation} but which we have argued against in Section \ref{sec:actors}.
    \item Sim 7: In addition to starting out in the fringes of the network as in Sim 6, bots take on the role of Likers instead of Repliers.
    \item Sim 8: Instead of being highly active Repliers, bots take on a variety of roles (Likers, Retweeters, etc.) with variable activity levels: low, medium, and high, just like human agents. The role and activity level distributions of the bots follow the distributions in \citep{antelmi2019characterizingbehavioralevolution}.
    \item Sim 9: Bots conserve a theme, Theme 7, instead of following the usual theme adoption process. Conservation involves a set of conserved themes that always have priority whenever the bot updates its list of themes to adopt and whenever the bot needs to decide which themes its tweet should be accompanied by.
    \item Sim 10: Just as in Sim 8, bots mimic human agents' roles and activity levels. Furthermore, the distribution of the bots' stances are the same as the humans' stance distribution from the March 2020 mask-wearing dataset.
    \item Sim 11: All agents in this scenario are humans, unlike other scenarios that contain 10\% bots.
    \item Sim 12: Bots are evenly distributed between supporters and detractors instead of being evenly distributed between neutrals and detractors, $1:0:1$ ($\text{negative}:\text{neutral}:\text{positive}$).
    \item Sim 13: Bot distribution follows a $1:0:0$ ($\text{negative}:\text{neutral}:\text{positive}$) ratio.
    \item Sim 14: Bot distribution follows a $1:1:1$ ($\text{negative}:\text{neutral}:\text{positive}$) ratio.
    \item Sim 15: Bot distribution follows a $2:2:1$ ($\text{negative}:\text{neutral}:\text{positive}$) ratio.
    \item Sim 16: Bot distribution follows a $4:0:1$ ($\text{negative}:\text{neutral}:\text{positive}$) ratio.
\end{enumerate}

\section{Results and discussion}
\label{sec:results}

\begin{table}[]
    \centering
    \resizebox{\textwidth}{!}{
        \begin{tabular}{lccccccccc}
        \toprule
         &  & Sim 1 & Sim 2 & Sim 3 & Sim 4 & Sim 5 & Sim 6 & Sim 7 & Sim 8 \\
        \cmidrule{3-10}
        Stance ratio (bot) &  & $1:1:0$ & $1:1:0$ & $1:1:0$ & $1:1:0$ & $1:1:0$ & $1:1:0$ & $1:1:0$ & $1:1:0$ \\
        Bot start &  & Embed & Embed & Embed & Embed & Embed & \textbf{Fringe} & \textbf{Fringe} & Embed \\
        Bot roles &  & Replier & Replier & Replier & \textbf{Liker} & \textbf{RTer} & Replier & \textbf{Liker} & \textbf{Human} \\
        Bot frequency &  & Hi-hum. & Hi-hum. & \textbf{Md-hum.} & Hi-hum. & Hi-hum. & Hi-hum. & Hi-hum. & \textbf{Human} \\
        Bot reply &  & Vuln. & \textbf{Engage} & Vuln. & Vuln. & Vuln. & Vuln. & Vuln. & Vuln. \\
        Bot adopts themes &  & Yes & Yes & Yes & Yes & Yes & Yes & Yes & Yes \\
        \cmidrule{3-10}
        Start/End N (Con) & 139 & 305 & 345 & 133* & 212 & 101* & 70* & 64* & 109* \\
        Start/End N (Neu.) & 129 & 568 & 551 & 400* & 391* & 159* & 74* & 67* & 124* \\
        Start/End N (Pro) & 732 & 127 & 105 & 468* & 397* & 740* & 856* & 870* & 767* \\
        \cmidrule{3-10}
        Start/End N (T1) & 158 & 232 & 223 & 277 & 246 & 236 & 224 & 404* & 248 \\
        Start/End N (T2) & 240 & 240 & 230 & 259 & 270 & 310* & 374* & 264 & 303* \\
        Start/End N (T3) & 549 & 288 & 317 & 329 & 360* & 430* & 592* & 589* & 439* \\
        Start/End N (T4) & 188 & 244 & 236 & 263 & 261 & 271 & 306 & 246 & 311 \\
        Start/End N (T5) & 34 & 204 & 195 & 189 & 184 & 181 & 126* & 155 & 180 \\
        Start/End N (T6) & 69 & 220 & 237 & 219 & 223 & 203 & 190 & 142* & 172* \\
        Start/End N (T7) & 1 & 60 & 56 & 87 & 36 & 34 & 20* & 15* & 54 \\
        Start/End N (T8) & 52 & 218 & 195 & 194 & 224 & 191 & 144* & 196 & 206 \\
        Start/End N (T9) & 26 & 193 & 221 & 186 & 176 & 172 & 126* & 101* & 151 \\
        Start/End N (T10) & 35 & 187 & 177 & 193 & 198 & 157 & 131* & 137 & 132* \\
        Start/End N (T11) & 26 & 194 & 207 & 184 & 185 & 180 & 108* & 125* & 144* \\
        Start/End N (T12) & 26 & 200 & 178* & 191 & 190 & 159* & 142* & 105* & 151* \\
        \bottomrule
        \end{tabular}
    }
    \resizebox{\textwidth}{!}{
        \begin{tabular}{lccccccccc}
        \toprule
         &  & Sim 9 & Sim 10 & Sim 11 & Sim 12 & Sim 13 & Sim 14 & Sim 15 & Sim 16 \\
        \cmidrule{3-10}
        Stance ratio (bot) &  & $1:1:0$ & \textbf{March} & \textbf{-} & $\mathbf{1:0:1}$ & $\mathbf{1:0:0}$ & $\mathbf{1:1:1}$ & $\mathbf{2:2:1}$ & $\mathbf{4:0:1}$ \\
        Bot start &  & Embed & Embed & \textbf{-} & Embed & Embed & Embed & Embed & Embed \\
        Bot roles &  & Replier & \textbf{Human} & \textbf{-} & Replier & Replier & Replier & Replier & Replier \\
        Bot frequency &  & Hi-hum. & \textbf{Human} & \textbf{-} & Hi-hum. & Hi-hum. & Hi-hum. & Hi-hum. & Hi-hum. \\
        Bot reply &  & Vuln. & Vuln. & \textbf{-} & Vuln. & Vuln. & Vuln. & Vuln. & Vuln. \\
        Bot adopts themes &  & \textbf{Consv.} & Yes & \textbf{-} & Yes & Yes & Yes & Yes & Yes \\
        \cmidrule{3-10}
        Start/End N (Con) & 139 & 320 & 25* & 12* & 196* & 760* & 115* & 213 & 540* \\
        Start/End N (Neu.) & 129 & 536 & 63* & 61* & 8* & 6* & 228* & 372* & 7* \\
        Start/End N (Pro) & 732 & 143 & 912* & 926* & 796* & 234* & 658* & 415* & 452* \\
        \cmidrule{3-10}
        Start/End N (T1) & 158 & 197* & 231 & 312 & 220 & 219 & 236 & 243 & 243 \\
        Start/End N (T2) & 240 & 218 & 347* & 367* & 246 & 237 & 248 & 229 & 241 \\
        Start/End N (T3) & 549 & 266 & 427* & 571* & 305 & 303 & 298 & 290 & 301 \\
        Start/End N (T4) & 188 & 214 & 283 & 355* & 220 & 232 & 218 & 238 & 231 \\
        Start/End N (T5) & 34 & 139* & 190 & 61* & 180 & 204 & 202 & 210 & 188 \\
        Start/End N (T6) & 69 & 181* & 182 & 151* & 198 & 205 & 214 & 221 & 200 \\
        Start/End N (T7) & 1 & 519* & 29 & 4* & 69 & 38 & 40 & 51 & 23 \\
        Start/End N (T8) & 52 & 175* & 184 & 138 & 196 & 190 & 204 & 206 & 194 \\
        Start/End N (T9) & 26 & 160* & 140* & 66* & 171 & 189 & 187 & 207 & 192 \\
        Start/End N (T10) & 35 & 152* & 153 & 56* & 195 & 181 & 182 & 212* & 194 \\
        Start/End N (T11) & 26 & 174 & 124* & 38* & 179 & 190 & 206 & 193 & 178 \\
        Start/End N (T12) & 26 & 157* & 144* & 81* & 175 & 191 & 192 & 173* & 188 \\
        \bottomrule
        \end{tabular}
    }
    \caption{The number of agents holding a stance and theme(s) at simulation end averaged over 20 runs for 16 scenarios with 1,000 agents. Agents can adopt a maximum of five themes. Stance ratio is $\text{negative}:\text{neutral}:\text{positive}$. An asterisk (*) indicates significant difference ($p<0.05$) compared to Sim 1. Starting number of agents for a stance or a theme is listed on the unnamed column on the left, between the column of row names and the named columns.}
    \label{tab:simulations}
\end{table}

\begin{figure}
    \centering
    \includegraphics[width=1.0\textwidth]{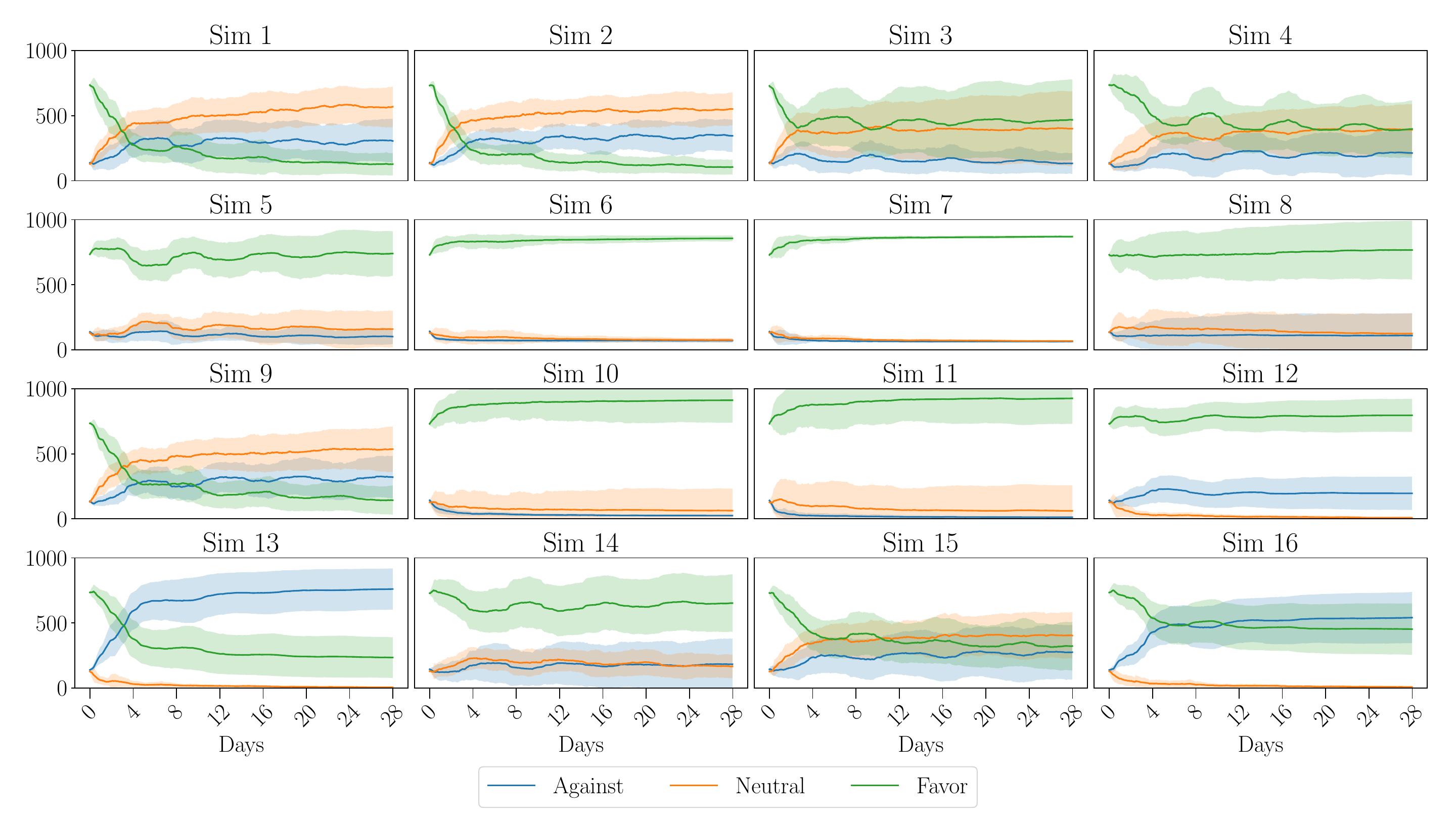}
    \caption{Temporal evolution of stance distribution among 1,000 agents. Values are from 20-run averages of different scenarios/simulations. Values at simulation end corresponds to the mean values reported in Table \ref{tab:simulations}. Descriptions of the scenarios are found in Table \ref{tab:simulations} and in Section \ref{sec:simulations}. }
    \label{fig:stance-time}
\end{figure}

\begin{figure}
    \centering
    \includegraphics[width=1.0\textwidth]{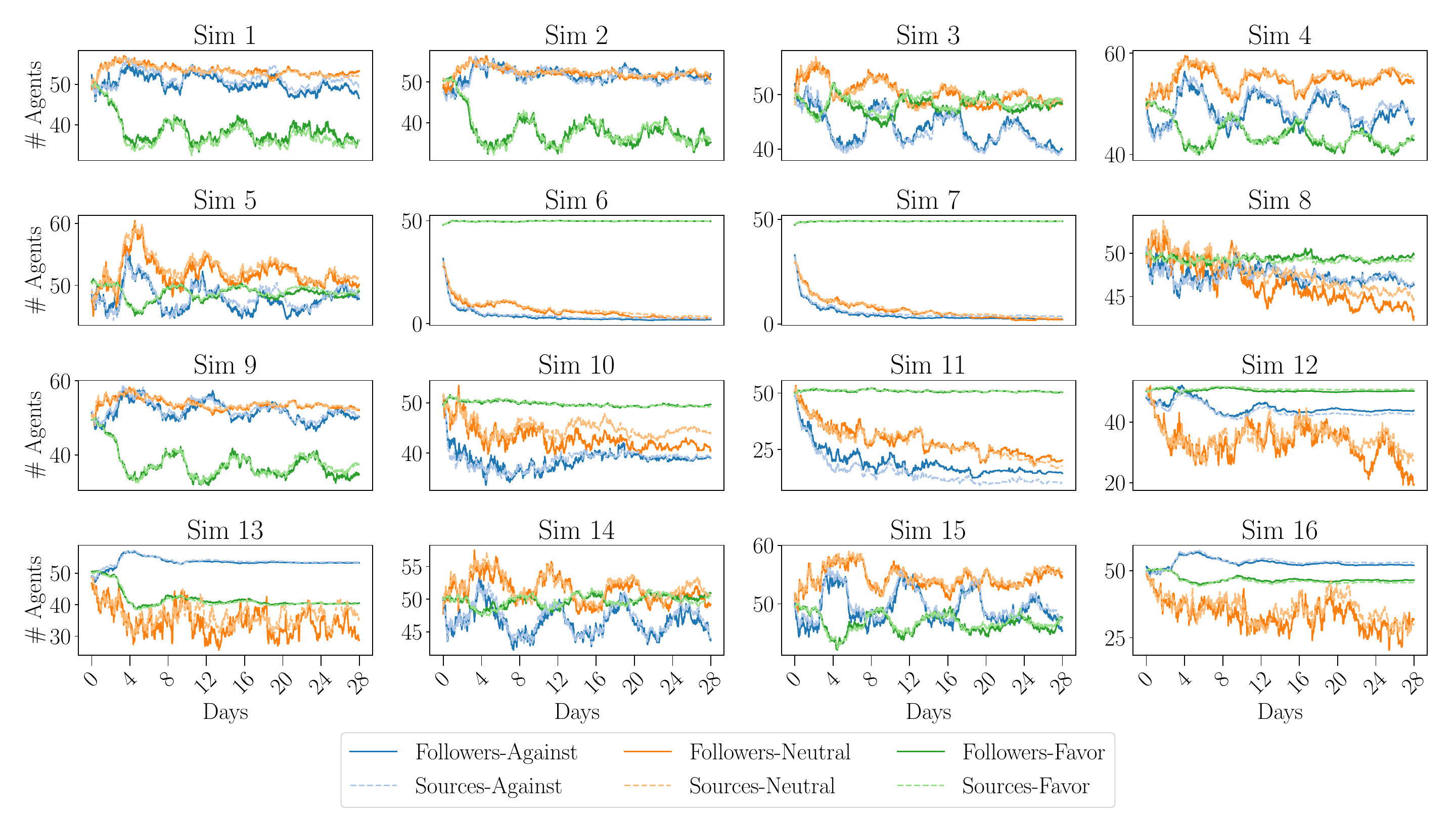}
    \caption{Temporal evolution of mean follower and followee (source) counts split by an agent's stance for 20-run averages of different scenarios/simulations. When an agent converts to a stance at a time-step, its follower and source counts are included in the average for the new stance and excluded for the old stance. Subplots do not share a common range for the y-axis due to significant variance across scenarios. Descriptions of the scenarios are found in Table \ref{tab:simulations} and in Section \ref{sec:simulations}.}
    \label{fig:follower-time}
\end{figure}

\sethlcolor{Black}

\begin{figure}
    \centering
    \includegraphics[width=1.0\textwidth]{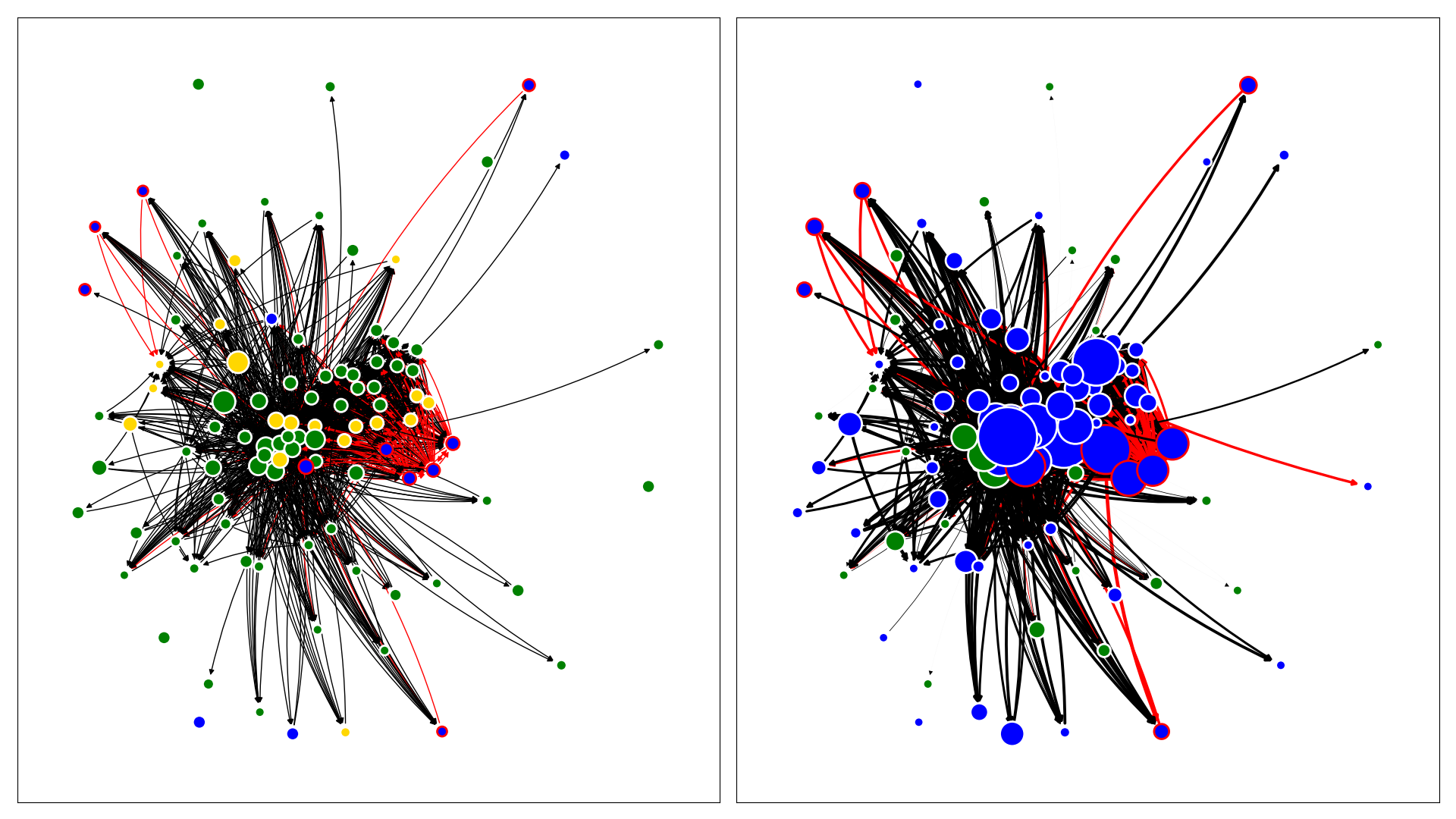}
    \caption{Embedded bots scenario. Each edge arrow represents the direction of information flow, i.e., the inverse of a directed follower-followee relationship, for a simulation with 100 agents in conditions equivalent to Sim 13 (Table \ref{tab:simulations}). In standard graph notation for an edge/link $(u, v)$, $u$ is a follower of $v$, so the follower-followee relationship is $u \rightarrow v$ but the information flow is reversed, $u \leftarrow v$. The image on the left shows the network at simulation start and has nodes sized according to the ratio of followers to followees. The image on the right shows the graph at simulation end and nodes are sized according to the average ratio of likes to replies received by the agent's tweets $\frac{1}{N} \sum_{N}^{i=1}\frac{\# \text{likes}_i}{\# \text{replies}_i}$, with higher values indicating that the agent's tweets tend to have a perception of being well-received. As likes and replies are accumulated over the course of a simulation, sizing the nodes based on engagement metrics is only possible post-simulation. The edge/link widths for the image on the left has a fixed value, while those on the right are based on stance and themes compatibility between two agents. Stance is indicated by each node's color. Agency is indicated in the edge/link color and the node's border color. \textcolor{Blue}{\textbf{Blue}} nodes are the opposition, \textcolor{Goldenrod}{\hl{\textbf{yellow}}} nodes are neutral, and \textcolor{ForestGreen}{\textbf{green}} nodes are supporters. A \textbf{\textcolor{Red}{red}} border indicates a bot while a \textcolor{White}{\hl{\textbf{white/invisible}}} border indicates a human.}
    \label{fig:flow-info-embed}
\end{figure}

\begin{figure}
    \centering
    \includegraphics[width=1.0\textwidth]{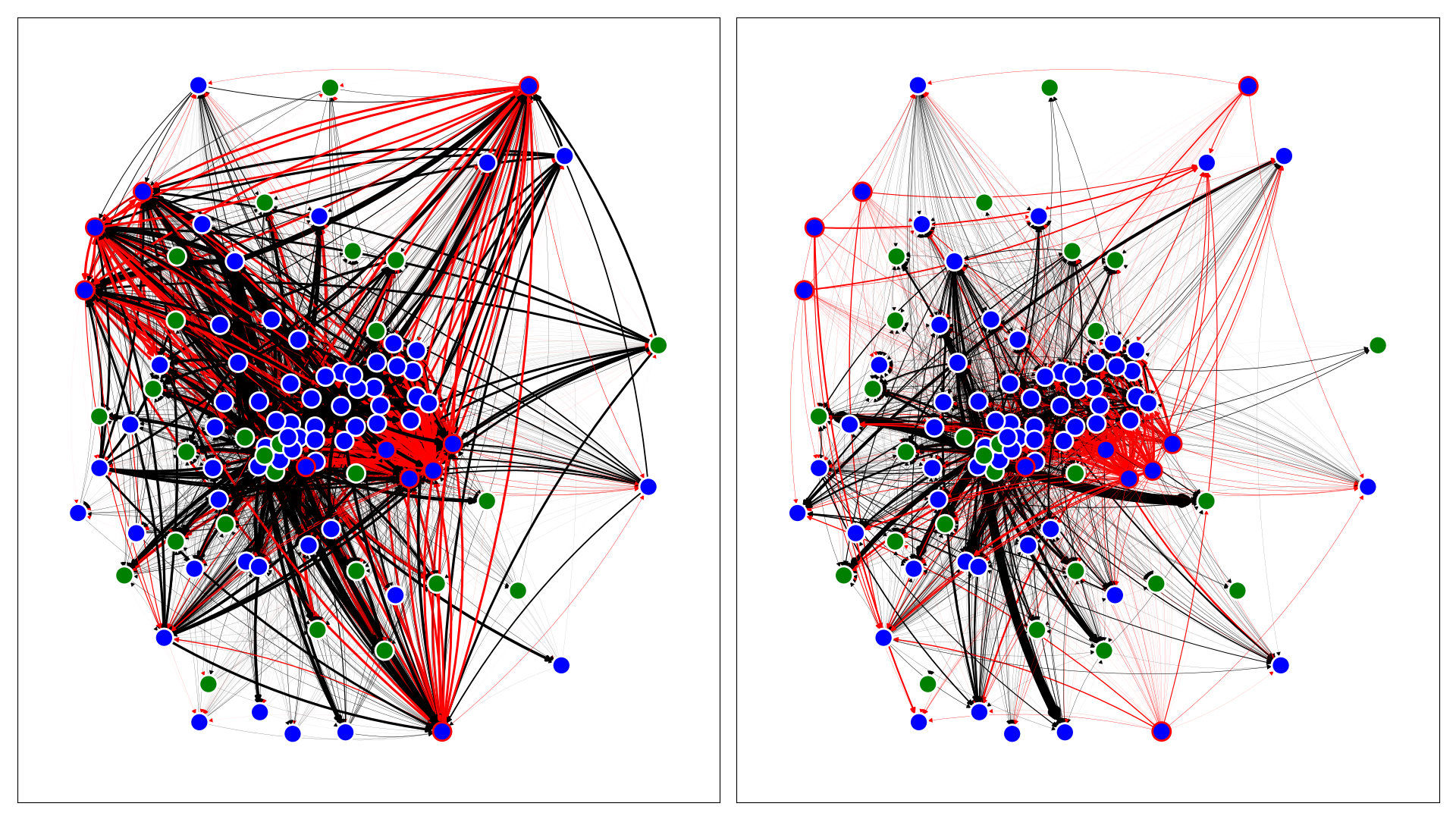}
    \caption{Embedded bots scenario. From the same simulation run as Figure \ref{fig:flow-info-embed}. The image on the left shows whose tweets a node has seen at the end of the simulation. The image on the right shows whose tweets contributed to a node's stance switching at the end of the simulation. For both images, the edge widths correspond to the frequency of the seen agent's (left)/the contributor's (right) appearance. Not all seen tweets led to stance conversion, hence the difference. The color of each node indicates its stance while the color of the node's edges indicates agency. \textcolor{Blue}{\textbf{Blue}} nodes are the opposition, \textcolor{Goldenrod}{\hl{\textbf{yellow}}} nodes are neutral, and \textcolor{ForestGreen}{\textbf{green}} nodes are supporters. A \textbf{\textcolor{Red}{red}} border indicates a bot while a \textcolor{White}{\hl{\textbf{white/invisible}}} border indicates a human. The edge/link colors distinguish between influence originating from bots, indicated in \textcolor{Red}{\textbf{red}}, and from humans, in \textcolor{Black}{\textbf{black}}.}
    \label{fig:seenhist-infectedby-embed}
\end{figure}

\begin{figure}
    \centering
    \includegraphics[width=1.0\textwidth]{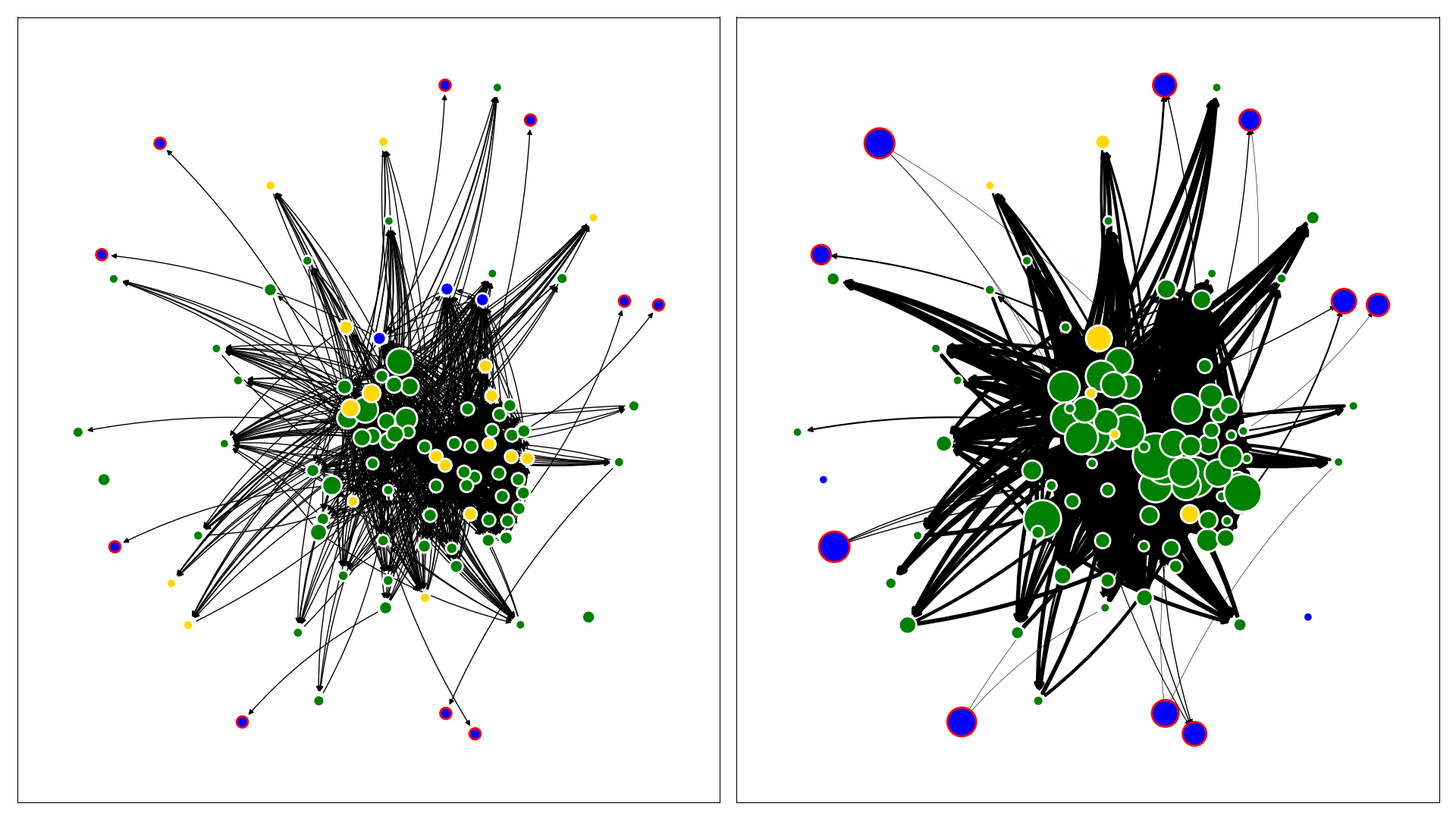}
    \caption{Peripheral bots scenario. Information flow for a simulation with 100 agents with all bots backing the negative stance (no equivalent in Table \ref{tab:simulations}, closest is Sim 6). This scenario differs from Sim 13 in Figures \ref{fig:flow-info-embed} and \ref{fig:seenhist-infectedby-embed} only in that each bot starts out in the periphery/fringes with a single outgoing edge (follower-followee) to a random human node instead of being embedded naturally in the network. This more challenging starting condition resulted in the non-bot-backed dominant opinion maintaining its dominance, although bots managed to convert two nodes in the fringes. Legend is the same as Figure \ref{fig:flow-info-embed}.}
    \label{fig:flow-info-fringe}
\end{figure}

\begin{figure}
    \centering
    \includegraphics[width=1.0\textwidth]{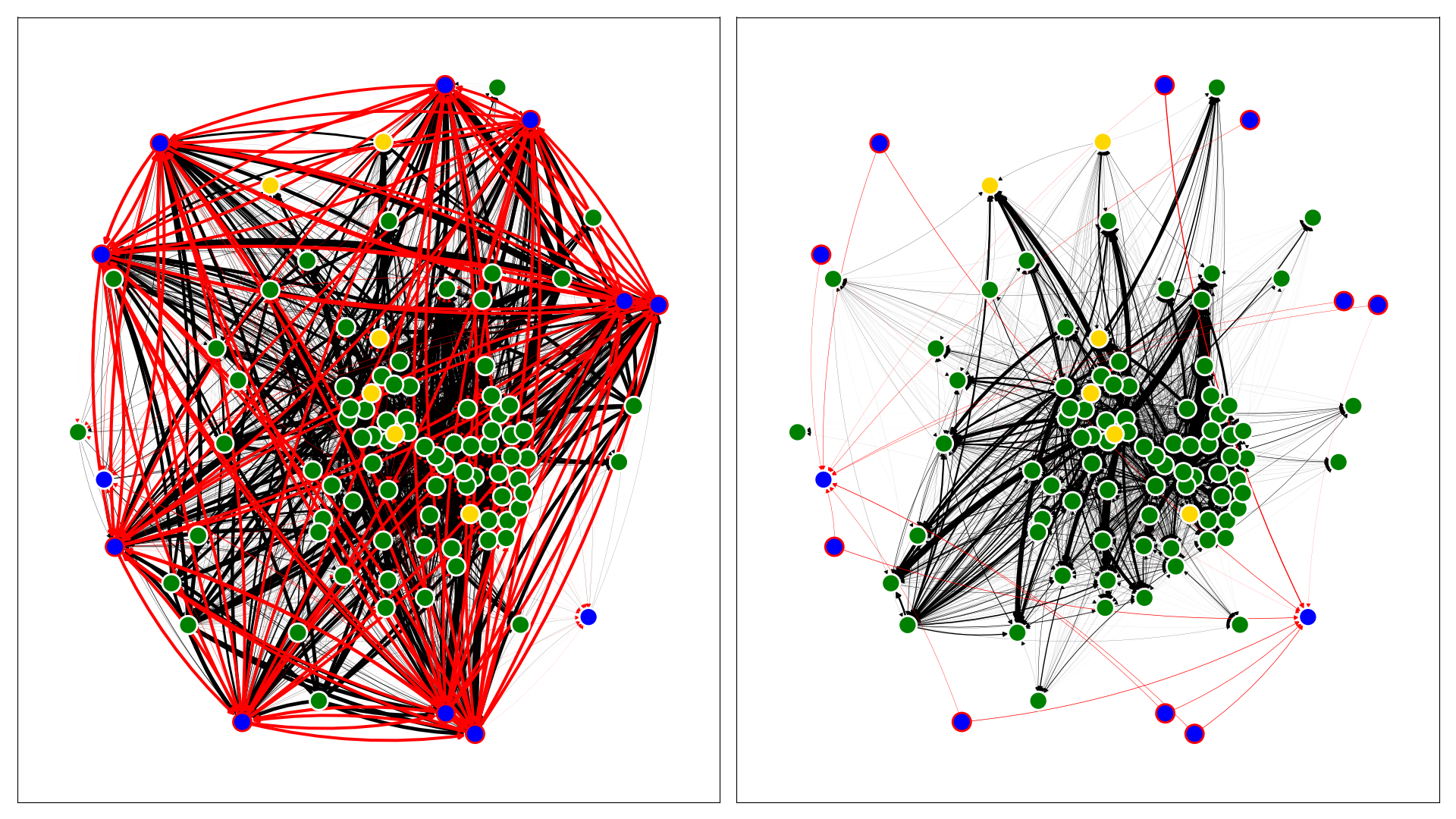}
    \caption{Peripheral bots scenario. From the same simulation run as Figure \ref{fig:flow-info-fringe}. Left image shows how often a different user's tweets were seen by an agent while right image shows how often a different user's tweets converted an agent's stance. Legend is the same as Figure \ref{fig:seenhist-infectedby-embed}.}
    \label{fig:seenhist-infectedby-fringe}
\end{figure}

\begin{figure}
    \centering
    \includegraphics[width=1.0\textwidth]{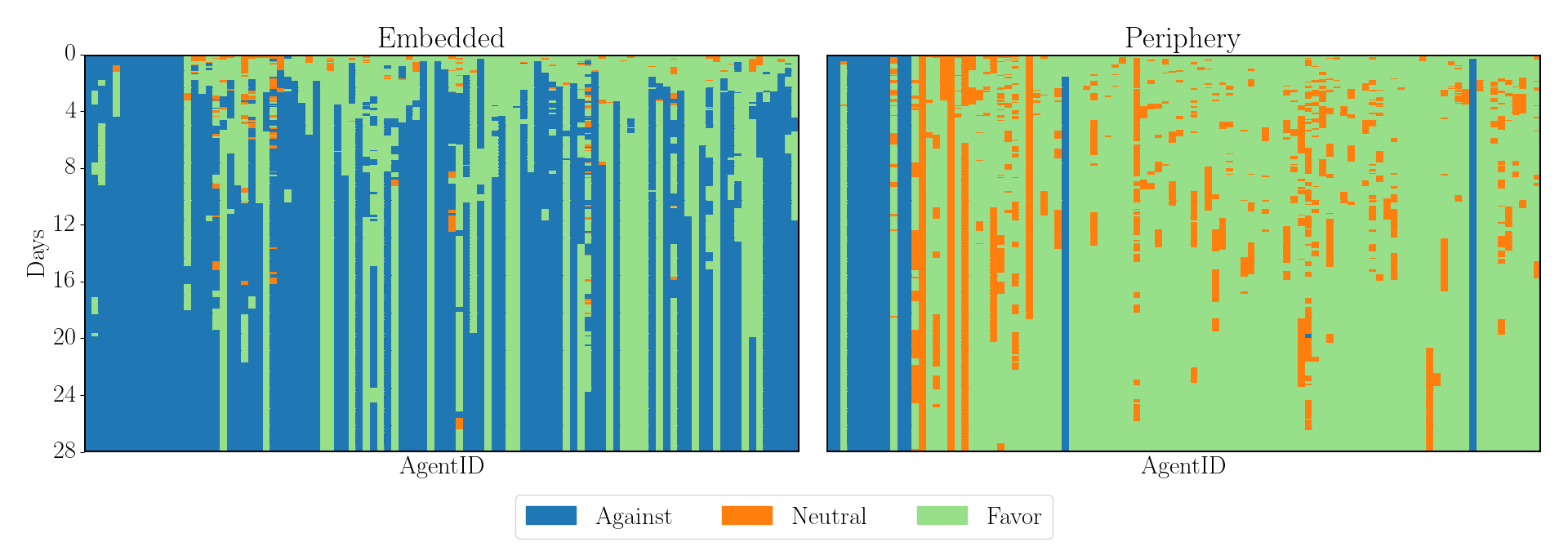}
    \caption{Each box in the heatmap represents the stance of an agent at a particular time-step. The evolution of an agent's stance can be tracked by following a column from top to bottom. Agents are sorted by their \textit{initial} stances, leftmost are negative agents and rightmost are positive, with neutrals in between. Rapid conversion at the early stage in the embedded scenario has rapidly eliminated neutral agents, hence their seeming absence. Data are from the same simulation runs as Figures \ref{fig:flow-info-embed}--\ref{fig:seenhist-infectedby-fringe}.}
    \label{fig:stance-evo}
\end{figure}

\par The simulation results are shown in Table \ref{tab:simulations}. Figure \ref{fig:stance-time} shows the temporal evolution of the stances for the simulations in Table \ref{tab:simulations} while Figure \ref{fig:follower-time} shows the temporal evolution for follower and followee counts, split by stances. Figures \ref{fig:flow-info-embed}, \ref{fig:seenhist-infectedby-embed}, \ref{fig:flow-info-fringe}, \ref{fig:seenhist-infectedby-fringe}, and \ref{fig:stance-evo} are for our discussion on the consequences of bots starting out embedded in the network versus its periphery (Section \ref{sec:result-embed-v-fringe}).
\par A note on the presentation of the results: not all models of information diffusion are strongly deterministic, which is why some authors, e.g., \citep{ross2019aresocialbots}, opted to instead show the number of simulations in which the bot-backed stance becomes dominant instead of showing the average number of agents per stance at the conclusion of a simulation as we have done in this paper. However, because our simulations all have a biased start where one stance is a clear majority that outnumbers the combined total of the minority stances by a factor of 2.7 to 1 (human agents' stances are evenly split in \citep{ross2019aresocialbots}) and bots can back more than one stance per scenario (bots all back the negative opinion in \citep{ross2019aresocialbots}), reporting our results as percentages of runs in which the bot-backed stance dominates is not viable.
\par When we speak of a ``majority'' or a ``minority'' stance, we are referring a stance's share among the agent population at the \textit{beginning} of the simulation. 
\par Since \textit{Diluvsion} is designed to support studies on info ops, discussion of simulation results will be interwoven with discussion of their relevance to info ops.

\subsection{Polarization}
\label{sec:result-polarization}
\par Winner-takes-all, i.e., no polarization/unipolar polarization, is the natural trend of a network where a stance has a numerical advantage over the combined total from minority stances, as witnessed in Sim 11 and Sim 10. This trend is not meaningfully altered whether the network is populated entirely by human agents (Sim 11) or the network contains 10\% bots which are allocated similarly as the human population's stance distribution (Sim 10). The slightly higher number of agents belonging to the minority stances in Sim 10 is likely a consequence of the bots' presence, whose stances can never be changed, slowing but not halting the network from reaching a consensus around the majority opinion. These are only trends, and not necessarily the final outcomes, as the simulations need to be run longer to observe whether all agents converge to a single opinion. And in the case where bots populate the simulation, Sim 10, the network will never fully converge as bots will never change their stance.
\par A winner-takes-all result \citep{prakash2012winnertakesall} has been found for networks with competing viruses where one virus is stronger than the other, and the strength of the virus is quantified by a single numerical parameter (derived from attack and recovery rates). While the viruses (opinions) in our model are equal in that they do not have an explicit parameter that offers one virus an advantage over others, a combination of factors enable the dominant stance in our work to replicate a similar advantage over competitors. The dominant stance in \textit{Diluvsion} is the stronger virus because homophily dictates that the stance has a more receptive audience, and their greater number means that their opinions are broadcasted and boosted more often, making others more susceptible to succumbing to peer pressure and adopting the majority stance.
\par Polarization can be induced if bots are present, and the stance distribution of bots differs from that of the humans. When half of the entire bot population holds a negative stance and the other half a neutral stance while the bots are restricted to human-like role and activity level distributions (Sim 8), the encroachment of the majority opinion is mostly halted. In Sims 12 and 16, where bots are only distributed between negative and positive stances with none for neutral, assigned a role more suitable for stance propagation (Replier), and are all as active as highly active humans, neutrals virtually disappear and agents congregate around the two polar opinions. In the case of Sim 13 where all agents start out belonging to only one minority stance, specifically the polar minority stance, a similar depopulation of neutral agents is observed. More time-steps are needed to determine whether Sim 12 ends with this polarized distribution or if all agents are eventually converted to the bot-backed negative stances.
\par In existing studies on binary opinion dynamics, consensus cannot be reached if stubborn agents are present in the network \citep{acemoglu2012opinionfluctuationsdisagreement, yildiz2013binaryopiniondynamics}. When the problem is extended to account for committed/stubborn agents and for multiple opinions (greater than binary), only opinions that are backed by committed people survive under the voter model \citep{hu2017competingopiniondiffusion}. The results from our simulations serve to further confirm the role played by stubborn agents, bots in our case, in driving polarization in a network.

\subsection{Impact of bots and bot ratios}
\label{sec:result-bot-impact}

\par Given that the distribution of roles and action frequencies among bots do not differ between the stances (i.e., bots all have the same role and frequency regardless of the stance that they back), the ratio of bots between the stances has a drastic impact on the temporal evolution of the stances, which can be seen in their impact on polarization in (Section \ref{sec:result-polarization}). If the stance ratio of bots follows that of humans, then the bots backing the majority stance would suppress the bots backing the minority stances, reproducing a dynamic very similar to the scenario where only humans presents while bots are absent. The bots' stance ratio must deviate from that of the humans to cause disequilibrium.
\par The co-opetitive relationship between the bots backing a minority polar stance (negative) and those backing a minority midpoint stance (neutral) benefits the midpoint stance the most. Across all scenarios in which the two minority stances are backed by an equal number of bots (Sims 1--9, 14, and 15), the number of neutral agents is always greater than that of negative agents. We suspect that this is due to the neutral agents' stance distance being the shortest for both polar stances (except for the distance between a polar stance and itself). Once neutral agents gain a critical mass through conversion, they have the numbers to peer pressure other agents in the network to adopt their opinion. But the strength of neutral agents is also their greatest vulnerability. Without agents immune to stance conversion, i.e., bots, in their midst, the group of neutral agents collapses quickly as they are the easiest targets to be converted by the majority stance (which does not need to be bot-backed) and the bot-backed minority stance. This phenomenon is evident in Sims 12, 13, and 16. 
\par Depending on the info ops goal of the bots' operators, one mixture of bots can be more optimal than the other. If the goal is to maximize the number of agents holding a stance that is the polar opposite of the majority, all bot resources should be devoted to backing that stance, as shown by the results of Sim 13. The cost of not having neutral bots is a slight inefficiency in reducing the number of people holding the majority stance as seen in the greater number of agents being positive at simulation end in Sim 13 compared to Sim 1. However, if one intends to purely reduce the share of the majority stance, Sims 1 and 2 show that having an equal number of negative and neutral bots is the most effective. The neutral agents' smaller stance distance with positive agents makes them more persuasive. A side effect of a bot operator employing the tactic of mixed neutral and negative bots is that the population of negative agents is ``cannibalized'', a consequence of the bots in our simulations acting independently to avoid being seen as a centrally-directed, coordinated info ops. In cases where the positive stance is also backed by bots, using a mix of neutral and negative bots to counter the positive bots (a $\text{pro}:\text{neutral}:\text{con}$ ratio of $2:2:1$ in Sim 15) still appears marginally better than relying purely on negative bots ($4:0:1$ in Sim 16).
\par If the goal is to drive polarization, bots can be assigned to both ends of the pole at equal ratios (Sim 12) or with a ratio that favors the minority polar stance (Sim 16). Without unwavering advocacy for neutrality, neutral agents quickly dwindle to nothing, as polar opinions compete to attract users to their stance. In situations where there are motivated neutral agents, they must be outnumbered to be overpowered. As Sim 10 shows, if the bots' starting stance distribution matches that of the humans from the March mask-wearing dataset, which implies that the numbers of bots representing the pro and neutral camps are roughly equivalent in most runs, the polar minority stance, negative, diminishes in size far more rapidly than the midpoint minority stance, neutral.
\par For stances starting out from a disadvantaged position (negative and neutral at 13.9\% and 12.9\% of the population), Sims 15 and 16 show that the combined number of bots backing the non-majority stances requires an approximate 4-to-1 numerical advantage over the bots backing the majority stance in order to reach influence parity, assuming bots for all stances are similarly capable (same roles and activity levels). Should bots only back the polar stances in equal numbers while neutral bots are absent, the majority stance maintains and extends its dominance, nearly eliminating neutrals but is unable to penetrate the enclave of the non-neutral minority stance (Sim 12). The source of the negative agents' resilience against conversion can potentially be explained by Figure \ref{fig:follower-time}, where the follower and followee counts for the bot-backed negative stance in Sim 12 remained steady as the simulation progresses (so did the counts for the bot-backed positive stance). The neutral stance which lacked bot backing showed greater volatility and trended downwards. Bots backing only one minority stance against the onslaught of the majority stance, which can be seen in Sim 13, also stabilizes the follower and followee counts of the minority stance. Bot-backing may be capable of creating echo chambers for minority stances.
\par Although echo chambers/filter bubbles are terrible for the purpose of gaining new recruits as the Islamic State's experience has shown (Section \ref{sec:result-stance}), echo chambers are excellent for the reinforcement and consolidation of opinions \citep{awan2017cyberextremismisispower}. A bot's broadcasted messages can reach people sharing the same stance, allowing bots to drive stance retention, i.e., keeping users with the same stance from being lost to other stances because those users are exposed to messages containing alternative stances. Bots play much of the same role as stubborn users in other information diffusion models; their unchanging stance making them ideal as the resilient core of a movement. Some have attributed Trump's electoral success to the presence of botnets ``spreading pro-Trump content in overwhelming numbers'' and ``digital echo chambers where users see content and posts that agree only with their preexisting beliefs'' \citep{difranzo2017filterbubblesfake}. Using bots to create echo chambers may even escape alerting social media monitors as certain platforms are already inclined towards echo chamber formation, with researchers finding that ``information diffusion is biased towards individuals who share a similar leaning in some social media, namely Twitter and Facebook'' compared to Reddit and Gab \citep{cinelli2021echochambereffect}.
\par When great political orators across the entire American political spectrum, e.g., Hillary Clinton \citep{lee2017hillaryclintonus} and George W. Bush, expressed sentiments along the lines of ``you are either with us, or against us'', that there is no middle ground, their words contain a kernel of truth. From the perspective of eroding the support base of a polar opinion (e.g., pro-war), the actions taken by neutrals and the opposition (e.g., situationally pro-war/war-ambivalent and anti-war) to draw people into supporting their respective stances have the same effect --- erosion. All our simulations featuring a $\text{pro}:\text{neutral}:\text{con}$ stance ratio at $1:1:0$ for bots provide evidence that support for the positive stance is inevitably eroded (Sims 6 and 7 are not exceptions as they are just examples of bots being ineffectual, as discussed in Section \ref{sec:result-embed-v-fringe}). In other words, if a predominant force supports one pole of a polarized issue, appealing to others not to pick either side can make one a collaborator for the ``enemy'' (supporters of the opposite pole). The slightest deviation from full, enthusiastic support of a stance is indeed ``treachery''.

\subsection{Stance conversion and retention strategies}
\label{sec:result-stance}
\par Repetition is the key to success for info ops. Repetition in this case means being extremely active in disseminating and boosting content. If the action frequency of the bots is cut in half, from 134 actions per week to 66, as seen in Sim 1 vs. Sim 3, then bots are substantially less effective at promoting minority stances against a non-bot-backed majority stance. This reduction in influence is also evident in other scenarios where bots went from being highly active to adopting activity levels that mimic the human distribution (low activity and medium activity users make up 67.4\% and 22.2\% of the human population respectively) which can be seen in Sim 1 vs. Sims 8 and 10. Our observation that repetition is important has support, e.g., simulations in \citep{kaligotla2020diffusioncompetingrumours} which showed that high energy agents contributed towards the prevalence of extreme opinions (polarization). Another study demonstrated that repetition can turn into influence: a bot in a social media experiment that had gained popularity simply by periodically visiting others' profile pages later recommended accounts for users to follow and succeeded in convincing a majority of them to take its recommendation \citep{aiello2012peoplearestrange}.
\par Not all forms of repetitive action elicit the same responses from the network. Sims 4 and 5 show that bots who favored liking or favored retweeting over all other actions are unable to convert as many users to their stance compared to bots which favored replying (Sim 1) despite having the same high activity level. In the case of Sim 5, the bots were only able to stop the majority opinion from gaining ground. There are several factors behind these outcomes. In a situation where the bots represent the minority stance, there are a greater number of tweets that express the majority stance and consequently more targets for negatory replies. On the other hand, bots that favor liking and retweeting, which are affirmatory actions, are likely to have run out of minority stance tweets to boost. As a small percentage of the population, bots are unlikely to occupy a majority of high-degree nodes, limiting the reach of their messages. Replier bots can sidestep the lack of an audience by borrowing the audience of others as a small portion of replies is seen by the followers of the parent tweet. Retweets, however, are only seen by the author of the parent tweet (and the bot's own followers). Liker bots do not have to find an audience as they are liking other users' tweets that already have audiences, namely those other users' followers, but the reach of liker bots is largely confined to these audiences as they seldom invade the conversations of other users compared to Replier bots. Since minority stances are minorities population-wise and Liker bots who back a minority stance will favor liking tweets advocating the minority stance, the audience of tweets liked by Liker bots are small too, although not as small as that of Retweeter bots.


\par Targeting vulnerable users --- friendless people, i.e., having a low number of followees --- for replies can result in a similar number of stance conversions as the alternative tactic of finding the largest platform to shout from (replying to tweets with high engagement), as demonstrated by Sim 1 vs. Sim 2. Also, in terms of decreasing the number of people holding the majority stance, both Sim 1 and Sim 2 are the most effective of all the tested tactics (Sim 9 will be discussed in Section \ref{sec:result-theme}). Our results are in accord with real-world observations. Bots finding success by targeting vulnerable users bears some similarities to how ``extremist'' groups have historically targeted loners, marginalized and ostracized groups, and other societal outcasts for recruitment/radicalization owing to these people's greater receptiveness towards extremist messaging \citep{torok2013developingexplanatorymodel, hales2018marginalizedindividualsextremism}. The tactic of maximizing audience reach is also favored by extremist groups, as seen in the Islamic State (IS) ``fighting for its propaganda to appear on search indexes and continued to launch `media invasion raids' to gain an even short-lived visibility on mainstream social media'' due to IS's favored social media platform, Telegram, having ``limited audience reach and echo-chamber effect, with the IS message being shared mostly among like-minded individuals'' \citep{weimann2022globaldigitalarenas}. Another example are BTC scam bots \citep{mazza2022readytousefake}, which are known to raid the replies sections of other tweets to maximize visibility.

\subsection{Themes conservation and propagation}
\label{sec:result-theme}
\par Themes are unique within our model in that agents do not guard against them unlike stances that have polarity, but their transmission is entirely dependent on the successful adoption of polarizable stances. Themes can improve the transmission probability of the stance that they are attached to if the targeted agent shares the same themes. Themes can therefore be thought of as having a symbiotic relationship with stances.
\par We tested a naive tactic for improving the propagation of themes --- saturation. By making all the bots conserve a particular theme (Theme 7), we aim to see if themes can spread beyond the bots themselves and into the human population. Conservation involves attaching the conserved theme(s) to every outgoing message from bots and also giving the conserved theme(s) a permanent place in a bot's set of themes. The results from Sim 9 show that not only are bots capable of making half the population of the simulation be infected by Theme 7, they were able to accomplish it without causing any appreciable change in the final distribution of the stances when compared to Sim 1, i.e., the bots still experienced similar success in stance propagation while conserving a theme. Because the number of agents holding the conserved stance, Theme 7, exceeds the bot population and an agent cannot hold duplicates in its theme set, we can conclude that the bots have succeeded in pushing the conserved theme onto human agents.
\par Even without an explicit theme conservation tactic, the bots' greedy strategy of keeping half of their own themes whenever they survey tweets for themes to adopt ensures that minority themes are given a boost whenever bots are involved in a simulation (Sims 1--16, except the humans-only Sim 11). In cases where bots failed to propagate their backed stances because the bots' starting positions are in the periphery of the network (Sims 6 and 7), the bots are still capable of elevating the presence of the initially obscure Theme 7. Bots can also suppress the growth of popular themes or even cause their decline, which can be observed by tracking the counts for the first and second highest frequency themes, Themes 3 and 2 across the simulations. The exceptions are when bots are exiled to the fringes, as in Sims 6 and 7, when bot activity levels are low, as in Sims 8 and 10, or when bots are completely absent, as in Sim 11, where a normative rich-gets-richer effect is at play where themes that were already popular mostly grew in popularity as the simulation progresses. 
\par Prior to obtaining our simulation results, our concern was that themes not matching fully will hamper stance adoption and therefore theme adoption; theme compatibility is  $c_t = \sfrac{|t_{\text{user}} \cap t_{\text{tweet}}|}{|t_{\text{user}} \cup t_{\text{tweet}}|}$, so deliberately including a theme that is virtually absent in the entire population would ensure that during the early phase of the simulation, the denominator of the theme compatibility equation will almost always be increased by one even in cases where all the non-conserved themes in a tweet matches with a user's theme. However, as the results from Sim 9 demonstrate, this is not a factor, at least not when the number of themes to be conserved is just one. The outcome might be different if more than one theme has to be conserved, as the reduced presence of compatible themes may be insufficient to compensate for the incompatibility arising from multiple conserved themes.
\par We theorize that engagement metrics, a high volume of replies representing a stance, and other effects that are within the ability of the bots in the simulation to manipulate are sufficient to overcome incompatibility in themes during stance adoption. Once a critical mass of agents are infected with the conserved theme, a synergistic effect begins to take hold. Human agents who have yet to be infected by the conserved theme find that their tweets are less effective at changing the stances of those who are infected with the conserved theme. This enables the number of agents with the conserved theme to grow without much interruption. These agents are also likely to share the same stance(s) as the bots because the conserved theme is almost exclusively possessed by the bots at simulation start, so acquiring that theme implies that the bots' stance(s) has infected the agent, at least in the early phase of the simulation. Once agents with the conserved theme become the majority, they can peer pressure others into accepting their stance. A theme can be viewed as performing a function similar to ``investments'' (sunk costs) that makes ``betraying''/leaving ``extremist'' groups \citep{altier2014turningawayterrorism, munden2023extremistgroupexits}, armed forces \citep{winslow1999ritespassagegroup}, religions \citep{davidman2007characterssearchscript}, and currency standards \citep{clark2005petrodollarwarfareoil, mastanduno2009systemmakerprivilege} difficult.
\par Bots being able to push a single theme without sacrificing progress on stance conversion, as results from Sim 9 demonstrated, is a non-trivial finding. There are implications for actors intending to conduct separate info ops on more than one issue. Such actors would ideally wish to succeed on all fronts without having to decide which info ops are the most important ones while making trade-offs and compromises for the rest. Our results show that not only can balancing acts be avoided, the actors can design an info ops in ways that can improve the outcomes of other related info ops. We illustrate this with an example concerning the issue of mask-wearing.
\par Under the guise of facilitating debate between the supporting, neutral, and opposing factions of mask-wearing, a motivated actor participating in more than one side of the debate via bots that always frame their messages with a theme of liberty (personal/individual freedom versus communal responsibilities and interests) is ensuring that the theme over time becomes the basis around which the issue is discussed as an increasing number of agents are infected by the theme. This mindset caused by the theme-infection can potentially be carried over into other issues, e.g., seatbelts, drunk driving, and climate change, causing the infected agents to frame those issues in terms of liberty. This is similar to the belief in the field of pedagogy that thinking habits can be taught and can be carried over across different situations (e.g., critical thinking \cite{seibert2021problembasedlearningstrategy}, thinking like an engineer \citep{lucas2016thinkingengineerusing}, fostering creativity in children \citep{ali2019canchildrenlearn}), exposing people who think in a certain way to cognitive blind spots \citep{oliveira2014itpsychologystupid} due to ``belief bias and the magnitude of framing effects'' \citep{west2012cognitivesophisticationdoes}. When others attempt to persuade these theme-infected agents to, for instance, not drink and drive by presenting arguments framed only in authoritarian terms (e.g., driving under the influence is against the law) or safety terms (e.g., drinking and driving often kills the driver), they would have difficulty dislodging the agent's stance.

\subsection{Embedded vs. peripheral bots}
\label{sec:result-embed-v-fringe}

\par Figures \ref{fig:flow-info-embed}, \ref{fig:seenhist-infectedby-embed}, \ref{fig:flow-info-fringe}, and \ref{fig:seenhist-infectedby-fringe} are visual illustrations of the differences between bots starting out embedded naturally within a network and bots starting out as outsiders with only one outgoing follower-followee link (i.e., the bots' tweets that are not replies will not be seen by anyone in the network). This challenging start is meant to mimic the ``time zero on the periphery'' with ``a single link to a random node'' start for bot disinformation maneuvers in \citep{beskow2019agentbasedsimulation}. Two other scenarios featuring bots in the periphery are Sims 6 and 7 in Table \ref{tab:simulations}. Although only 100 agents are used for plotting these graphs, the embedded scenario is representative of the 1,000 agent simulation results in Table \ref{tab:simulations} for Sim 13. The isolated/fringe scenario with 100 agents has no equivalent in Table \ref{tab:simulations}, but it can be considered a variant of Sim 6 where the stance ratio for bots is $1:0:0$ instead of $1:1:0$. The influence of bots is easier to see visually if the bots all belonged to one stance, hence the chosen configurations.
\par Embedded bots (Sim 13 and Figures \ref{fig:flow-info-embed} and \ref{fig:seenhist-infectedby-embed}) succeeded in converting a majority of the network population to its stance but the bots that started out in the network's fringes (Sims 6 and 7, and Figures \ref{fig:flow-info-fringe} and \ref{fig:seenhist-infectedby-fringe}) could only convert other users who also exist on the fringes. Cut off from the influence of negative bots existing on the periphery, the positive human agents in the central clusters rapidly converted the few neutral and negative human agents into positive agents (Figures \ref{fig:flow-info-fringe} and \ref{fig:seenhist-infectedby-fringe}, and Sims 6 and 7). Although bots did not actively initiate the severing of links --- they start the simulation already isolated from the central clusters --- the peripheral bots scenarios serve as an \textit{in situ} demonstration of the effects of a community (central clusters) shielding itself from outside influence (bots). The results from these simulations of ours are in accord with the model in \citep{sasahara2021socialinfluenceunfollowing} which produced predictions of social influence (homophily in opinion) and unfollowing being drivers behind the emergence of echo chambers, consistent with the empirical Twitter data they collected.
\par The fierceness of ideological competition is evident from comparing both halves of Figure \ref{fig:seenhist-infectedby-embed}, which represents the embedded bots scenario. While a large share of outgoing messages were read by users in the network, as indicated by the thick lines of the edges in the graph on the left, the messages that convinced another user to switch their stances are a smaller proportion of those that have been read, indicated by the thinner influence lines connecting nodes on the right. The same observation applies even more so to the peripheral bots scenario in Figure \ref{fig:seenhist-infectedby-fringe}. Bot messages lack the consistent persuasiveness of more centrally located humans, as the bots' red influence lines are comparatively thin next to the thick black influence lines emanating from humans. One shared trait of the winning sides in both the embedded and fringe cases is that their tweets receive a lot of affirmation, i.e., likes received outnumbered the replies, indicated by their greater node size in the right halves of Figure \ref{fig:flow-info-embed} (blue negative agents) and Figure \ref{fig:flow-info-fringe} (green positive agents).
\par Figure \ref{fig:stance-evo} shows that stance conversion happens early in both the embedded and fringe scenarios. While there are occasional bouts of rapid stance switches, most agents tend to hold a consistent stance for long periods of time. In the fringe scenario, there is a greater occurrence of positive agents converting to neutrality for short stretches of time in the early phase of the simulation. We speculate that this is because of negative bots having numbers that are sufficient to dampen positive tweets with their actions but are insufficient to boost negative tweets past the point where their stance incompatibility with positive agents is minimal, leaving neutrality to benefit from the struggle\footnote{An example with a positive agent undergoing stance conversion after having read a tweet each from negative, neutral, and positive stances: Recall that a stance's conversion score $w_{r}$ is made up of two components, engagement $r_{g}$ and compatibility $r_{c}$, $w_{r}=m^{-1}r_{g}+mr_{c}$. The $m^{-1}$ and $m$ terms are weights used to de-emphasize engagement and emphasize compatibility, respectively. Let themes be ignored in the compatibility calculation and assume that the two components will \textit{not} be weighted when summed. Assume that in the absence of bot-backing, the three tweets receive zero engagement. Assume that in the presence of bot-backing for the negative stance, the backing involves only actions that impact the engagement component of the conversion weight --- the positive tweet receiving one reply (assume that the reply was not seen by the agent experiencing this conversion event). For the non-bot backed case, $r_{g\text{-con}}=r_{g\text{-neu}}=r_{g\text{-pro}}=0$ and $r_{c\text{-con}}=0, r_{c\text{-neu}}=0.33, r_{c\text{-pro}}=0.66$. For the bot-backed case, $r_c$ remains the same but $r_{g\text{-pos}}=-0.5$ due to receiving a reply. Engagement components for negative and neutral remain the same, $r_{g\text{-con}}=r_{g\text{-neu}}=0$, because they received no engagement. The conversion weights without bot-backing are $w_{r\text{-neg}}=0, w_{r\text{-neu}}=0.5, w_{r\text{-pro}}=1$ and with bot-backing, they are $w_{r\text{-neg}}=0, w_{r\text{-neu}}=0.5, w_{r\text{-pro}}=0.16$. In percentages, the chances of a stance being selected, from negative to neutral to positive, are 0\%, 33\%, and 66\% for the case without bot-backing, and are 0\%, 76\%, and 24\% with bot-backing. By ``attacking'' positive tweets, the bots' actions ended up making neutrality a more attractive option. In the actual simulations, the $m^{-1}$ and $m$ weights are meant to ensure that the influence from engagement metrics does not easily overpower homophily for stance and themes.}. The outcomes for Sim 6 and Sim 7 in Table \ref{tab:simulations} provide support for this observation in Figure \ref{fig:stance-evo}. Despite the absolute counts at simulation end dropping for both negative and neutral, neutral has a higher count relative to negative despite starting with a comparatively lower count.
\par The peripheral bots' tactic of targeting low-followee accounts for replying and quote-tweeting is unlikely to be the cause of the bots' failure in combating the majority stance, because replying and quote-tweeting retweeting is limited to once each per tweet, so bots would switch to other targets once they have hit their interaction quota with agents on the fringes. For liking and retweeting, bots target tweets with high engagement written by agents with high follower counts. Incidentally, bots frequently liking other tweets to boost their infectivity appears to be a slightly inferior tactic compared to frequently replying to messages (Sim 6 vs. Sim 7). This difference between the Liker and Replier bots are less pronounced compared to the embedded scenarios (Sim 1 vs. Sim 4). 
\par The difficulty faced by bots under the fringe scenario in overturning the predominant stance can perhaps be attributed to the follow and unfollow mechanism in \textit{Diluvsion}. The probability for an agent to follow and unfollow accounts are based on data of user behavior on social media (Section \ref{sec:real-params}) that likely was collected under ``peacetime'', i.e., when a community is not vigorously debating a topic. Therefore, they may not reflect a greater tendency to cluster into echo chambers/hives during ``wartime'' by rapidly establishing social ties with like-minded users and severing ties with non-believers. However, as we aim to ground our model in as much empirical data as possible and we did not encounter data of heightened follow/unfollow behavior on ``war footing'', we opted to keep probabilities low. The model in \citep{beskow2019agentbasedsimulation}, \texttt{twitter\_sim}, strongly incentivizes the creation of new links, which explains the difference between their simulation results and ours, where they demonstrated that bots are successful in backing a stance and in bridging two separate communities despite bots starting in the fringes. Under \texttt{twitter\_sim}, link creation probability for humans is low at 5\% but is higher than our model's $0.\overline{6}\%$, the check for link creation is triggered every time a user wakes instead of being a separate event that consumes a user's action budget under our model (compounding the higher link creation probability), and links in \texttt{twitter\_sim} are never broken.
\par Figure \ref{fig:follower-time} shows the crippling effect of being deprived of an audience. Unlike other scenarios where the minority stances' mean follower and followee counts are at parity with the majority stance's at simulation start, they are only about three-fourths of the majority's in Sims 6 and 7. Once agents begin renouncing a minority stance for the majority stance, the minority stance's follower and followee counts begin experiencing a precipitous decline that they never recover from. 
\par Our results here are corroborated by the details of real info ops on social media that have emerged (Section \ref{sec:actors}), showing that great effort is invested in equipping many bots with extensive social ties, going as far as having real humans work under a bot. No matter how persuasive a bot is constructed to be, its effect is limited if it does not have a ready audience that is always exposed to the messages that the bot broadcasts. Bombarding the replies section of other users to persuade them and their neighbors into changing their stances cannot succeed if the bots are just 10\% of the population, as our simulations have shown. Resorting to other tactics to improve the performance of bots without increasing the number of bots, e.g., explicitly coordinating with other bots and increasing the frequency of activity, runs the risk of the info ops being discovered.

\section{Limitations and future work}
\par At the outset of this paper, we have stated that no model is perfect and that they are all reflections of some aspect of real life that the modelers wish to bring to attention. There is room for future work because we made several simplifying assumptions in our model. For instance, we accounted for the bandwagon effect but did not implement a mechanism for the spiral of silence. Our simulations start out with one dominant and two minority stances, which do not cover scenarios where the distributions are balanced. Some other assumptions involved picking one side of an ongoing debate (e.g., whether resistance increases with repeat infections) and not having the opportunity to explore the alternative side in-depth in this paper.
\par One promising route for future investigation is to allow a reinforcement agent to control the bots. The reinforcement learning agent AlphaStar has been demonstrated to perform at the Grandmaster level for the real-time strategy (RTS) game StarCraft II \citep{vinyals2019grandmasterlevelstarcraft}; an info ops problem bears strong resemblance to an RTS game. Just as playing an RTS game involves controlling groups of hundreds of units with specific roles (attackers, defenders, healers, etc.), so too can the managing of an info ops on social media involve controlling hundreds of bots with specific roles (Repliers, Likers, etc.). However, a central controller dictating the actions of multiple bots does violate the emergent/uncoordinated coordination condition that this paper is predicated upon and could potentially make this proposed form of bot-driven info ops easier to detect, so an alternative scenario where there are multiple reinforcement agents that each control one bot should also be explored.
\par Greater heterogeneity in terms of bots should also be investigated. While we have varied the distribution of roles and activity levels, we have kept other aspects homogeneous throughout a simulation run, such as the process by which a bot chooses a tweet to like. And when deciding how bots should act, e.g., the per-week frequency of actions such as retweeting, we have relied on pre-defined roles derived from empirical observations of Twitter user behavior. All this was done to facilitate comparison between scenarios. There may be more optimal action distributions for maximizing stance propagation. Heterogeneity would also enable us to explore more complicated info ops scenarios. An example would be explicit cooperation between neutral and opposition bots, where neutral bots would only target agents who are supporters for conversion to neutrality while opposition bots would target both neutral and supporter agents. In this example, opposition bots would effectively function as a conversion pipeline for opposition bots. Compared to opposition bots attempting to convert supporters into detractors, the lesser difference in stance between supporters and neutrals means that neutral bots would have an easier time converting supporters to neutrality. Once converted to neutrality, opposition bots can more easily convert the former supporters into agents of the opposition.
\par Agent heterogeneity should include class and role transitions, e.g., deepening radicalization turning human agents into bots and de-radicalization turning bots to humans. Bots switching roles as the info ops situation demands should also be included. Role transition in an ABM for information diffusion has been explored before in \citep{sobkowicz2021agentbasedmodel} where radicalized patients become staunch anti-vaccine propagandists. Our model's radicalization and de-radicalization involved only an agent switching stances. In addition, human and bot users entering and exiting the network should be accounted for. The dynamics of a social network in which the ratio between bots and humans is ever shifting may provide additional info ops insights.


\section{Conclusion}
\par In this paper, we developed and empirically validated an agent-based information diffusion model for bot-driven information operations on social media, \textit{Diluvsion}, that incorporated crucial but oft-neglected real-world aspects of diffusion. Chief among them is indirect influence, consisting of two forms, the exposure to a message's engagement metrics inducing bandwagon/conformity effects within a user and the ability to transmit information to another node in the absence of a social tie linking two nodes together. Other key innovations of the model are the designation of neutrality as an independent infection-capable stance instead of a transition point between the two discrete polar stances (opinions) and the introduction of the non-polarizable, homophily-inducing, infection-capable, and stance-symbiotic themes as a counterpart to stances.
\par Through simulations, we demonstrated the different routes by which motivated agents, i.e., bots, can exploit these aspects of information diffusion to manipulate the information environment. The simulations showed that highly active Replier bots representing minority stances are capable of overturning the dominance of a non-bot-backed majority stance, and that a naive saturation tactic can not only keep a minority theme, e.g., ``conspiracy'', alive but also further its adoption over time without compromising stance propagation. Neutrality is also shown to be non-neutral in its threat to the continued dominance of the majority stance. Overall, our model shows that info ops originating from agents embedded in the network who are not counterbalanced by a similarly motivated oppositional force can easily pollute the information environment and induce different outcomes, such as increased polarization.
\par Although much of our paper has been steeped in the language of war, the model we introduced here can be used towards beneficial, pacific ends. Reducing polarization on social media through the introduction of bots that stubbornly advocate a neutral stance, for instance, can be studied through \textit{Diluvsion}, in the tradition of the numerous works that have been published on depolarization, e.g., \citep{vinokur1978depolarizationattitudesgroups, moore-berg2020neuralpolarizationroutes, currin2022depolarizationechochambers, sobkowicz2023socialdepolarizationdiversity}. The assumption here is that polarization reduction is prosocial, an assumption that some may not agree with as they may believe that there is no peaceful coexistence with certain opinions or ideas, e.g., Nazism. On the other hand, there are real-life examples --- the US, for one --- of comfortable coexistence being possible between ``extreme'' ideologies and the ideals of liberal democracy, the best governing system ever and which Churchill, the widely venerated British prime minister who presided over the Bengal famine \citep{mishra2019droughtfamineindia, safi2019churchillpoliciescontributed, kuchay2019churchillpoliciesblame}, wryly described as ``the worst form of Government except for all those other forms that have been tried from time to time''. The US never lost its global status (global as in encompassing only the ``garden'' \citep{majumdar2022eutopdiplomat}) as the shining beacon of democracy despite being the direct progenitor of Nazism via cherished American ideals such as Manifest Destiny \citep{whitman2017hitleramericanmodel, miller2020nazigermanyracea} and despite fascism resurging in modern-day America \citep{davidson2006globalisationtheocracynew, robinson2012globalcapitalismtwentyfirst, bhatt2021whiteextinctionmetaphysical, richards2022ecofascismonlineconceptualizing}.
\par Or, perhaps, certain states, systems, ideas, etc., maintaining an untarnished reputation as a force for good are not due to their inherent merits but due to having waged successful info ops campaigns, and maintaining that reputation in the age of social media requires active effort in maintaining an edge in tools for wargaming info ops on social media, in tools such as \textit{Diluvsion}.

\section{Acknowledgments}

This research is supported by NSERC Discovery Grants (RGPIN-2018-03872), NSERC CREATE Grants (CREATE-554764-2021) and Canada Research Chairs Program (CRC-2019-00041).

\bibliographystyle{unsrtnat}
\bibliography{twitter-abm}

\end{document}